\documentclass[%
 reprint,
superscriptaddress,
 amsmath,amssymb,
 aps,
]{revtex4-1}

\usepackage{graphicx}
\usepackage{dcolumn}
\usepackage{bm}
\usepackage{subfiles}
\usepackage{float}

\begin{document}

\preprint{APS/123-QED}

\title{Latent dynamical variables produce signatures  of spatiotemporal criticality in large biological systems}

\author{Mia C. Morrell}
\altaffiliation[Current address: ]{%
Los Alamos National Laboratory, XCP-8, Los Alamos, NM 87545, USA
}%
\affiliation{%
 Department of Physics, Emory University, Atlanta, GA 30322, USA
}%

\author{Audrey J.\ Sederberg}%
\affiliation{%
 Department of Physics, Emory University, Atlanta, GA 30322, USA
}%
\affiliation{%
 Initiative in Theory and Modeling of Living Systems, Emory University, Atlanta, GA 30322, USA
}%


\author{Ilya Nemenman}
\affiliation{%
 Department of Physics, Emory University, Atlanta, GA 30322, USA
}%
\affiliation{%
 Department of Biology, Emory University, Atlanta, GA 30322, USA
}%
\affiliation{%
 Initiative in Theory and Modeling of Living Systems, Emory University, Atlanta, GA 30322, USA
}%


\date{\today}

\begin{abstract}
Understanding the activity of large populations of neurons is difficult due to the combinatorial complexity of possible cell-cell interactions. To reduce the complexity, coarse-graining had been previously applied to experimental neural recordings, which showed over two decades of scaling in free energy, activity variance, eigenvalue spectra, and correlation time, hinting that the mouse hippocampus operates in a critical regime. We model the experiment by simulating conditionally independent binary neurons coupled to a small number of long-timescale stochastic fields and then replicating the coarse-graining procedure and analysis. This reproduces the experimentally-observed scalings, suggesting that they may arise from coupling the neural population activity to latent dynamic stimuli. Further, parameter sweeps for our model suggest that emergence of scaling requires most of the cells in a population to couple to the latent stimuli, predicting that even the celebrated place cells must also respond to non-place stimuli. \\
\end{abstract}

\maketitle

A key problem in modern biological physics is extracting useful knowledge from massive data sets enabled by high-throughput experimentation. For example, now one can record simultaneous states of thousands of neurons \cite{Segev2004, Nguyen2015, Gauthier2018, Schwarz2014, Lin2020} or gene expressions \cite{Zheng2017, Cao2017, Gierahn2017}, or the abundances of species in microbiomes \cite{Martin2006, Palmer2007, Vega2017}. Inferring and interpreting the joint probability distributions of so many variables is infeasible. A promising resolution to the problem is to adapt the Renormalization Group (RG) \cite{Goldenfeld2018} framework for coarse-graining systems in statistical physics to find relevant features and large-scale behaviors in biological data sets as well. Indeed, recently, RG-inspired coarse-graining  showed an emergence of nontrivial scaling behaviors in neural populations \cite{Meshulam2019, Meshulam2018}. Specifically, the authors analyzed the activity of over 1000 neurons in the mouse hippocampus as the animal repeatedly ran through a virtual maze. Their coarse-graining scheme involved combining the most correlated neurons into neural clusters by analogy with Kadanoff's hyperspins \cite{Kadanoff1966}, while using cluster-cluster correlations as a proxy for locality. Various correlation functions of neural clusters exhibited self-similarity for different cluster sizes, suggestive of criticality. Further analysis inspired by Wilson's momentum space approach to renormalization \cite{Wilson1983}  revealed that the joint distribution of cluster activities flowed to a non-trivial, non-Gaussian fixed point. Mechanisms responsible for these behaviors remain unknown. Thus it is unclear which other systems may exhibit them.

Observation and interpretation of signatures of criticality in high-throughput biological experiments is a storied field \cite{Mora2010,Socolar2003, Nykter2008,Mora2010,Touboul2017, Barton2015, Chialvo2010}. As a specific example, one commonly observed signature is the Zipf's law, which describes a power-law relation between the rank and the frequency of a system's states. It has been explained by the existence of stationary latent (unobserved) fields (such as stimuli or internal states) that couple neurons (spins) over long distances \cite{Latham2016, Schwab2014}. Similarly, here we show that the observations of Ref.~\cite{Meshulam2018}, including scaling properties of the free energy, the cluster covariance, the cluster autocorrelations, and the flow of the cluster activity distribution to a non-Gaussian fixed point can be explained, within experimental error, by a model of non-interacting neurons coupled to latent {\em dynamical} fields. This is the first model to explain such a variety of spatio-temporal scaling phenomena observed in large-scale biological data.

Below we introduce the model, implement the coarse-graining of Ref.~\cite{Meshulam2018} on data generated from it and compare our findings with experimental results. We conclude by discussing which other experimental systems may exhibit similar scaling relations under the RG procedure.

{\em The model. ---}
To understand how scaling relationships could arise from coarse-graining data from large-scale systems, we study a model of $N$ binary neurons (spins) $s_i \in \{0, 1\}, i \in [1,N]$, where $s_i=0$ or $1$ corresponds to a neuron being silent or active. The neurons are conditionally independent and coupled only by $N_{\rm f}$  fields $h_{m}(t)$, $m \in [1,N_{\rm f}]$ such that the probability of a population being in a certain state $\{s_i\}$ is 
\begin{equation}\label{eq:prob}
    P(\{s_i\}| \{h_m\})=\frac{1}{Z(\{h_m\})}e^{-H(\{s_i\}, \{h_m\})},
\end{equation}
where $Z$ is the normalization, and $H$ is the ``energy'': 
\begin{equation}\label{eq:energy}
    H=\eta\left[\sum^{N,N_{\rm f}}_{i,m=1} h_m(t)W_{im}s_i + \epsilon s_i \right].
\end{equation}
Here $\epsilon$ is the bias toward silence, $\eta$ controls the variance of individual neuron activity, and $W_{im}$ are the coupling constants that link neurons to fields. The model includes two types of fields (place and latent), explained below. 
 
\begin{figure*}
\includegraphics[width=1.0\textwidth]{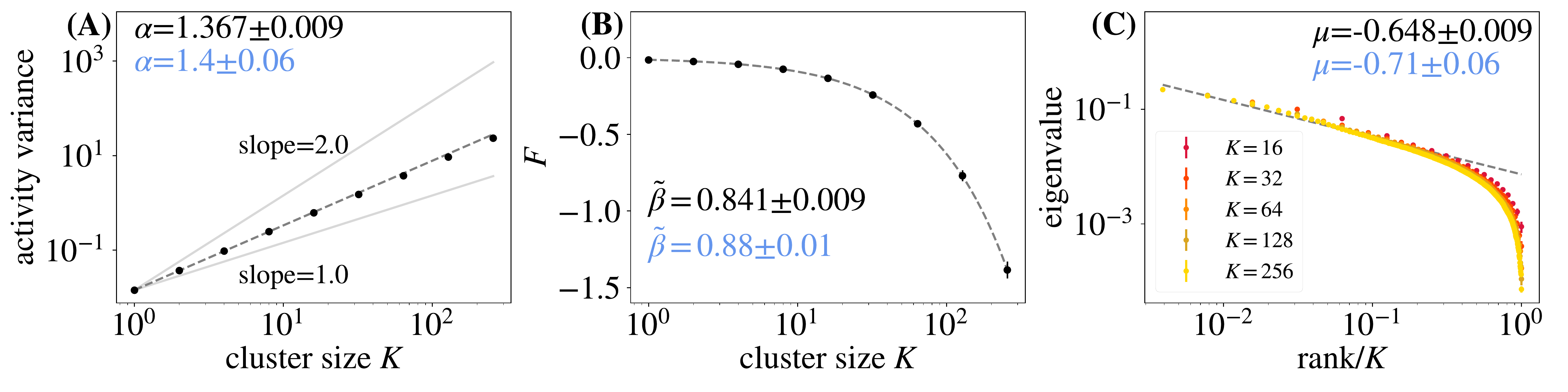}
\caption{\label{fig:fig_1} \textbf{(A)} Activity variance of coarse-grained variables at each coarse-graining iteration, fit to $\propto K^{\alpha}$, $\alpha = 1.37 \pm 0.01$. This is within the error of the experimental observation $\alpha=1.4 \pm 0.06$ \cite{Meshulam2018}, shown in blue. \textbf{(B)} Average free energy, Eq.~\ref{eq:free}, at each coarse-graining iteration, fit to  $\propto K^{\tilde{\beta}}$, $\tilde{\beta} = 0.84 \pm 0.01$, again close to the experimentally found $\tilde{\beta}=0.88 \pm 0.01$ \cite{Meshulam2018}. \textbf{(C)}  Eigenvalue spectrum of cluster covariance for cluster sizes $K=32,64,128$ against the scaled rank, averaged over clusters. We observe a scaling, as in Eq.~(\ref{eq:eig}), for about 1.5 decades with $\mu=0.65 \pm 0.01$, within error  of the experimental $\mu=0.71 \pm 0.06$ \cite{Meshulam2018}. For all panels, the error bars are standard deviations over randomly selected contiguous quarters of the simulation.}
\end{figure*}

\begin{table}[b]
\caption{\label{tab:tab_1} Simulation parameters for Figures \ref{fig:fig_1}-\ref{fig:fig_3}.}
\begin{tabular}{ |p{1.7cm}||p{3.5cm}|p{3.0cm}|  }
 \hline
 Parameter& Description & Value\\
 \hline
 $\phi$   & latent field multiplier&$\phi = 1.0$\\
 \hline
 $\epsilon$ &   bias towards silence  & $\epsilon=-2.67$\\
 \hline
 $\eta$ &variance multiplier & $\eta = 6.0$ \\
 \hline
 $q$   &probability of coupling to latent field & $q= 1.0$\\
 \hline
 $N_{\rm f}$ &   number of latent fields  & $N_{\rm f}=10$\\
 \hline
 $\tau$ & latent field time constant & $\tau=0.1$\\
 \hline
  $h_{m}^{({\rm place})}$ & presence or absence of place fields& all cells couple to latent fields, half couple to place fields\\
  \hline
 \end{tabular}
 \end{table}

In the experiment analyzed in Ref.~\cite{Meshulam2018}, a mouse ran on a virtual track repeatedly, while neural activity in a population of hippocampal neurons was recorded. A subset of these neurons, called {\em place cells}, are activated when the mouse is at certain points on the track. To capture this structure, we define place fields distributed along a virtual track of length $X$. We simulate 200 repetitions of a run along a track of length $X$ with an average forward speed $v$.  As in the experiments, at the end of each run, the mouse is transported instantaneously to the beginning of the track. Thus the mouse position is $x(t) = v (t\mod T)$, where $T=X/v = 1$ is the time to run a track length. The place fields $h_{m}^{({\rm place})}(x)$ are modeled as Gaussians with centers $\mu_{m} \sim {\rm unif}(0,X]$ and standard deviations $\sigma_m \sim \Gamma(4, X/40) $ drawn from the $\Gamma$-distribution with shape $4$ and scale $X/40$. Coupling between a spin and its place field $W_{im}^{({\rm place})}$ is nonzero with probability $q$, with its value drawn from the standard $\Gamma$ distribution, $\Gamma(1,1)$. We include place fields in our model to match the observed data, but we reproduce the  scaling results within error bars whether or not place cells are modeled (see {\em Discussion} and {\em Online Supplementary Materials}).

The second type of field is a latent field, which we interpret as processes, such as head position or arousal level, known to modulate neural activity, but not directly controlled or measured by the experiment \cite{McGinley2015}. We model each latent field $h_m^{({\rm latent })}$ as an Ornstein-Uhlenbeck process with zero mean, unit variance, and the time constant $\tau$. We model the couplings to the latent fields as 
\begin{equation}
W_{im}^{({\rm latent})}= \phi\times
    \begin{cases}
        \sim\mathcal{N}(0,1) & \text{if $i$  couples to latent fields,}\\
          0, & \text{otherwise.}\\
    \end{cases}
\end{equation}
Here $\sim{\cal N}(0,1)$ denotes sampling from the standard normal distribution, and $\phi$ controls the relative strength of the latent fields compared to the place fields in driving the neural activity. We present results with all latent fields $h_m^{({\rm latent})}$ possessing the same time constant $\tau$ (see Tbl.~\ref{tab:tab_1} for parameters), so that the temporal criticality cannot be attributed to the diversity of time scales in the fields driving the neural activity. 

While we explored many different parameter choices (see Tbl.~\ref{tab:tab_2}), we  present results largely with $N=1024$ \cite{Meshulam2018}, and $N_{\rm f}=10$. Consistent with Ref.~\cite{Meshulam2018}, we choose $p=50$\% of neurons to be place cells, each coupled to its own place field $(\mu_m,\sigma_m)$.  Each latent field is coupled to every neuron. Thus in our typical simulations, about 512 neurons respond to place and latent stimuli, and about 512 are exclusively latent-stimuli neurons.

\begin{figure*}
\includegraphics[width=1.0\textwidth]{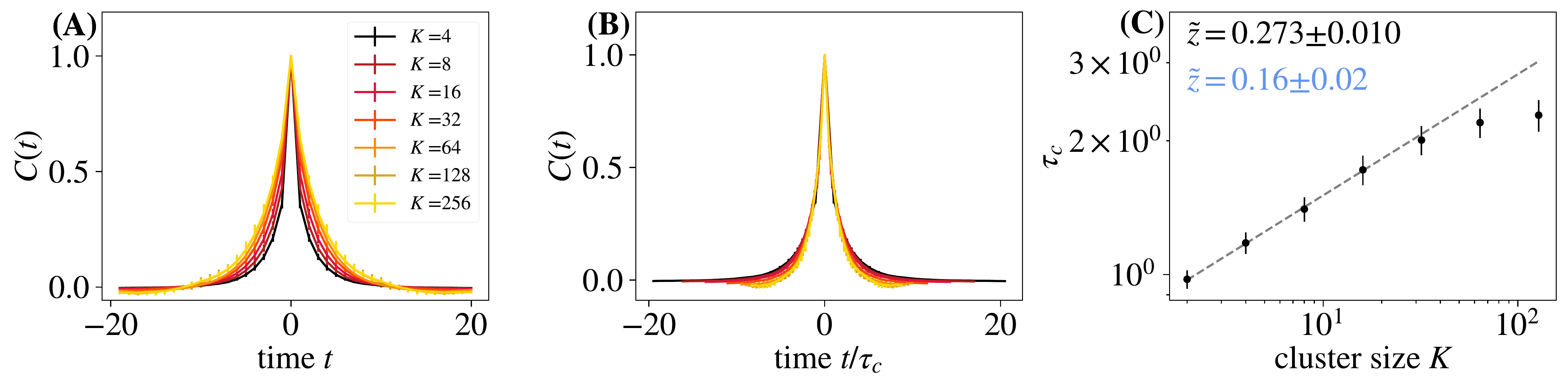}
\caption{\label{fig:fig_2} \textbf{(A)} Average autocorrelation function for cluster sizes $K=2, 4, ..., 256$ as a function of time. \textbf{(B)} Same data, but with time rescaled by the appropriate $\tau_c$ for each cluster size. \textbf{(C)} Time constants $\tau_c$ extracted from each curve in \textbf{(A)} obey $\tau_c \propto K^{\tilde{z}}$, $\tilde{z}=0.27 \pm 0.01$, for roughly 1 decade. Experimentally found $\tilde{z}$ is shown in blue \cite{Meshulam2018}. Error bars are standard deviations over randomly selected contiguous quarters of the simulation.  }
\end{figure*}

{ \em {Results.} ---}
In the following, we simulate random neural activity according to Eq.~(\ref{eq:prob}) and then we replicate the real-space and momentum-space coarse-graining schemes of Ref.~\cite{Meshulam2018}, while tracking the  distributions of variables within clusters as we iterate the coarse-graining algorithms. Briefly, in each iteration of the real-space coarse-graining scheme, pairs of highly correlated neurons are combined into clusters. The cluster activity is the sum of the activity of the pair. At each iteration step, the population size is therefore halved.  In the momentum-space coarse-graining scheme, neural activity fluctuations  are projected onto the eigenvectors of the covariance matrix of the population activity, selecting the $K$ eigenvectors with the largest eigenvalues, and then projected back to the original system size, $N$. All results of  Ref.~\cite{Meshulam2018} can be {\em quantitatively} reproduced by our model, and we include corresponding experimental results in blue on each figure when appropriate. Several scaling exponents were not included or were only reported for a single recording in Ref.~\cite{Meshulam2019}, and therefore we refer to Ref.~\cite{Meshulam2018}.

{\em 1. Scaling of the activity variance.}  Real-space coarse-graining of experimental data \cite{Meshulam2018} reported that the variance of the cluster variables scaled with the cluster size $K$ as $K^{\alpha}$, $\alpha = 1.40 \pm 0.06$, in one experiment. In our simulations, the coarse-grained activity variance scales as $K^{\alpha}$,  $\alpha = 1.36 \pm 0.01$, over more than two decades in $K$ (Fig.~\ref{fig:fig_1}A), within error bars of the experimental value. This indicates that the microscopic variables are not fully independent (which would be $\alpha = 1$), nor are they fully correlated (which would be $\alpha = 2$). 

{\em 2. Scaling of the free energy.} The effective free energy is related to the probability of silence in a cluster, and is expected to scale as a power of cluster size, $K^{\tilde{\beta}}$ \cite{Meshulam2018}. Specifically, we marginalize Eq.~(\ref{eq:prob}) over all fields:
\begin{equation}
    P(\{s_i\})= \int d\{h_m\} P(\{h_m\})P(\{s_i\}|\{h_m\})
\end{equation}
and compute $\ln P(\{s_i = 0\})= \ln P(\{s_i=0\}|\{h_m\}) + \ln \sum_{\{h_m\}} P(\{h_m\})$, where $P(\{s_i=0\})$ is the probability that all neurons $\{s_i\}$ are silent. This defines 
\begin{equation}\label{eq:free}
    F(\{s_i\})=-\ln P(\{s_i=0\}|\{h_m\}),
\end{equation}
where $F(\{s_i\})$ is effective free energy. 
In Fig.~\ref{fig:fig_1}B, we observe that the average free energy at each coarse-graining scales, with a scaling exponent of $\tilde{\beta} = 0.84 \pm 0.01$, within error bars of experimental results, $0.88 \pm 0.01$ \cite{Meshulam2018}.

{\em 3. Scaling of the eigenvalue spectra.} We expect the eigenvalues of the covariance matrix of microscopic variables within each cluster to scale as a power law of the scaled eigenvalue rank \cite{Meshulam2018}. Thus there are two scalings: the rank by the cluster size, and the eigenvalue by the scaled rank. Specifically, the $R^{\rm th}$ eigenvalue $\lambda_R$ of a cluster of size $K$ was shown in \cite{Meshulam2018} to follow
\begin{equation}\label{eq:eig}
    \lambda_R \propto \Bigg (\frac{K}{R} \Bigg )^{\mu}.
\end{equation}
In Fig.~\ref{fig:fig_1}C, we plot the average eigenvalue spectrum of the covariance matrix for each coarse-grained variable for cluster sizes $K=16,32,64,128, 256$. We observe scaling according to Eq.~\ref{eq:eig} for roughly 1.5 decades, with the scaling exponent $\mu=-0.65\pm0.01$, within error bars of the experimental value of $\mu=-0.71\pm 0.06$. 

{\em 4. Scaling of the correlation time.} Another signature of critical systems is that the timescale of cluster autocorrelation $\tau_c$ is a power law of length scale (cluster size $K$) with exponent $\tilde{z}$. In Fig.~\ref{fig:fig_2}A we plot the average autocorrelation function for $K=4,8, ..., 256$. In Fig.~\ref{fig:fig_2}B, we show the same data  as a function of the rescaled time, $\tau/\tau_c$, where $\tau_c$ is calculated by fitting the correlation function to the exponential form. The collapse shown in Fig.~\ref{fig:fig_2}B suggests that $C(t/\tau_c)$ is scale invariant. We then observe a power law relation between the time constant $\tau_c$ and the cluster size $K$ for roughly 1.5 decades in Fig.~\ref{fig:fig_2}C, with a scaling exponent $\tilde{z}=0.27\pm 0.01$. For the recording  reported in Ref.~\cite{Meshulam2018}, the exponent was somewhat different, $\tilde{z} =0.16 \pm 0.02$, but the value over three different recordings,  $\tilde{z}=0.22 \pm 0.08 \pm 0.10$ (mean, individual recording rms errror, standard deviation across recordings) again matches our result.

\begin{figure}
\includegraphics[width=0.33\textwidth]{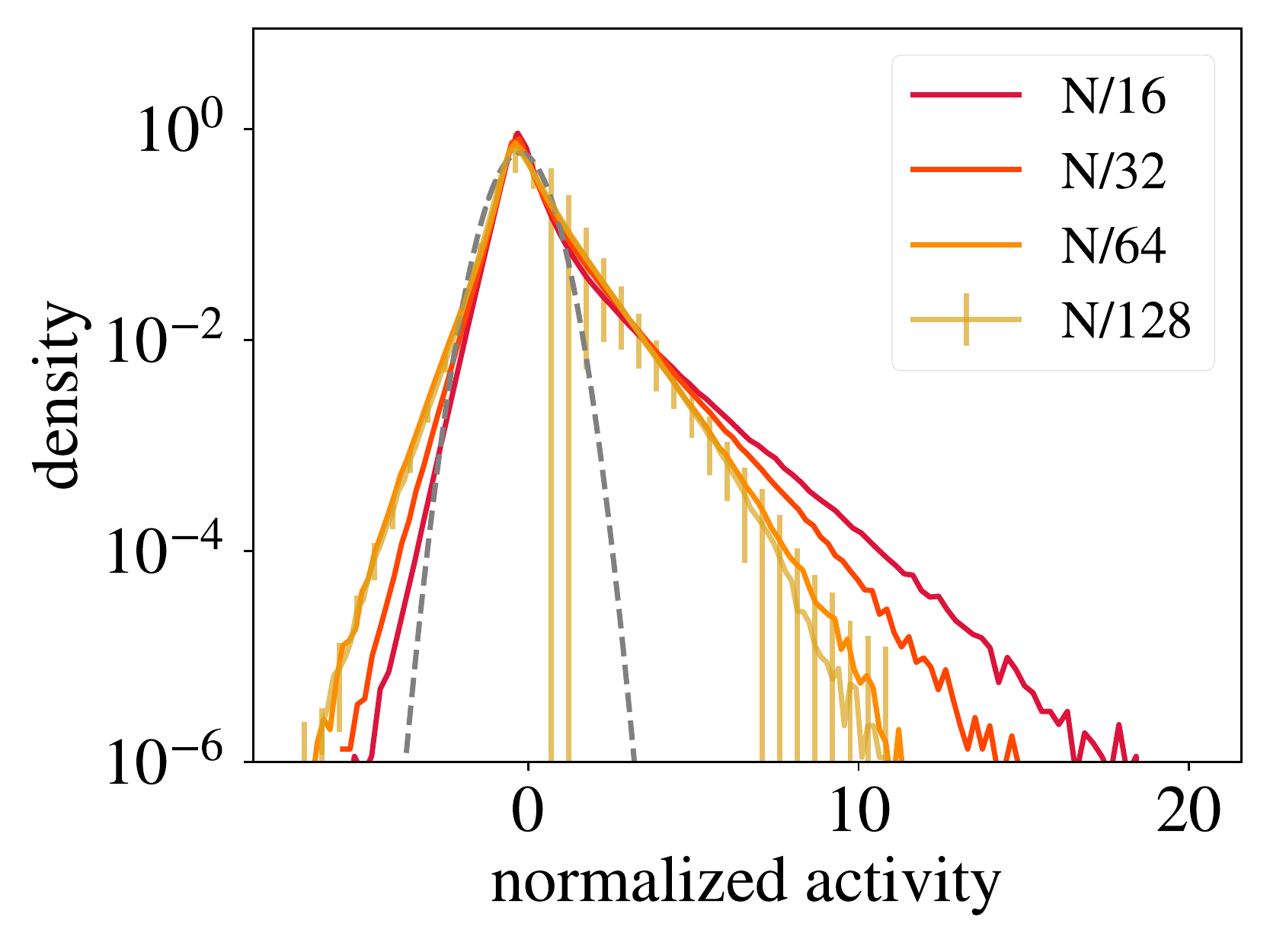}
\caption{\label{fig:fig_3} Distribution of coarse-grained variables for $k= N/16, N/32, N/64, N/128$ modes retained under momentum-space coarse-graining, with a Gaussian distribution (gray dashed line) shown for comparison. The distribution of coarse-grained variables approaches a non-Gaussian limit as $k$ decreases. Error bars are standard deviations over randomly selected contiguous quarters of the simulation.}
\end{figure}

{\em 5. Flow to a non-Gaussian fixed point.}
We replicated the momentum space coarse-graining analysis of   Ref.~\cite{Meshulam2018}. For this, we first calculated the covariance matrix $\Gamma_{ij}$ of the neural activity fluctuations matrix $\Phi_{it}=S_{it} -\langle S_{it} \rangle_{t}$, where $i$ indexes neurons and $t$ indexes time step. We then calculated the eigenvalues and eigenvectors of $\Gamma_{ij}$ and constructed a matrix $\tilde{S}_{ij}$ containing the eigenvectors in its columns, ordered by the corresponding eigenvalues, from largest to smallest. Summing over the first $k$ modes, we calculated the coarse-grained variable 
\begin{equation}
	S_{it}^{(k)} = z_i \sum_{l, j'}^{N, k} \tilde{S}_{ij'}\Phi_{lt} \tilde{S}_{lj'} ,
\end{equation} 
where we set $z_i$ such that $\langle [S^{(k)}_{it}]^2 \rangle_{t} = 1$ \cite{Meshulam2018}.

In Fig.~\ref{fig:fig_3}, we follow the distribution of $S_{it}^{(k)}$ over coarse-graining cut-offs $k$. As the coarse-grained variables are linear combinations of the original variables, if the correlations between the original variables are weak, the distribution will approach a Gaussian due to the central limit theorem. However, close to criticality, the system may flow to a non-Gaussian fixed point. We show these distribution of coarse-grained variables $S_{it}^{(k)}$ for $k= N/16, N/32, N/64, N/128$ modes retained, observing the flow to a non-Gaussian limit  as $k$ decreases: the limit distribution retains a sharp peak at 0 and a heavy positive tail, similar to the experiments \cite{Meshulam2018}.

\begin{figure*}
\includegraphics[width=0.9\textwidth]{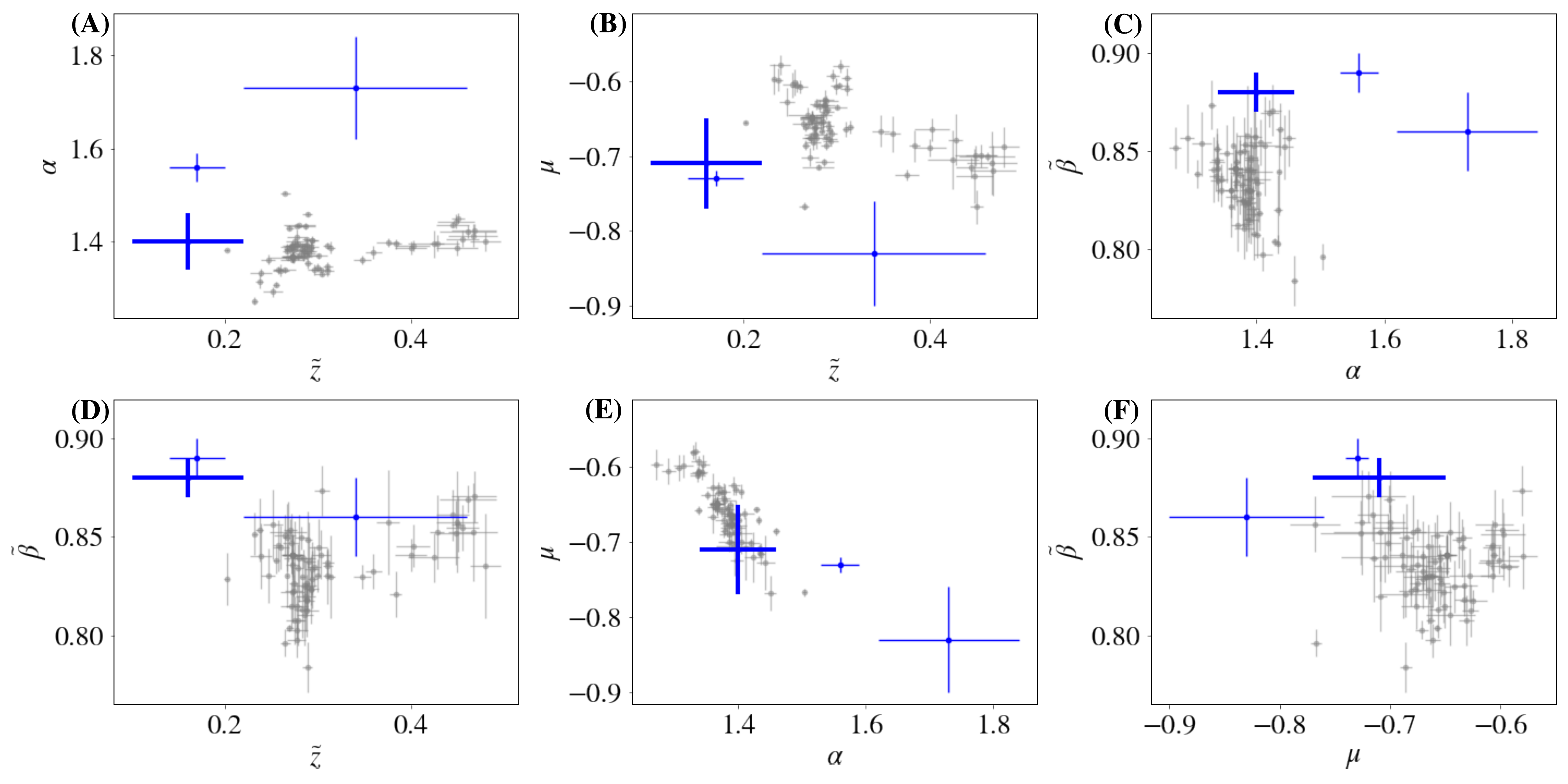}
\caption{\label{fig:fig_4} \textbf{(A)-(F)} Scatter plots of scaling exponents $ \alpha, \tilde{\beta}, \mu, \tilde{z}$ generated by simulations with varying parameters. Experimental results are in blue \cite{Meshulam2018}, with the result highlighted in previous figures in bold. Error bars are standard deviations over randomly selected contiguous quarters of the simulation.}
\end{figure*}

\begin{table*}
\caption{\label{tab:tab_2} Results from parameter sweeps over $\eta$, $\phi$, and $\epsilon$ in Eq.~(\ref{eq:energy}). We vary these parameters one at a time while keeping all others at default values.} 
\begin{tabular}{ |p{1.3cm}||p{2.6cm}|p{2.6cm}|p{10.8cm}|  }
 \hline
 Param.& Sweep range & Critical values& Comments\\
 \hline
 $\phi$   &   $[0.8,1.5]$     &$\phi \in [0.9, 1.2]$&   Weak latent fields: damaged variance scaling (Fig.~S53).\\
 \hline
 $\epsilon$ &   $[-2.91,-1.33]$  & $\epsilon<-1.92$ & $\epsilon > -1.92$: damaged eigenvalue scaling (Fig.~S27); $\epsilon > -1.5$: flow to a Gaussian fixed point (Fig.~S33)\\
  \hline
 $\eta$ &  $[2.8,6.6]$ & $\eta \in [2.8,6.6]$ &  Does not impact existence of scaling (Fig.~S43-S49)\\
  \hline
 $q$   &  $q=[0.25,1.0]$ & $q \geq 0.5$&  $q<0.5$ is deleterious to variance scaling (Fig.~S20)\\
  \hline
 $N_{\rm f}$ &   $[1,20]$  & $N_{\rm f} \geq 5$& $\geq 5$ latent fields needed for scaling (Fig.~S11)\\
  \hline
 $\tau$ &  $[0.05,1.2]$  & $\tau \in [0.05,1.2]$   & No significant impact on scaling (Fig.~S35-S41)\\
  \hline
 $h_{m}^{\rm (place)}$ & presence / absence & -- & Place fields only: no scaling behavior (Fig.~S3-S9) \\
 \hline
\end{tabular}
\end{table*}

{\em Experimental agreement.} To investigate which parameter regimes give rise to scaling in our model, we vary the parameters $\eta$, $\phi$, and $\epsilon$ in Eq.~(\ref{eq:energy}), the latent field correlation time $\tau$, the number of latent fields $N_{\rm f}$, and the probability that a neuron couples to a latent field $p$. We vary them one at a time, while keeping other parameters at values in Tbl.~\ref{tab:tab_1}. We also run simulations with only nonplace fields $h_m^{\rm (latent)}$ included, or with only place fields $h_m^{\rm (place)}$. We record parameters whose simulations display eigenvalue spectra collapse for at least 1.5 decades, as in Fig.~\ref{fig:fig_1}D, and activity variance scaling for over 2 decades, as in Fig.~\ref{fig:fig_1}A. Parameter regimes leading to scaling behaviors are summarized in Tbl.~\ref{tab:tab_2}, with detailed plots shown in {\em Online Supplementary Materials} \cite{supplement}. We also provide scatter plots of pairs of scaling exponents (if scaling is observed) in Fig.~\ref{fig:fig_4}, compared to the values from three different experiments as reported in Ref.~\cite{Meshulam2018}, highlighting the experiment we used as a benchmark in the previous figures. Our simulations show that a broad range of parameters lead to scaling exponents in a quantitative agreement with the experiments. 

{\em Discussion. ---} When the number of activity variables is large, working with their joint probability distributions is infeasible, and one need to coarse-grain to develop interpretable models of the data. We have shown that, under two different coarse-graining schemes, a model of a neural population in which neurons (spins) are randomly coupled to a few slowly varying latent stimuli or fields (certainly fewer than would be needed to overfit the data) replicates power law scaling relationships as well as the flow of activity distributions to a non-Gaussian fixed point, reported for the mouse hippocampus experiments \cite{Meshulam2019,Meshulam2018}. Other models, such as a randomly connected rate network \cite{Vreeswijk1996}, or a spiking Brunel neural network in the synchronous irregular regime \cite{Brunel2000}, cannot reproduce these results \cite{Meshulam2018}. In the latter case, one can approximate the network by a population of uncoupled neurons driven by a single common time-varying input \cite{Touboul2017}, but we show that the scaling does not appear for fewer than about five latent processes, explaining why these previous models failed to match experiments.

Our parameter sweeps show that emergence of scaling in the model is robust to parameter changes. The existence of scaling is most sensitive to nearly all cells having significant latent field coupling, irrespective of whether they additionally couple to place fields. This is especially clear in Fig.~S7, where only simulations with widespread latent field coupling reproduce the autocorrelation time collapse \cite{supplement}. This allows us to make an interesting biological prediction that even place cells in hippocampus must be driven not solely by the animal's position. This is consistent with the observations that place cells carry information about activity of other cells in the population \cite{Meshulam2017}. Further, since it is difficult to reproduce temporal scaling over many decades using latent fields with a single time constant, we suggest that this  may be easier with latent fields with diverse time scales.

More broadly, we have shown that the surprising spatio-temporal scaling results of Ref.~\cite{Meshulam2018} can be explained by the presence of multiple unknown, time-varying driving fields (possibly with just a single  time constant). Further,  these latent fields {\em necessarily} result in scale free activity. To our knowledge, our mechanism is the first one to explain these  results. While here we have focused on neural data, our results show that the signatures of criticality  discussed in this {\em Letter} will emerge from {\em any} sparsely active multivariate system (whether biological, inanimate, social, or human-made) driven by several latent dynamical processes.

\begin{acknowledgments}
We thank L.~Meshulam and W.~Bialek for helping us to understand their work, and S.~Boettcher and G.~Berman for valuable feedback. This work was supported in part by NIH Grants R01NS084844 (AS and IN), R01EB022872, and R01NS099375 (IN), and by NSF Grant BCS-1822677 (IN). 
\end{acknowledgments}

\bibliography{bib}

\clearpage

\setcounter{figure}{0}
\makeatletter
\renewcommand{\thefigure}{S\arabic{figure}}

\appendix
\section{Pairwise correlations and place cell activity}

References \cite{Meshulam2018,Meshulam2019} reported two additional observations: the first and second moments of the cell activity were recorded and the effect of coarse-graining on place cell activity was tracked. We did not address these observations in the Main Text, but we report similar results here. 

\begin{figure}[H]
\centering
\includegraphics[width=0.75\columnwidth]{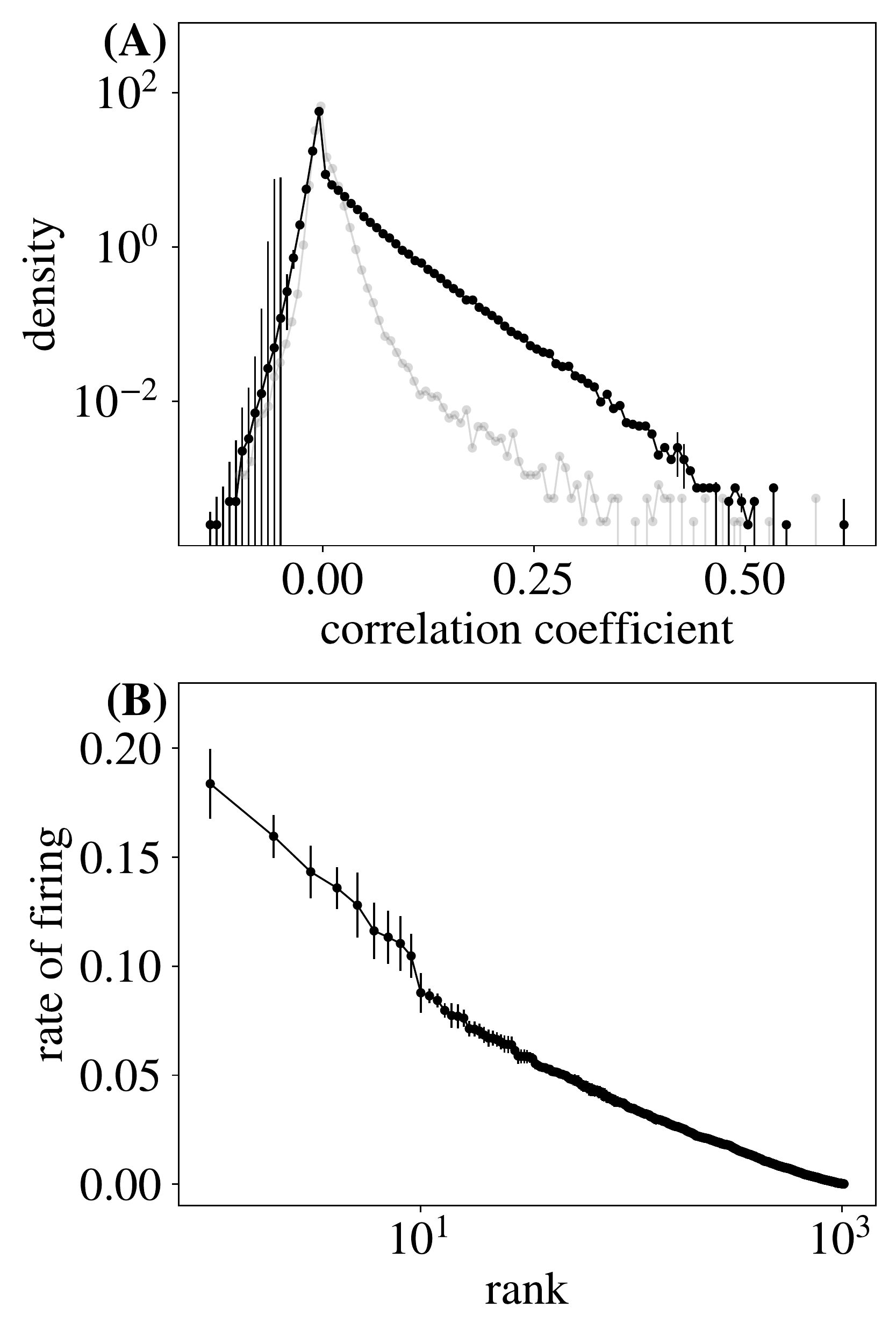}
\caption{\label{fig:fig_5} \textbf{(A)} Distribution of pairwise correlation coefficients. \textbf{(B)} Rate of firing vs.\ rank of neuron. For \textbf{(A)} and \textbf{(B)}, error bars are standard deviations over randomly selected contiguous quarters of the simulation.}
\end{figure}

\begin{figure*}
\includegraphics[width=1.0\textwidth]{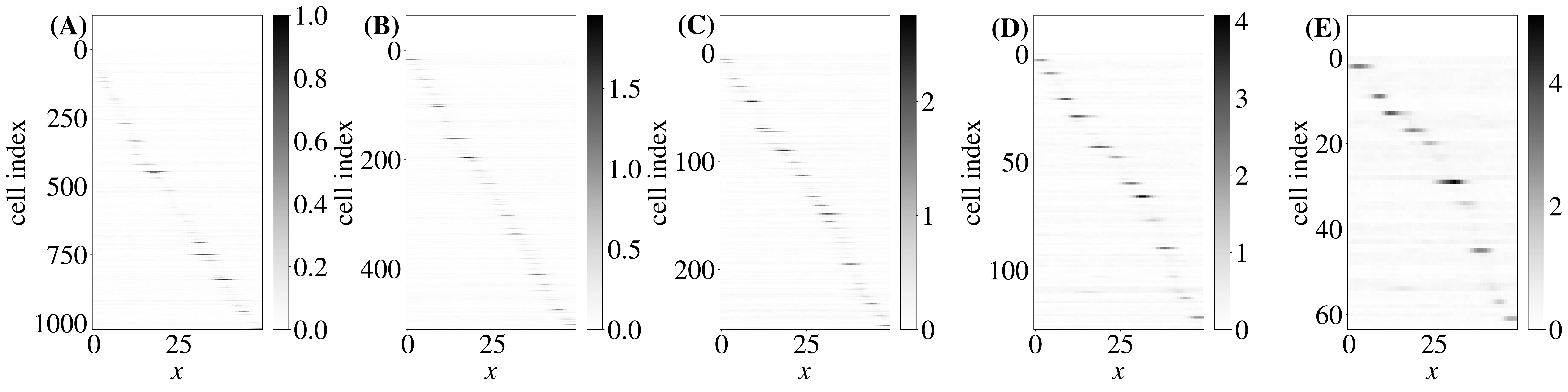}
\caption{\label{fig:fig_6} Average activity at spatial location $x$ for each neuron or coarse-grained variable. Simulation parameters are tabulated in Tbl.~\ref{tab:tab_1}. Coarse-graining steps 0,1,2,3 and 4 are displayed in \textbf{(A)}, \textbf{(B)}, \textbf{(C)}, \textbf{(D)}, and \textbf{(E)}, respectively. Observe that place cell activity remains strong even over many coarse-graining iterations.} 
\end{figure*}

In Fig.~\ref{fig:fig_5}A, we plot the probability distribution of the pairwise correlation coefficients of our simulated neurons. This is qualitatively similar to the experimental results \cite{Meshulam2018}: the distribution has a sharp peak in density just to the right of 0 (small positive correlations), a short left tail, and a long right tail, ending at correlation coefficients greater than 0.6. In Fig.~\ref{fig:fig_5}B we plot neuron firing rate vs.~its rank. Again, this is similar to the experiments \cite{Meshulam2018}, including a maximum firing rate of less than 0.2, and a slight elbow in the otherwise near-straight rate vs.~rank curve.

Further, place cells are visible in our simulations across the coarse-graining, as in Ref.~\cite{Meshulam2018}. Indeed,  Fig.~\ref{fig:fig_6} shows the activity of coarse-grained variables vs.~position on the track, which shows the characteristic localized place cell bump over many coarse-graining iterations. 

\section{Parameter sweeps}

Most plots in the Main Text use the default parameters listed in Tbl.~\ref{tab:tab_1}. In order to further investigate the behavior of our model, we perturb each parameter around its default value, while holding all others fixed. Below we include detailed results of each of these parameter sweeps. A summary is tabulated in Tbl.~\ref{tab:tab_2}. 

\subsubsection{Varying how cells couple to stimuli and latent fields}

\begin{figure}[H]

\includegraphics[width=0.47\textwidth]{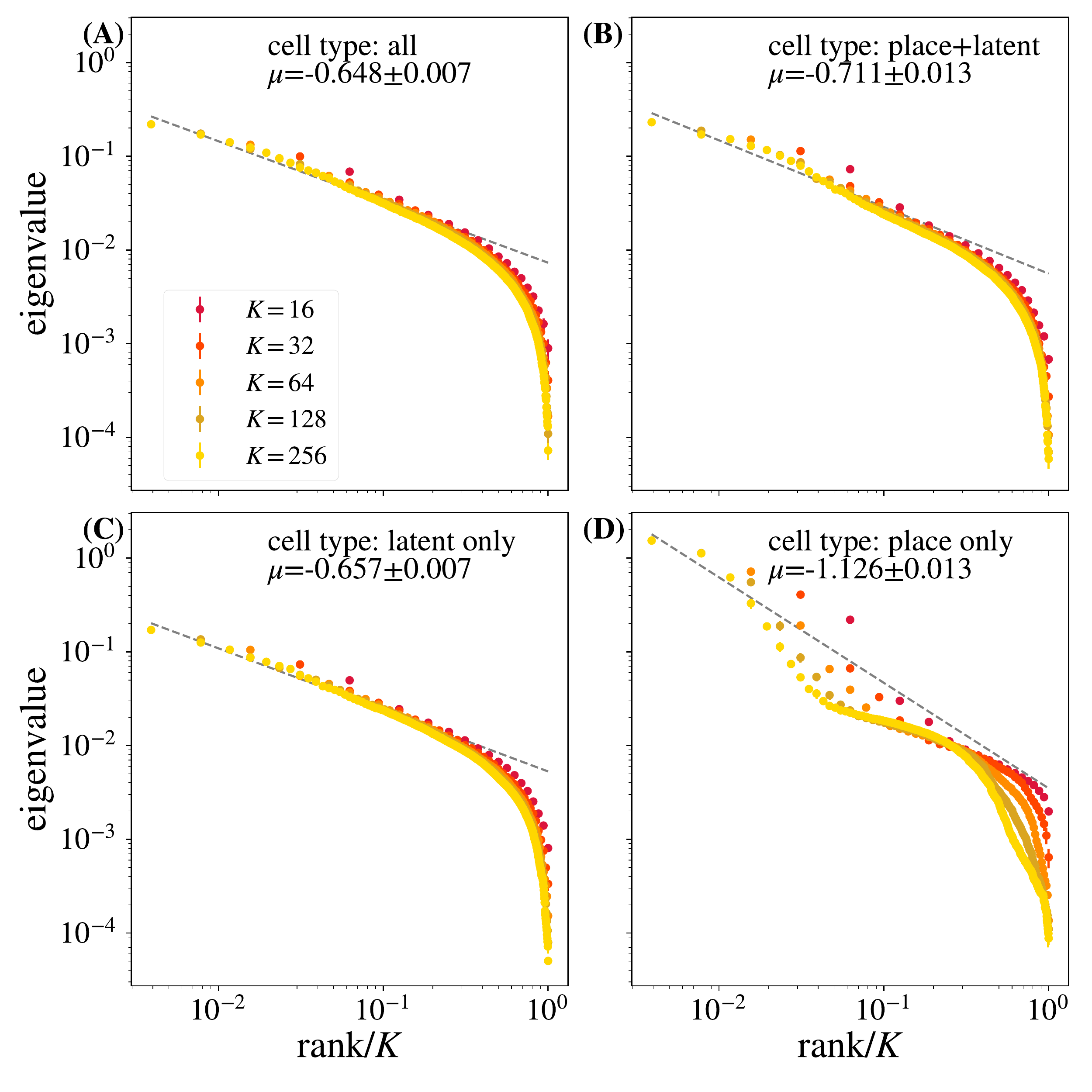}
\caption{\label{fig:fig_7} Average eigenvalue spectrum of cluster covariance for cluster sizes $K=32,64,128$. Cell types present in the simulation are labeled. No significant difference in quality of scaling is observed for panels (A)-(C), while simulations with only place cells have a lower quality eigenvalue collapse. Error bars are standard deviations over randomly selected contiguous quarters of the simulation.}
\end{figure}

\begin{figure}[H]
\includegraphics[width=0.47\textwidth]{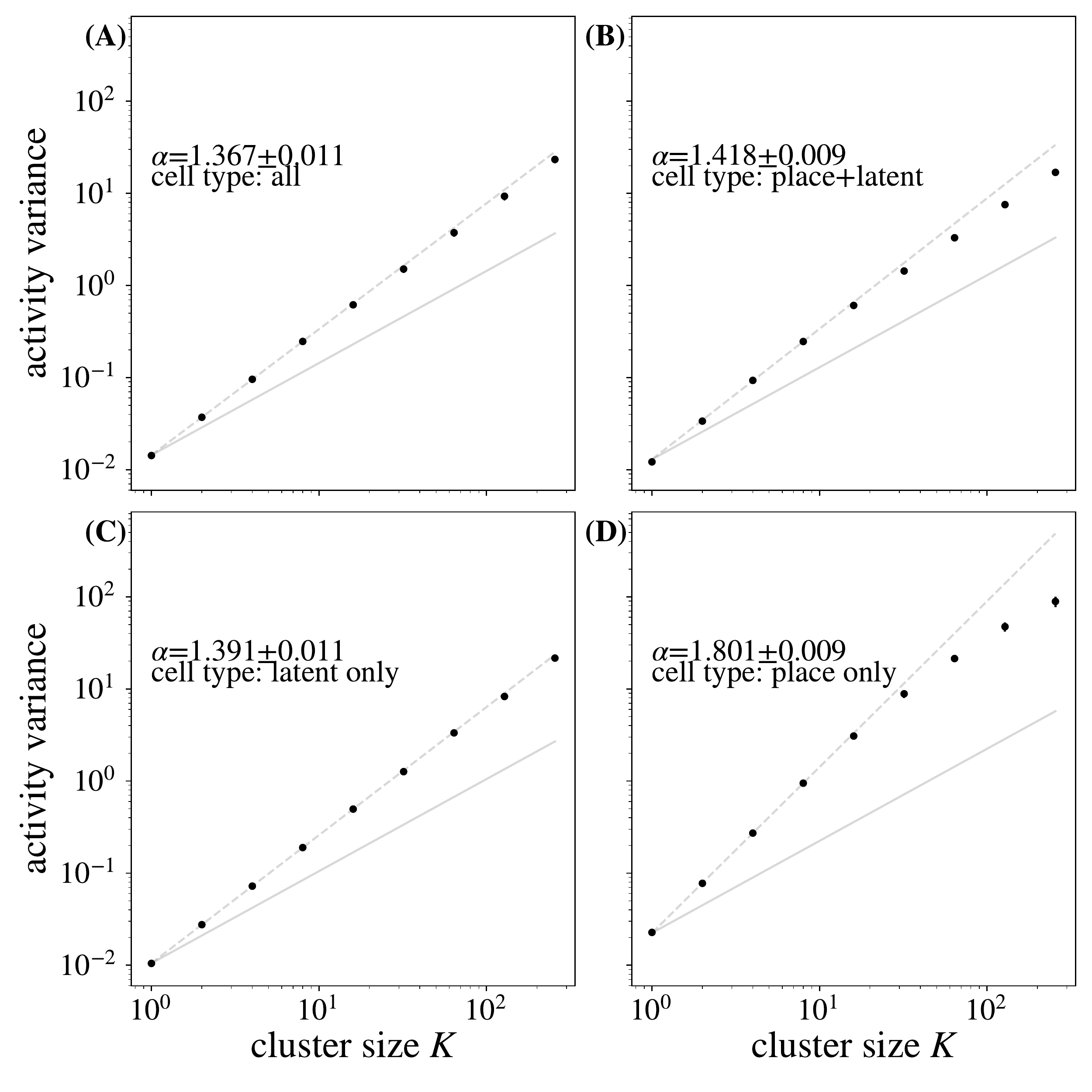}
\caption{\label{fig:fig_8}Activity variance over coarse-grained variables at each coarse-graining iteration.  Cell types present in the simulation are labeled. No significant difference in quality of scaling is observed for panels (A)-(C),  while simulations with only the place cells have a lower quality variance scaling. Error bars are standard deviations over randomly selected contiguous quarters of the simulation.}
\end{figure}

\begin{figure}[H]
\includegraphics[width=0.47\textwidth]{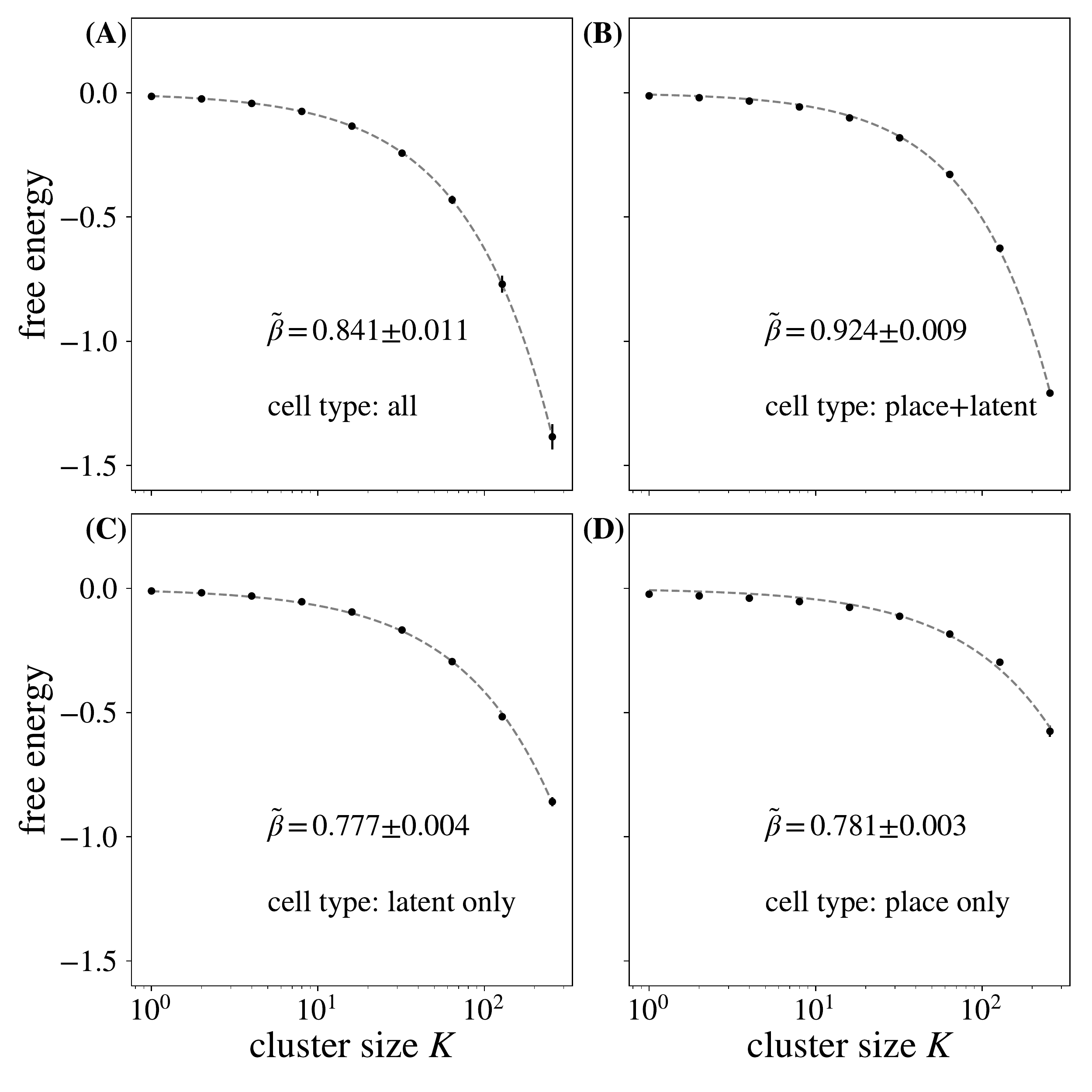}
\caption{\label{fig:fig_9} Average free energy at each coarse-graining iteration. Cell types present in the simulation are labeled. No significant differences in quality of scaling between simulations is observed. Error bars are standard deviations over randomly selected contiguous quarters of the simulation.}
\end{figure}

\begin{figure}[H]
\includegraphics[width=0.47\textwidth]{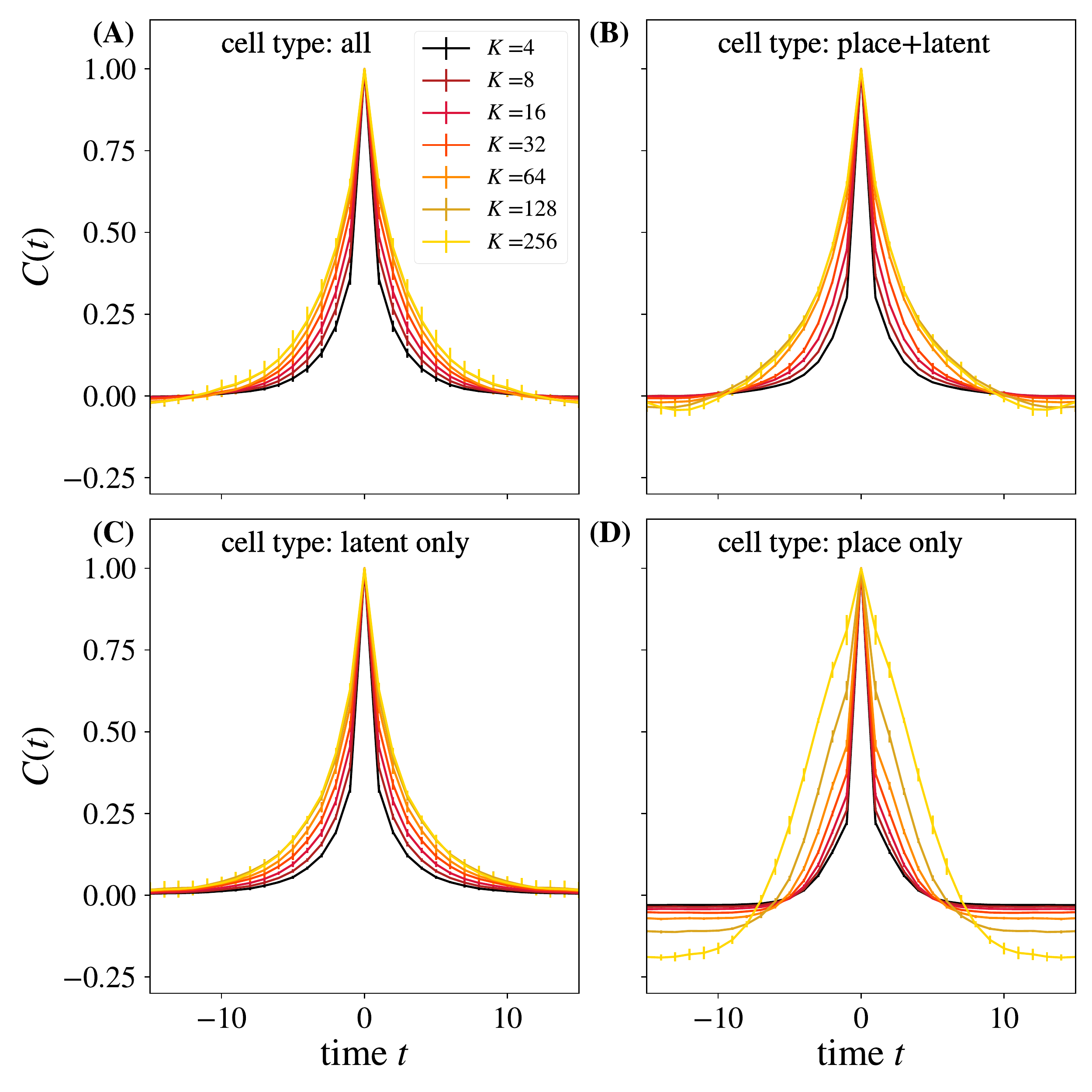}
\caption{\label{fig:fig_10} Average autocorrelation function for cluster sizes $K=2, 4, ..., 256$ as a function of time, cluster size indicated by legend. Cell types present in the simulation are labeled. Panel (D)  with place cells only substantially differs from the rest. Error bars are standard deviations over randomly selected contiguous quarters of the simulation.}
\end{figure}

\begin{figure}[H]

\includegraphics[width=0.47\textwidth]{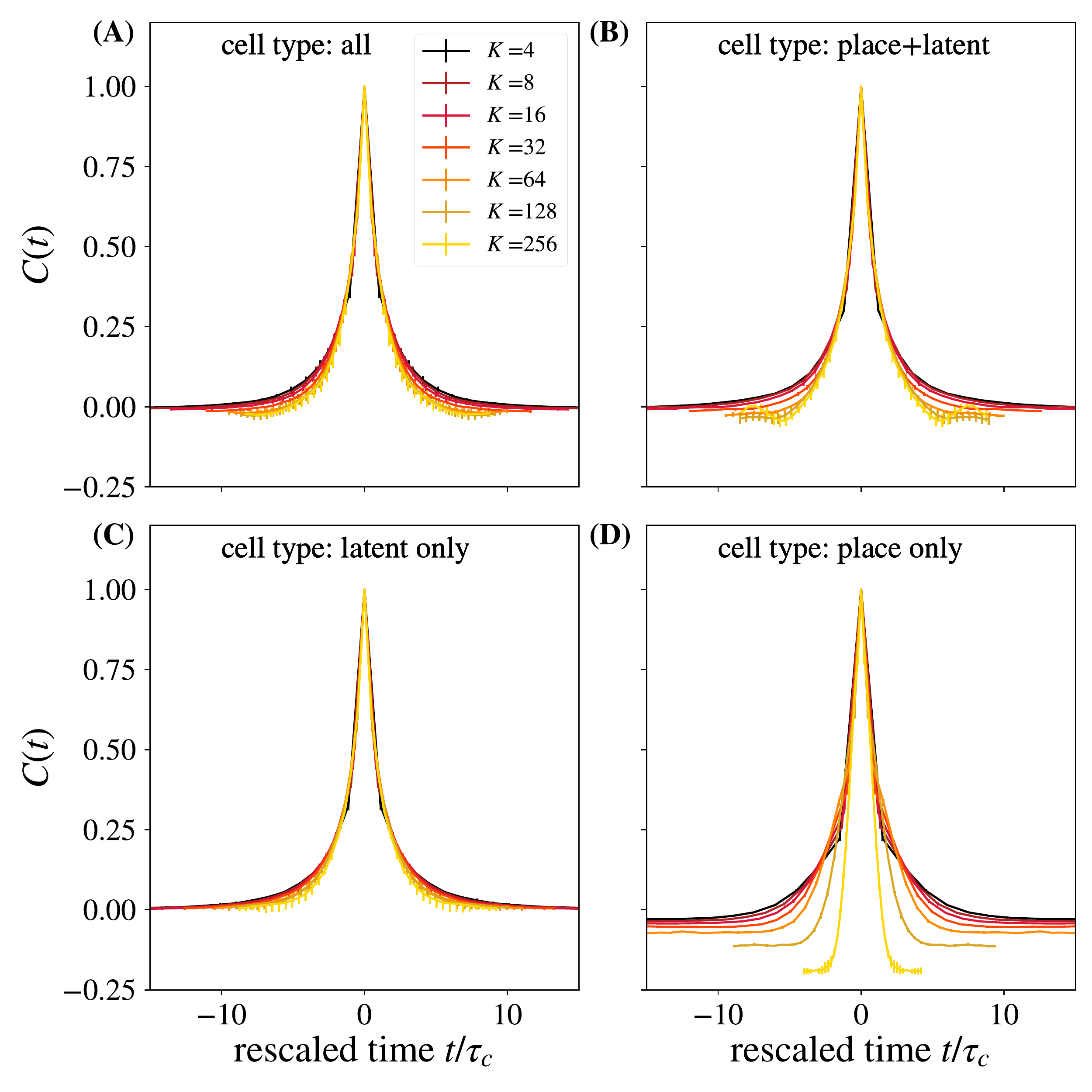}
\caption{\label{fig:fig_11} Average autocorrelation function for cluster sizes $K=2, 4, ..., 256$, where time is rescaled by the appropriate $\tau_c$ for that coarse-graining iteration. Panel (D) with place cells only substantially differs from the rest. The quality of collapse and agreement with experiments are worse in panels (B) and (D), where some cells have no latent field coupling. Error bars are standard deviations over randomly selected contiguous quarters of the simulation.}
\end{figure}

\begin{figure}[H]

\includegraphics[width=0.47\textwidth]{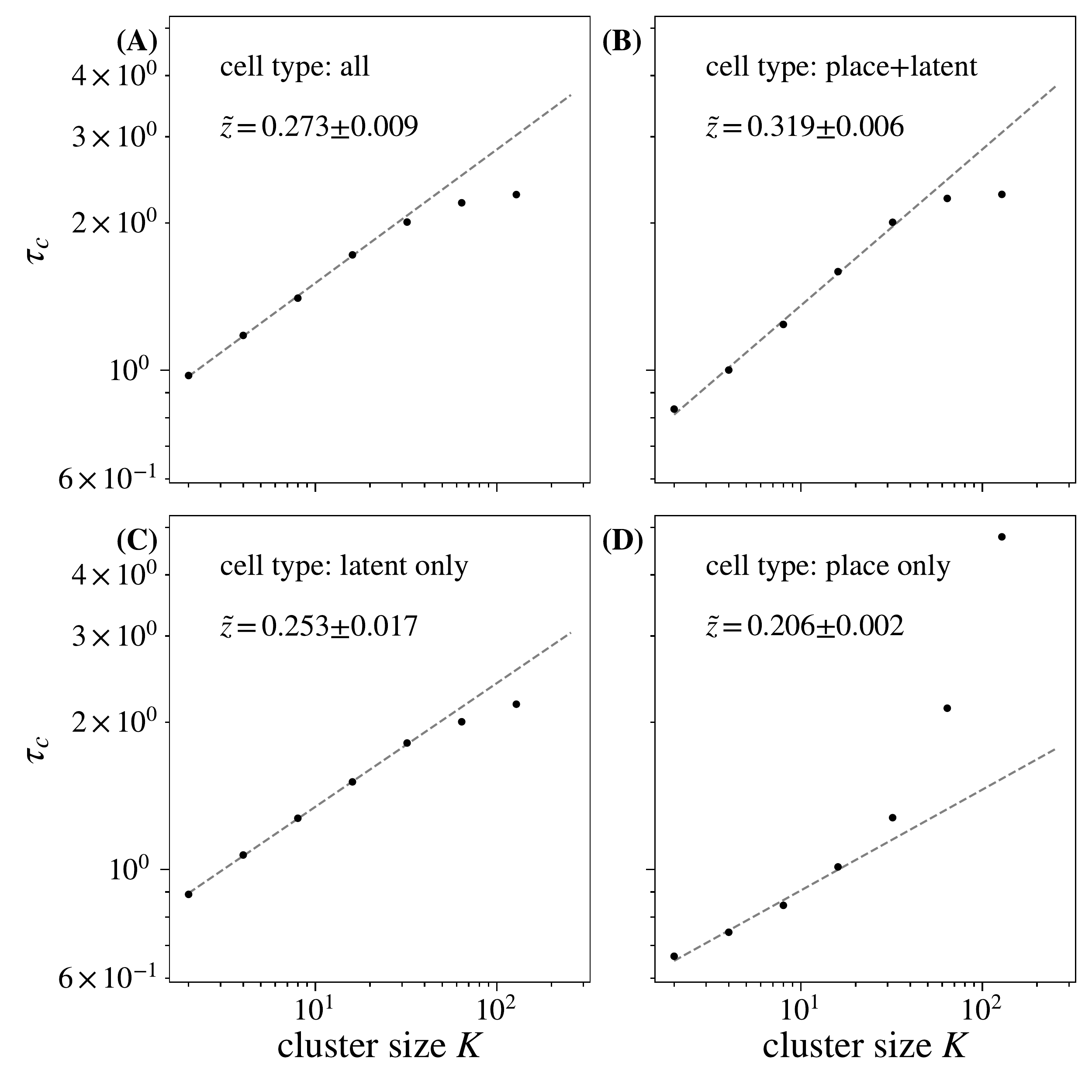}
\caption{\label{fig:fig_12} Time constants $\tau_c$ extracted from each curve in in Fig.~\ref{fig:fig_10}, and fits to  $\tau_c \propto K^{\tilde{z}}$.  Cell types present in the simulation are labeled. Panel (D) with place cells only shows no scaling. Error bars are standard deviations over randomly selected contiguous quarters of the simulation.}
\end{figure}

\begin{figure}[H]

\includegraphics[width=0.47\textwidth]{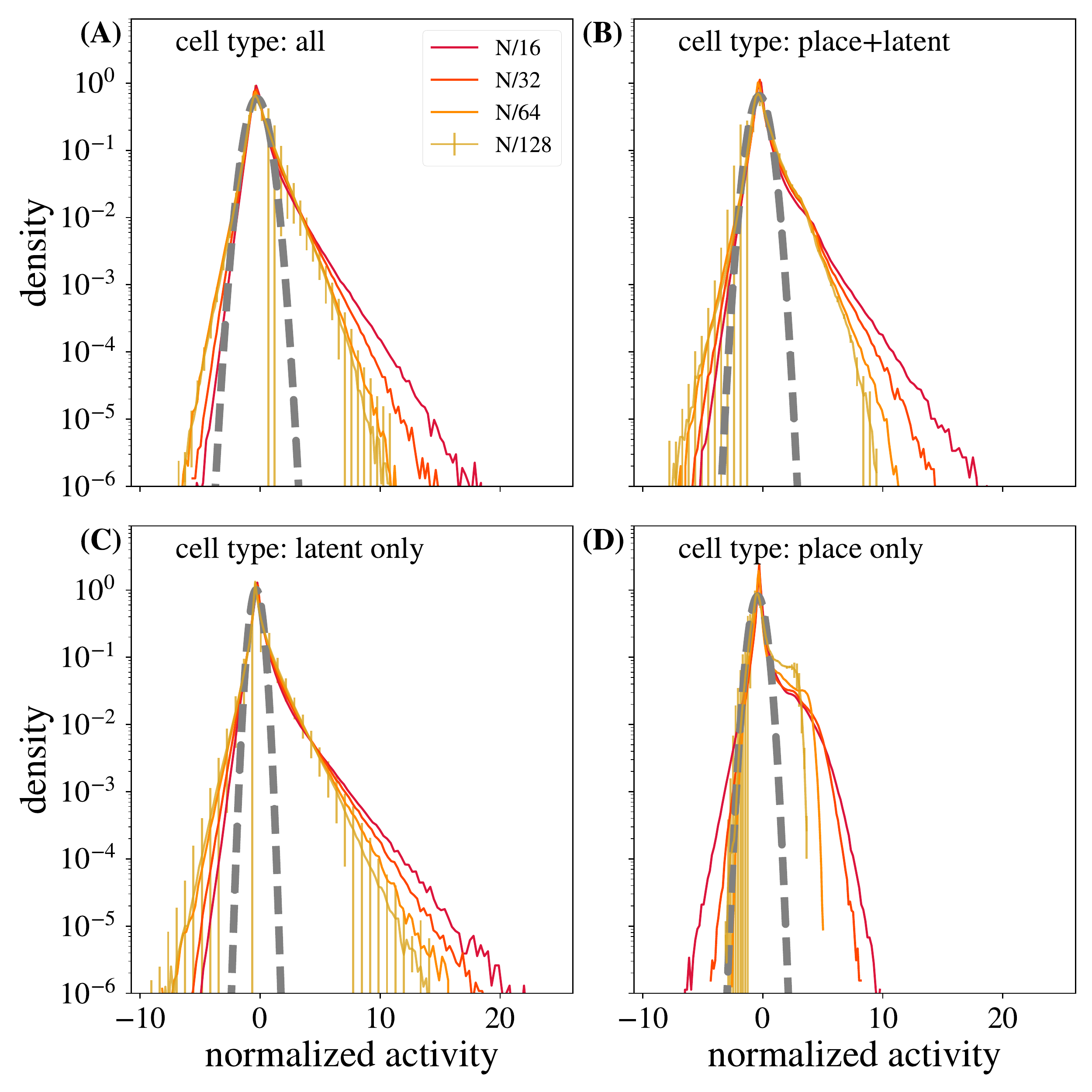}
\caption{\label{fig:fig_13} Distribution of coarse-grained variables for $k= N/16, N/32, N/64, N/128$ modes retained.  Cell types present in the simulation are labeled. No significant differences in approach to a non-Gaussian fixed point are observed between simulation types except for panel (D), where latent fields are present. Error bars are standard deviations over randomly selected contiguous quarters of the simulation.}
\end{figure}

In the model in the Main Text, all cells coupled to latent fields, and half of the cells coupled to place fields. Here we show the effects of changing this. We complete the following 4 simulations, analyzed as in the Main Text:
\begin{enumerate}
    \item $N/2$ cells couple only to latent fields, $N/2$ cells couple to both latent fields and place fields (Main Text).
    \item $N/2$ cells couple only to place fields, $N/2$ cells couple  only to latent fields. 
    \item $N$ cells couple only to latent fields.
    \item $N$ cells couple only to place fields.
\end{enumerate}
We refer to these simulations as ``all'', ``place + latent'', ``latent only'', and ``place only'', respectively. Note that for the ``place only'' simulation, we increased $\epsilon$ from default value $\epsilon=-2.67$ to $\epsilon=-1.33$ to compensate for the omission of latent fields and the resulting decrease in the activity.  In Figs.~\ref{fig:fig_7}-\ref{fig:fig_13} we show that including place fields in our simulations together with latent fields does not significantly alter free energy scaling (Fig.~\ref{fig:fig_9}), correlation time scaling (Figs.~\ref{fig:fig_10}-\ref{fig:fig_12}), or approach to a non-Gaussian fixed point (Fig.~\ref{fig:fig_13}). However, including place fields in simulations with latent fields creates slight deviation from power law scaling in variance at large cluster size, Fig.~\ref{fig:fig_8}) and is damaging to the eigenvalue collapse, Fig.~\ref{fig:fig_7}. In contrast, omitting latent fields from simulations has a disastrous effect on scaling.

By examining Figs.~\ref{fig:fig_7}-\ref{fig:fig_13}, we conclude that the presence of scaling behavior does not depend on the presence of place fields, but does depend on whether {\em all} (or, at least, nearly all) cells also couple to latent stimuli. In fact, the presence of place fields is deleterious to scaling and does not yield scaling behavior without the inclusion of latent fields. Thus existence of scaling in experimental data suggests that most cells (including place cells) in the mouse hippocampus, in fact, are also coupled to latent fields.

\subsubsection{Varying the number of latent fields $N_{\rm f}$}

\begin{figure}[H]
\includegraphics[width=0.47\textwidth]{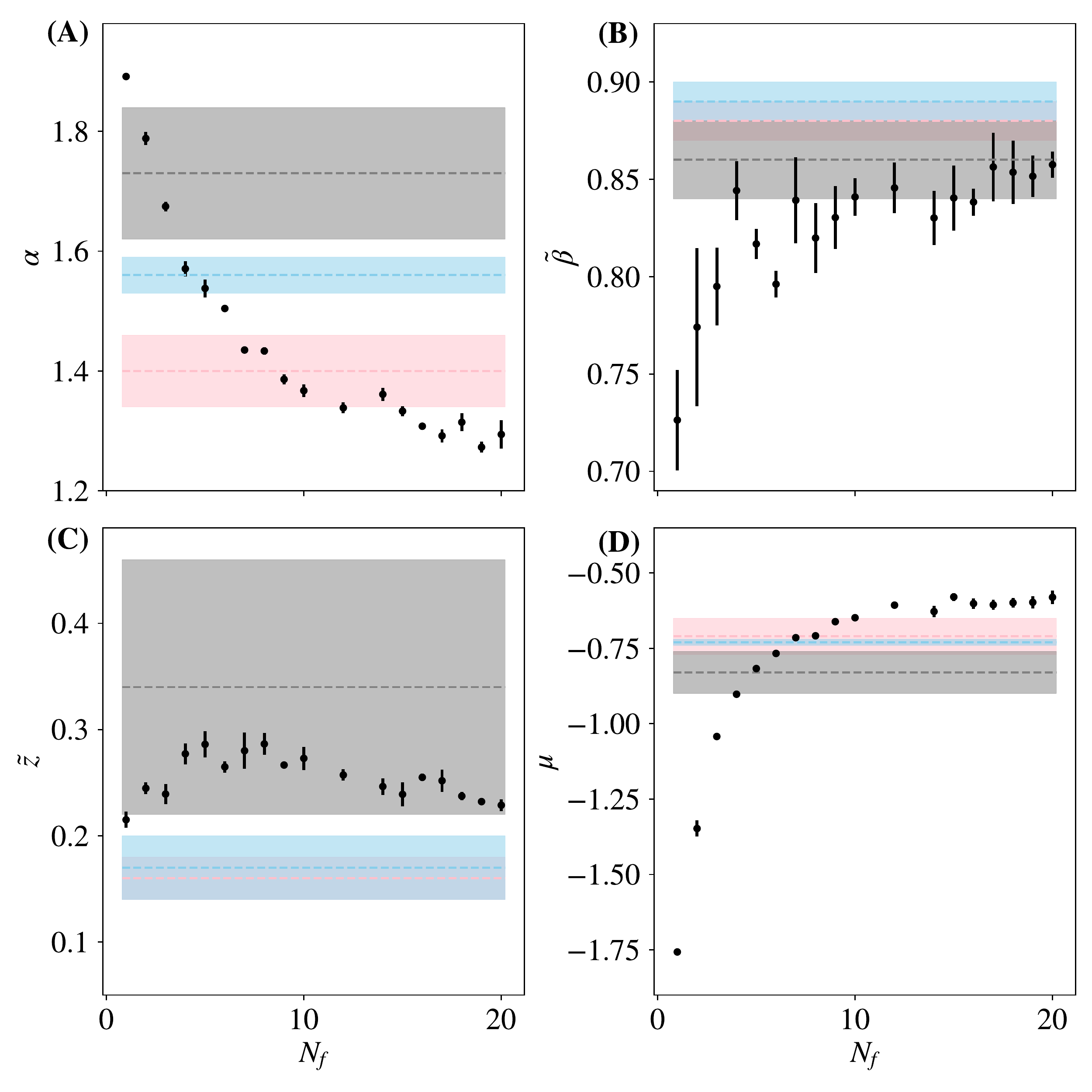}
\caption{\label{fig:fig_14} The critical exponent, $\alpha, \tilde{\beta}, \tilde{z}, \mu$ vs.~number of latent fields $N_{\rm f}$. Results from three different experiment \cite{Meshulam2018} are shown in gray, pink, and blue bands. Error bars are standard deviations over randomly selected contiguous quarters of the simulation.}
\end{figure}

We will now consider the effects of varying the number of latent fields $N_{\rm f}$ in our simulation. We perform simulations with the default parameters sweeping over values of $N_{\rm f}=1,\dots,20$. In Fig.~\ref{fig:fig_14}, we note that our simulations include a regime that quantitatively matches experimental results \cite{Meshulam2018}.

\begin{figure}[H]
\includegraphics[width=0.47\textwidth]{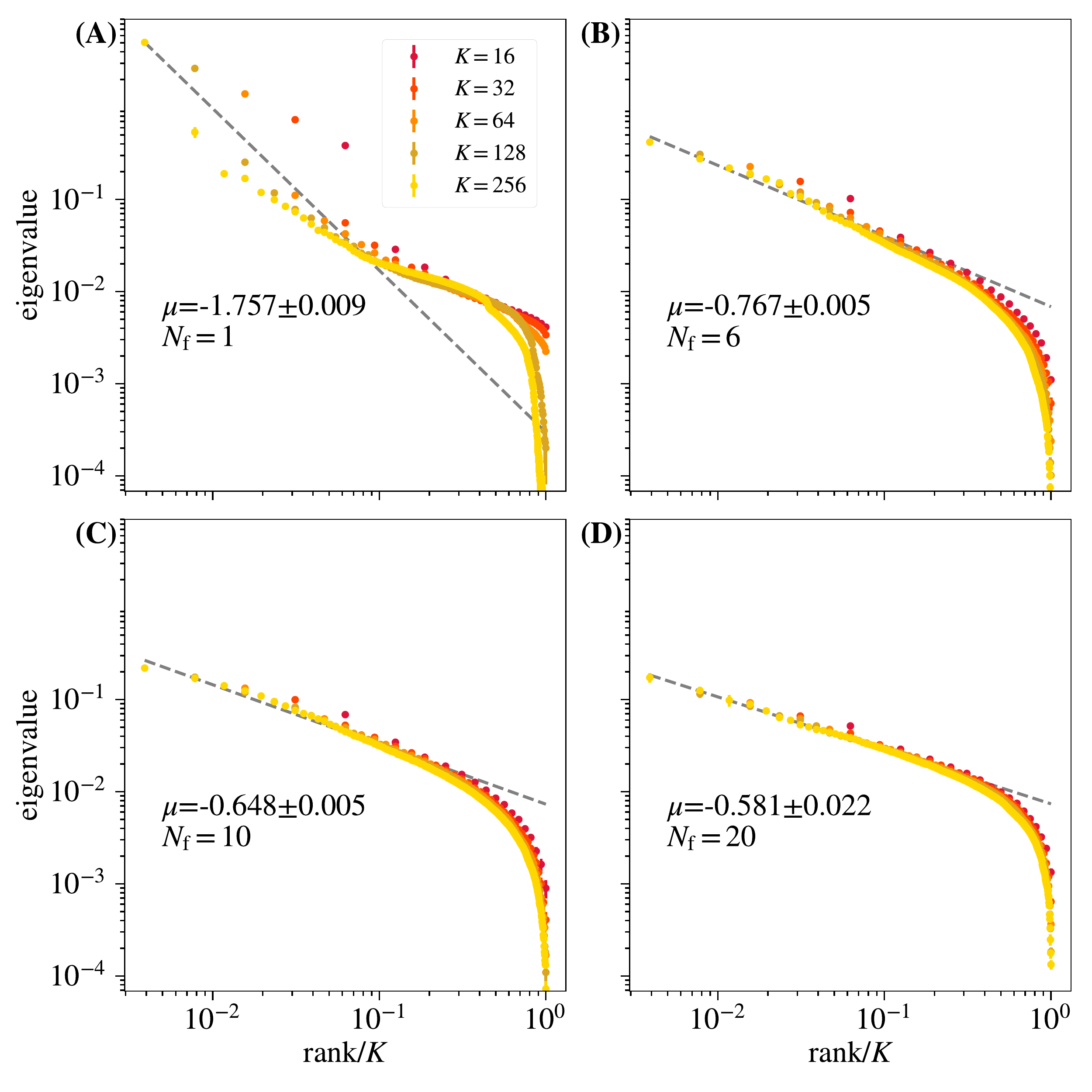}
\caption{\label{fig:fig_15} Average eigenvalue spectrum of cluster covariance for cluster sizes $K=32,64,128$ for different $N_{\rm f}$. Too few latent fields are insufficient to reproduce the scaling. Error bars are standard deviations over randomly selected contiguous quarters of the simulation.}
\end{figure}

\begin{figure}[H]
\includegraphics[width=0.47\textwidth]{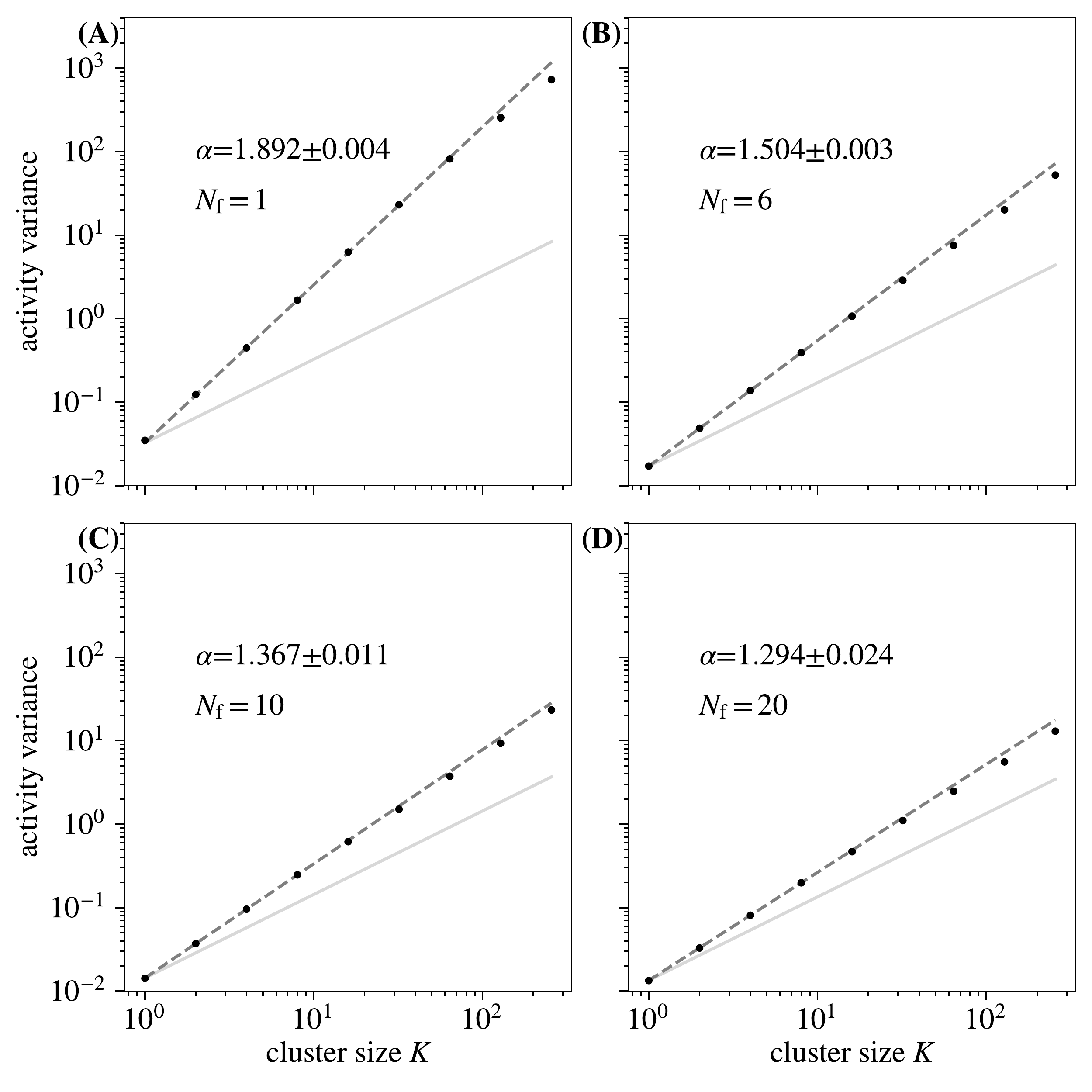}
\caption{\label{fig:fig_16} Activity variance over coarse-grained variables at each coarse-graining iteration  for different $N_{\rm f}$. While changing $N_{\rm f}$ changes the value of $\alpha$, the scaling persists for all explored $N_{\rm f}$. Error bars (too small to be seen) are standard deviations over randomly selected contiguous quarters of the simulation. }
\end{figure}

\begin{figure}[H]
\includegraphics[width=0.47\textwidth]{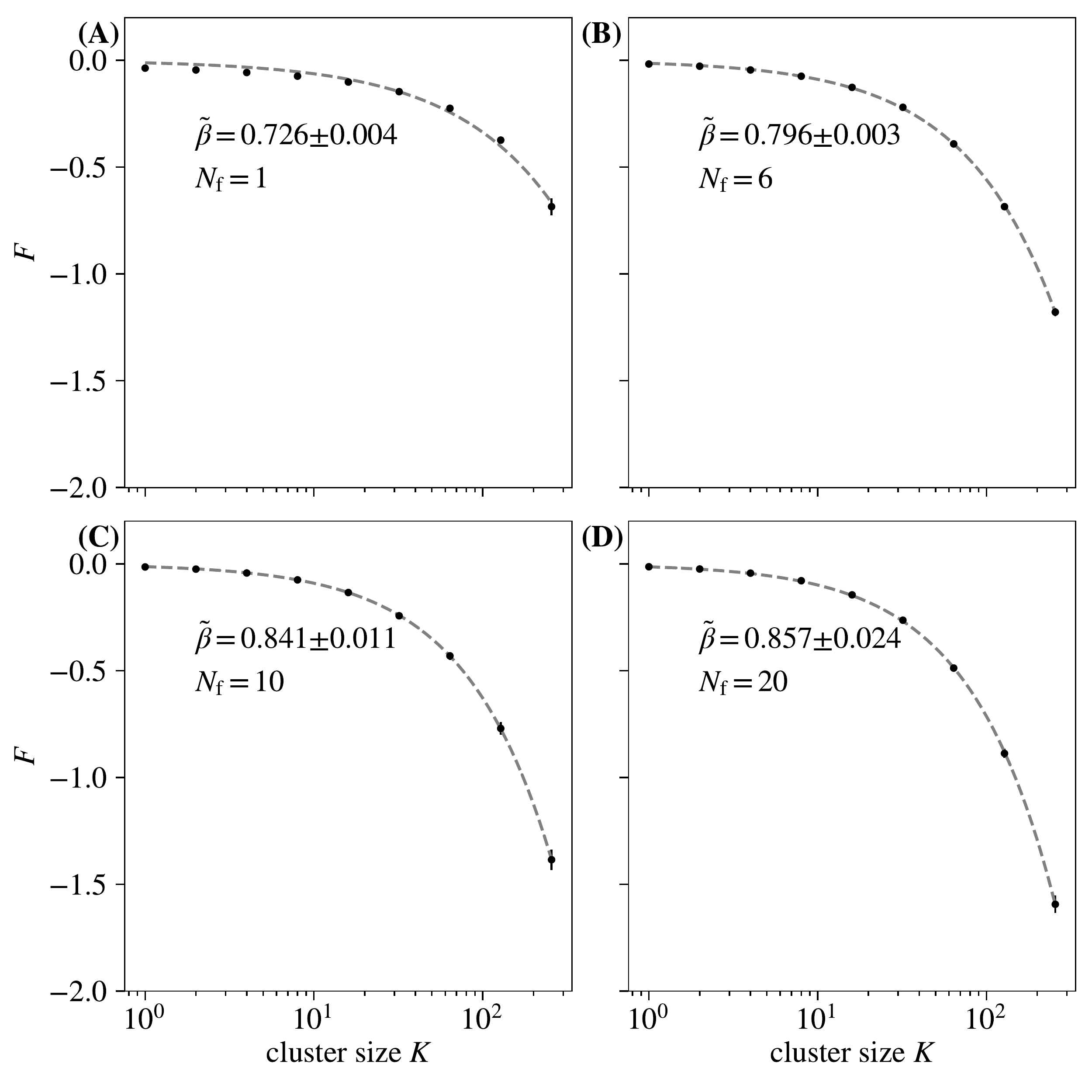}
\caption{\label{fig:fig_17} Average free energy at each coarse-graining iteration for different $N_{\rm f}$. While changing $N_{\rm f}$ changes the value of $\tilde{\beta}$, the scaling persists for all explored $N_{\rm f}$. Error bars (too small to be seen) are standard deviations over randomly selected contiguous quarters of the simulation.}
\end{figure}

\begin{figure}[H]
\includegraphics[width=0.47\textwidth]{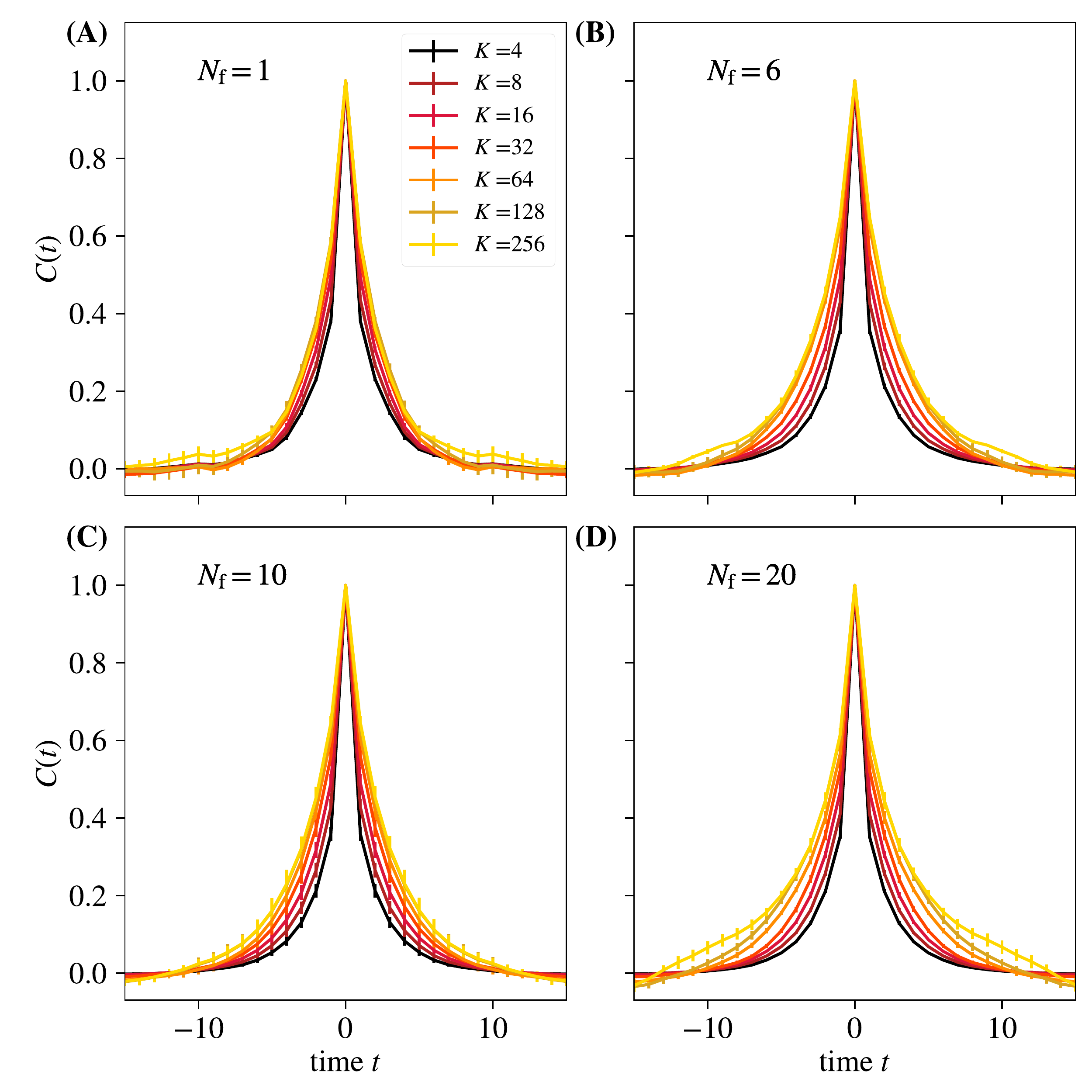}
\caption{\label{fig:fig_18} Average autocorrelation function for cluster sizes $K=2, 4, ..., 256$  for different $N_{\rm f}$.  Error bars (too small to be seen) are standard deviations over randomly selected contiguous quarters of the simulation.}
\end{figure}

\begin{figure}[H]
\includegraphics[width=0.47\textwidth]{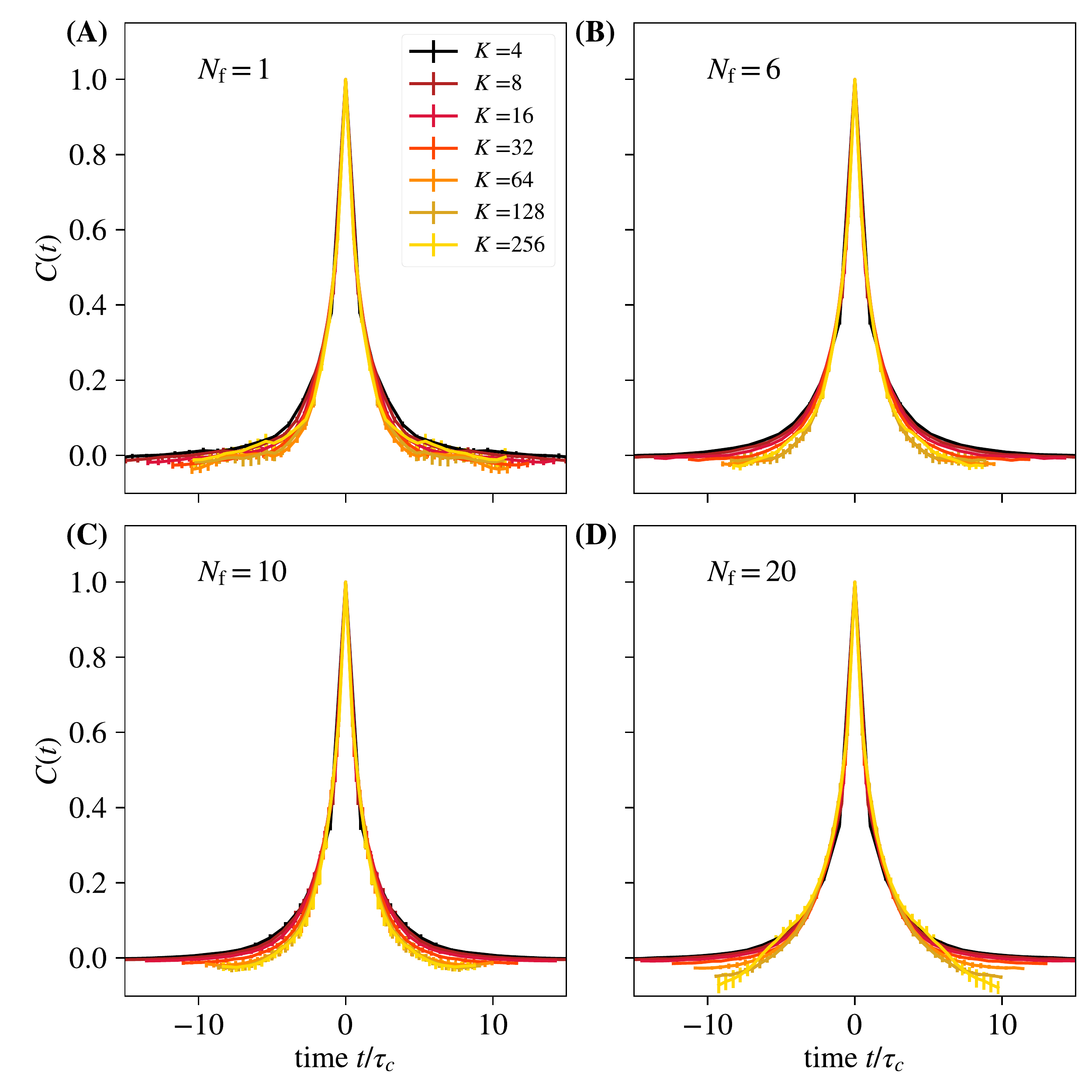}
\caption{\label{fig:fig_19} Average autocorrelation function for cluster sizes $K=2, 4, ..., 256$, where time is rescaled by the appropriate $\tau_c$ for that coarse-graining iteration, for different $N_{\rm f}$. Too many scaling fields produce negative lobes in the correlation functions, which are not observed experimentally. However, it is not clear how significant this is. Error bars (too small to be seen) are standard deviations over randomly selected contiguous quarters of the simulation. Default simulation parameters with labeled values of $N_{\rm f}$.}
\end{figure}

\begin{figure}[H]
\includegraphics[width=0.47\textwidth]{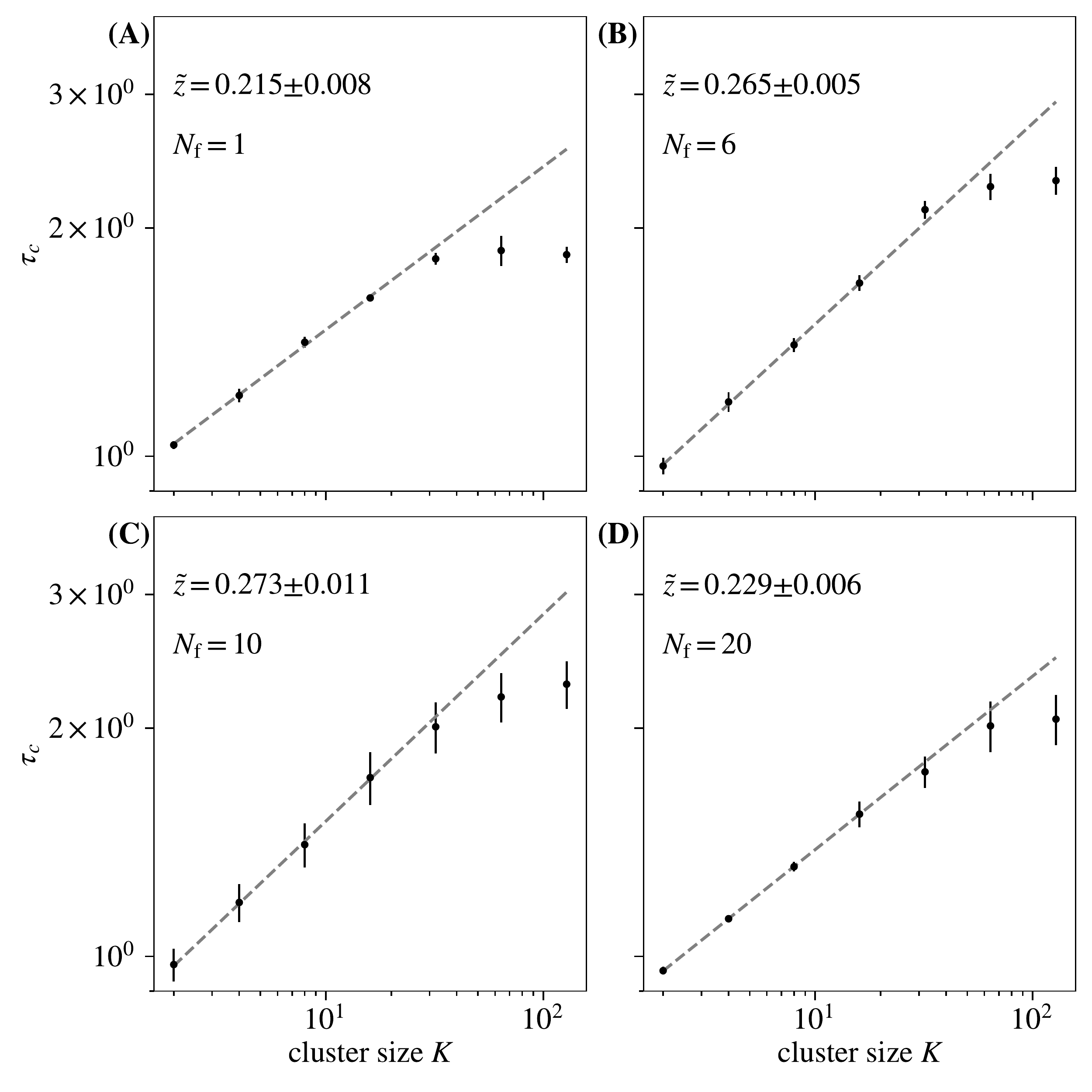}
\caption{\label{fig:fig_20} Time constants $\tau_c$ extracted from each curve in Fig.~\ref{fig:fig_19}, fitted to $\tau_c \propto K^{\tilde{z}}$,  for different $N_{\rm f}$. While changing $N_{\rm f}$ changes the value of $\tilde{z}$, the scaling persists for all explored $N_{\rm f}$. Default simulation parameters with labeled values of $N_{\rm f}$.}

\end{figure}

\begin{figure}[H]
\includegraphics[width=0.47\textwidth]{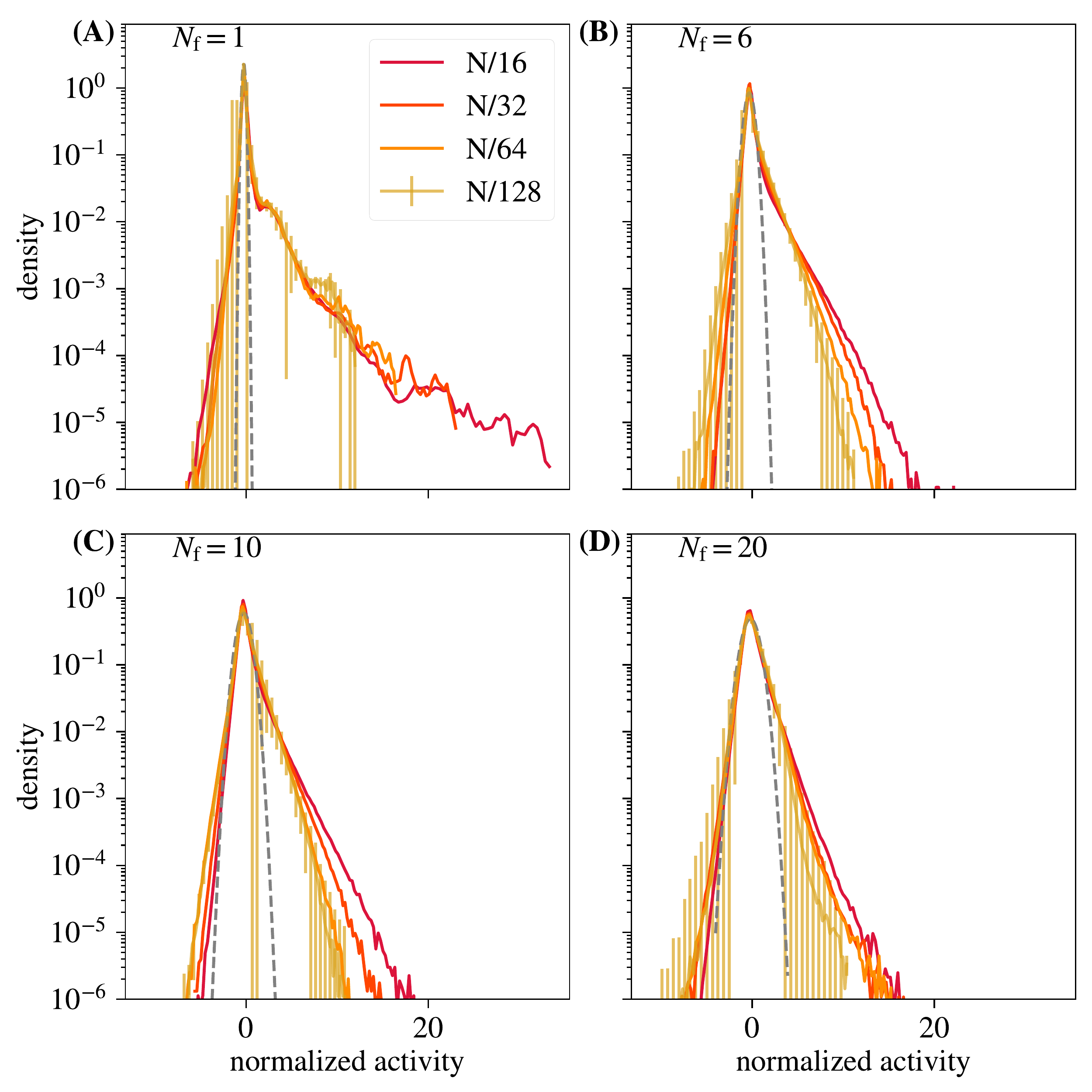}
\caption{\label{fig:fig_21} Distribution of coarse-grained variables for $k= N/16, N/32, N/64, N/128$ modes retained,  for different $N_{\rm f}$. Panel (A), with just a single latent field, shows substantial deviations from the rest. Error bars are standard deviations over randomly selected contiguous quarters of the simulation.}
\end{figure}

We find that for $N_{\rm f} < 5$, eigenvalue scaling, Fig.~\ref{fig:fig_15}, and variance scaling, Fig.~\ref{fig:fig_16}, are damaged. However, free energy scaling, Fig.~\ref{fig:fig_17}, and correlation time scaling, Fig.~\ref{fig:fig_18}-\ref{fig:fig_20}, are not significantly affected by variation in $N_{\rm f}$. Figure~\ref{fig:fig_15} through Fig.~\ref{fig:fig_21} suggest that 5 or more latent fields are required to observe scaling. 

There are hints of an upper limit of $N_{\rm f}$ for a simulation to display critical behavior. As $N_{\rm f}$ increases, distributions of coarse-grained activity become increasingly short-tailed (Fig.~\ref{fig:fig_21}). In addition, Fig.~\ref{fig:fig_14}A shows variance scaling exponent $\alpha$ approaching 1.2, and the autocorrelation starts having large negative lobes, Fig.~\ref{fig:fig_19}. It is thus possible that, for some $N_{\rm f}> 20$, the system will stop exhibiting nontrivial scaling, but additional analysis is need to confirm this.
\subsubsection{Varying the probability of coupling to a latent field $q$}

\begin{figure}[H]
\includegraphics[width=0.47\textwidth]{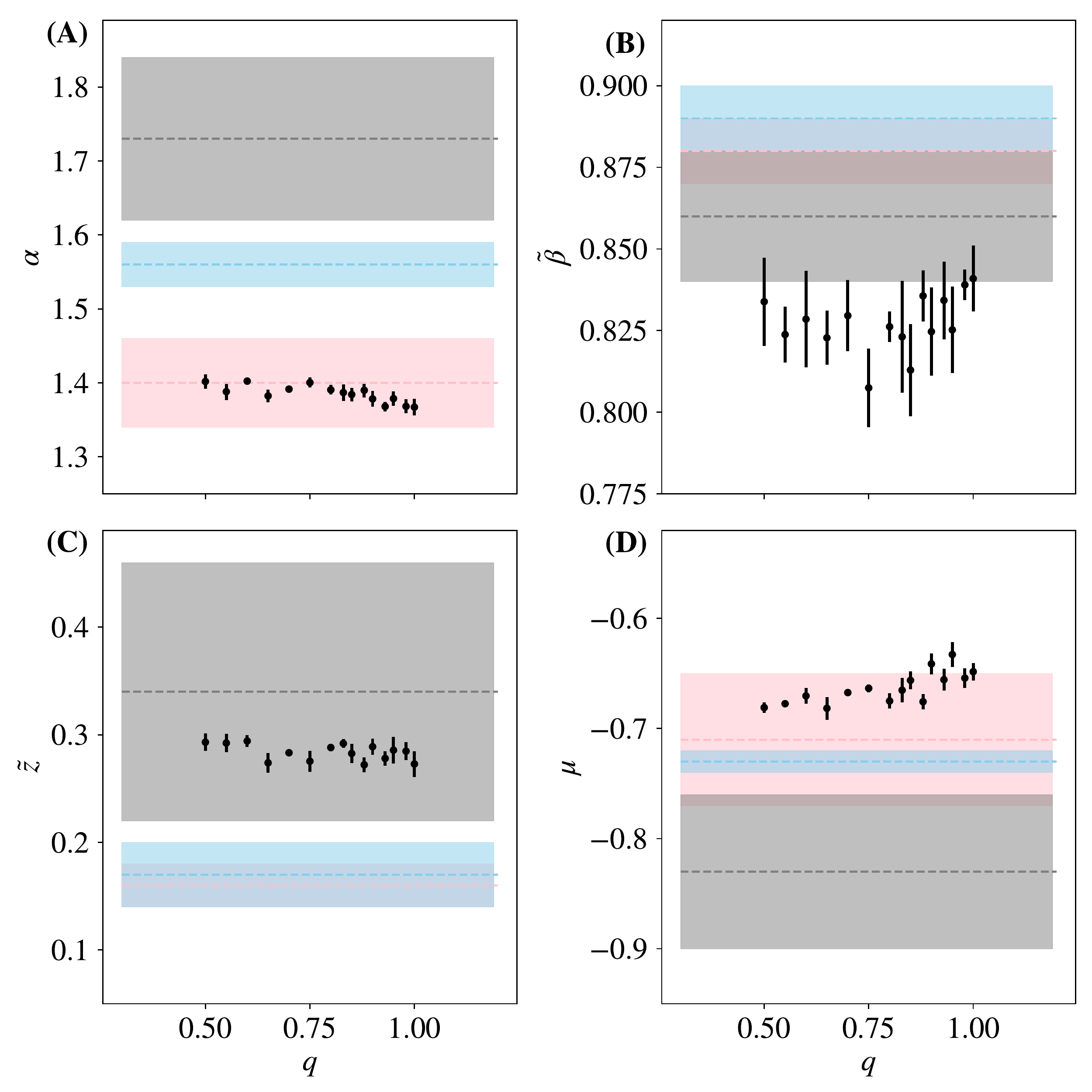}
\caption{\label{fig:fig_22} Each critical exponent, $\alpha, \tilde{\beta}, \tilde{z}, \mu$ vs probability of coupling to a latent field $q$. Results from \cite{Meshulam2018} marked and shaded in gray, pink, and blue. Error bars are standard deviations over randomly selected contiguous quarters of the simulation.}
\end{figure}

We vary the probability $q$ of coupling to a latent field in our simulation. We perform simulations with all other parameters set to the default values while sweeping over $q=0.25, \dots ,1.0$. Our simulations include a regime (Fig.~\ref{fig:fig_22}) which quantitatively matches experimental results \cite{Meshulam2018}. 

\begin{figure}[H]
\includegraphics[width=.5\textwidth]{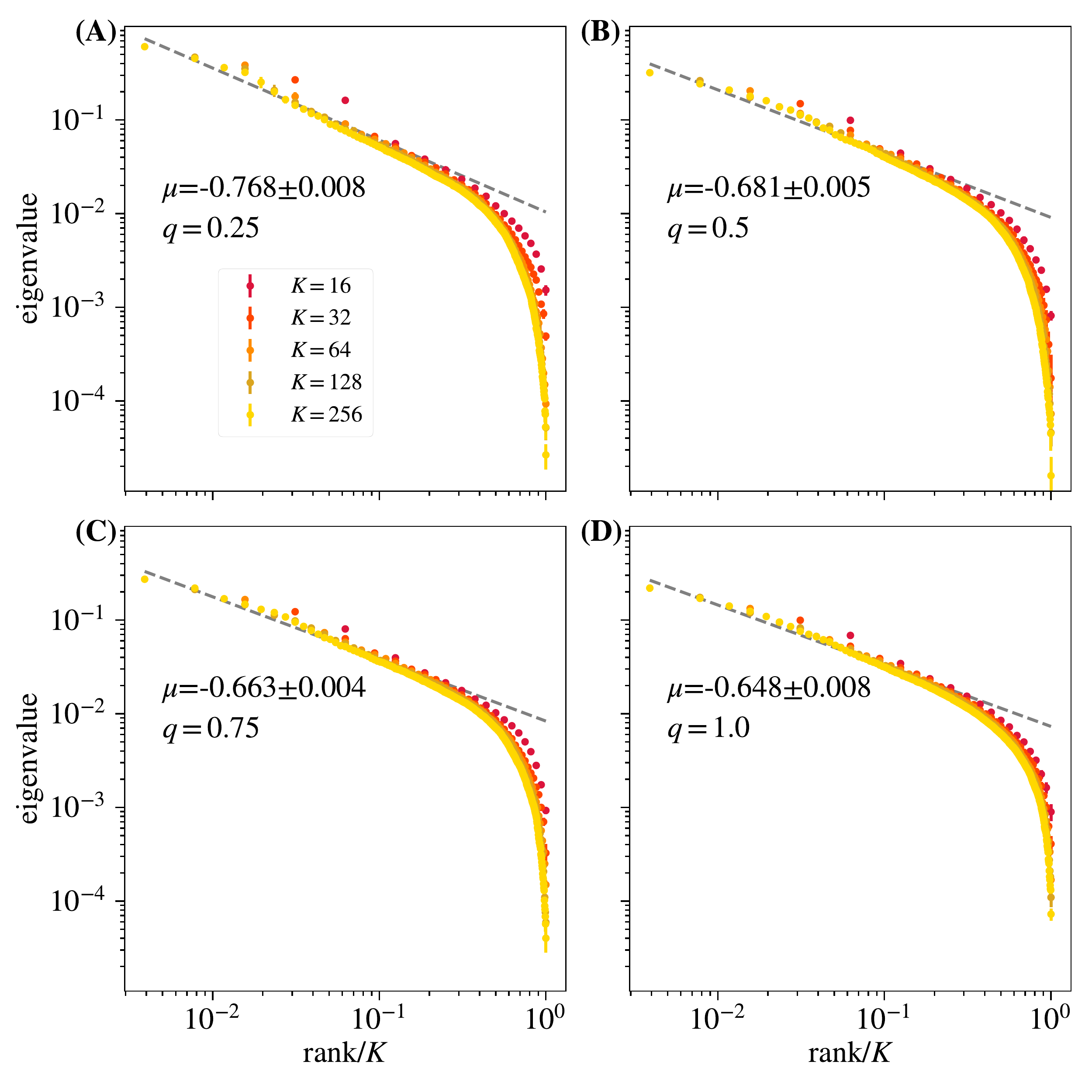}
\caption{\label{fig:fig_23} Average eigenvalue spectrum of cluster covariance for cluster sizes $K=32,64,128$ for different $q$. Varying $q$ produces no noticeable differences in scaling quality. Error bars are standard deviations over randomly selected contiguous quarters of the simulation.}
\end{figure}

\begin{figure}[H]
\includegraphics[width=0.47\textwidth]{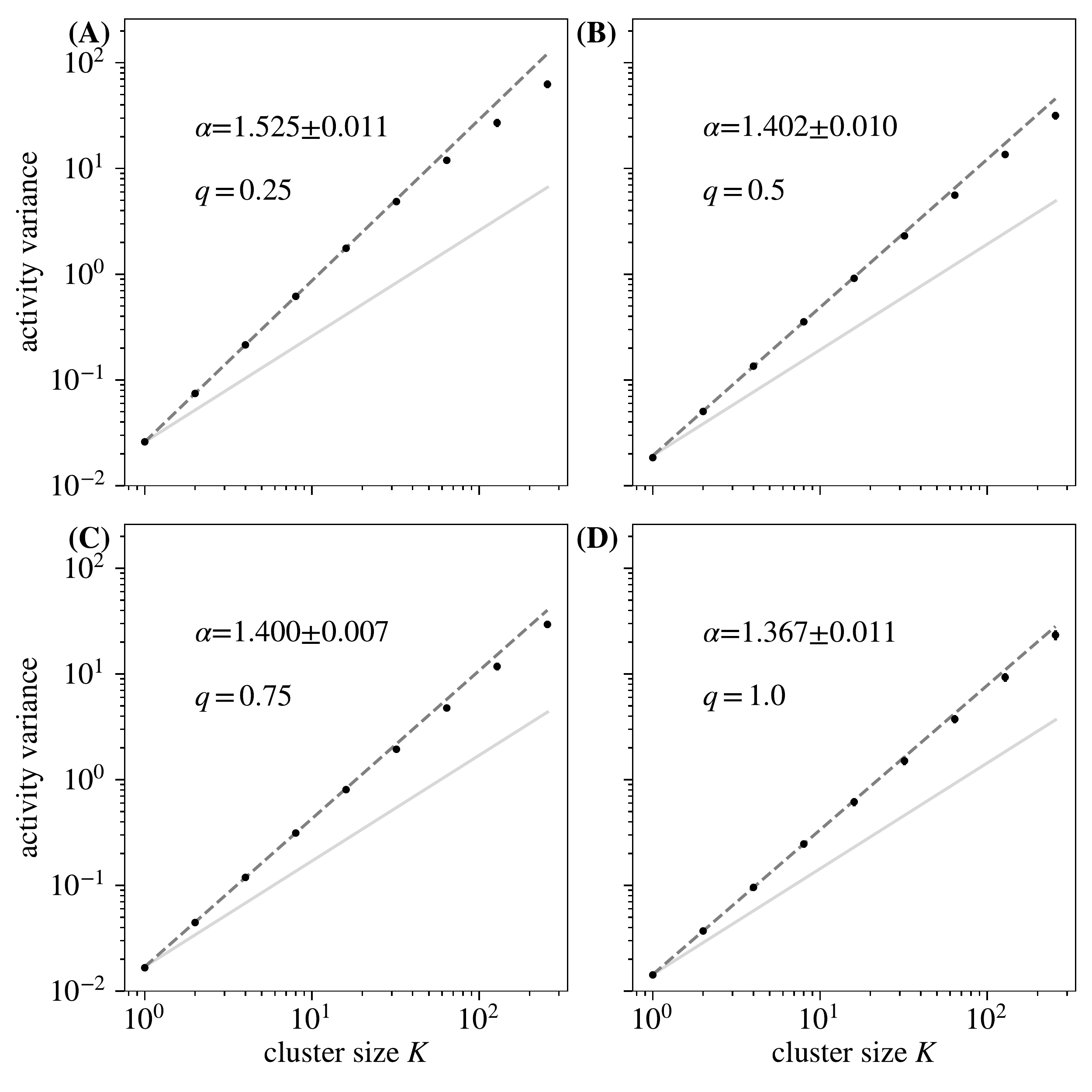}
\caption{\label{fig:fig_24}Activity variance over coarse-grained variables at each coarse-graining iteration for different $q$. For $q = 0.25$, deviations from a scaling relationship appear at large cluster sizes. Error bars are standard deviations over randomly selected contiguous quarters of the simulation.}
\end{figure}

\begin{figure}[H]
\includegraphics[width=0.47\textwidth]{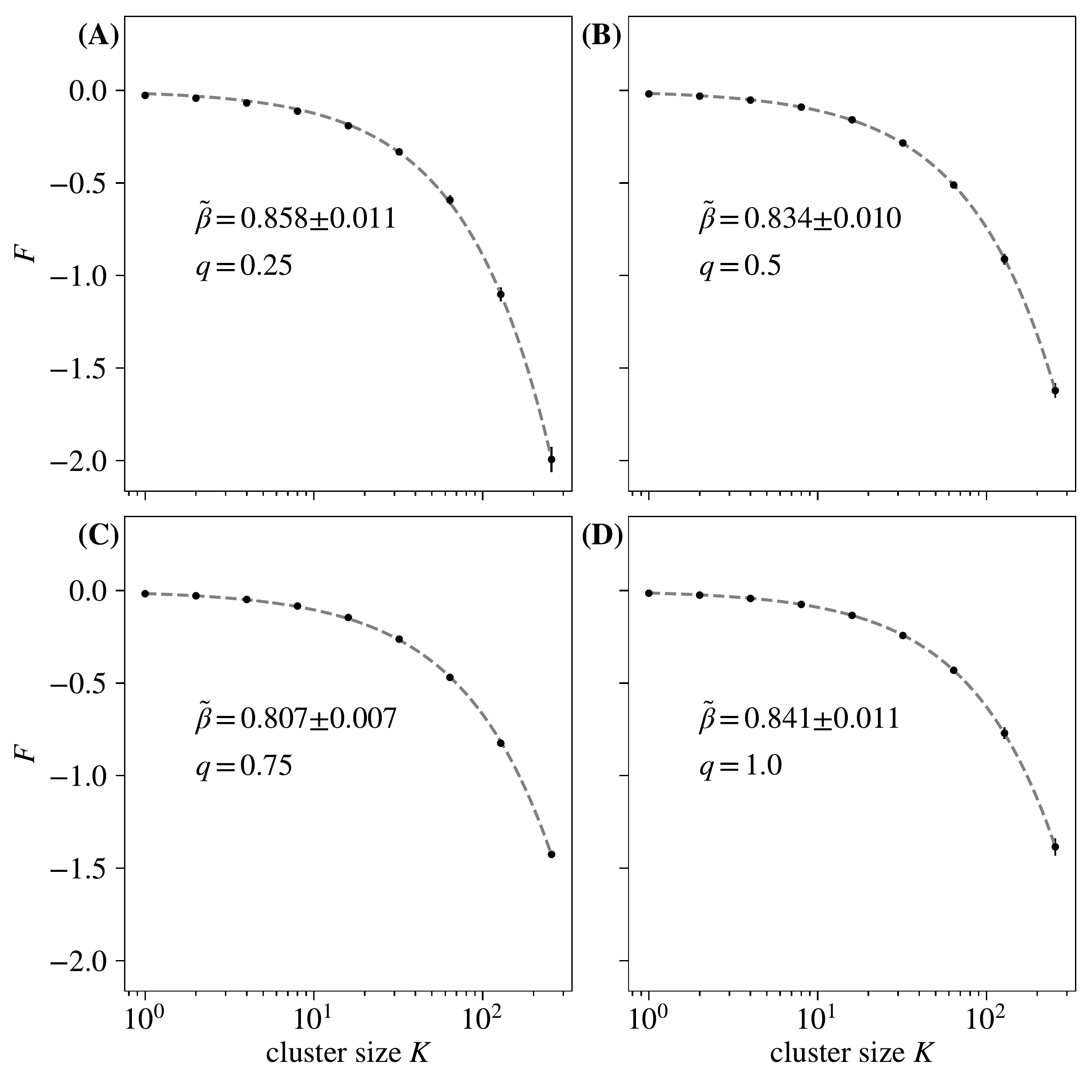}
\caption{\label{fig:fig_25} Average free energy at each coarse-graining iteration for different $q$. While changing $q$ changes the value of $\tilde{\beta}$, the scaling persists for all explored $q$. Error bars are standard deviations over randomly selected contiguous quarters of the simulation.}
\end{figure}

\begin{figure}[H]
\includegraphics[width=0.47\textwidth]{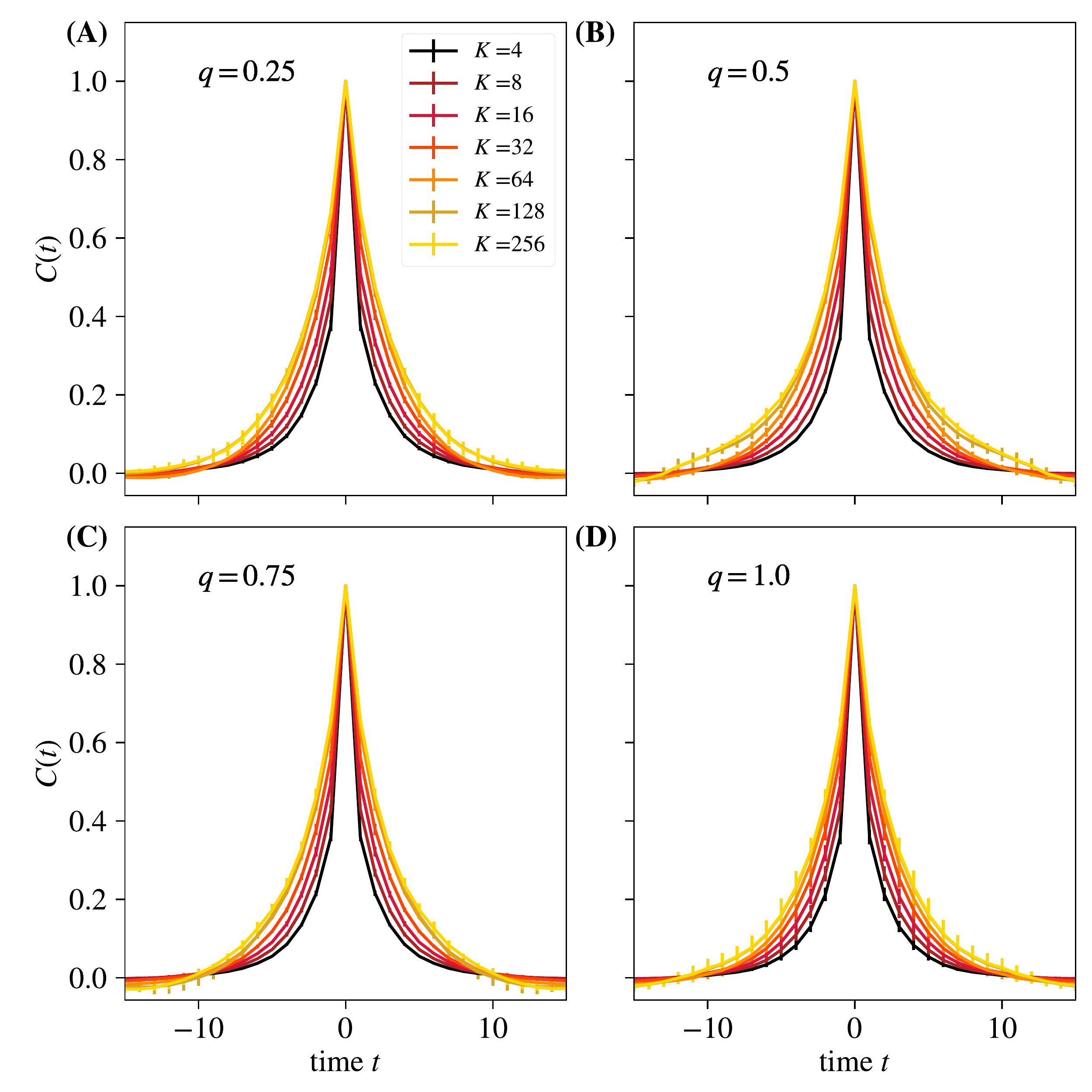}
\caption{\label{fig:fig_26} Average autocorrelation function for cluster sizes $K=2, 4, ..., 256$ for different $q$. Error bars are standard deviations over randomly selected contiguous quarters of the simulation.}
\end{figure}

\begin{figure}[H]
\includegraphics[width=0.47\textwidth]{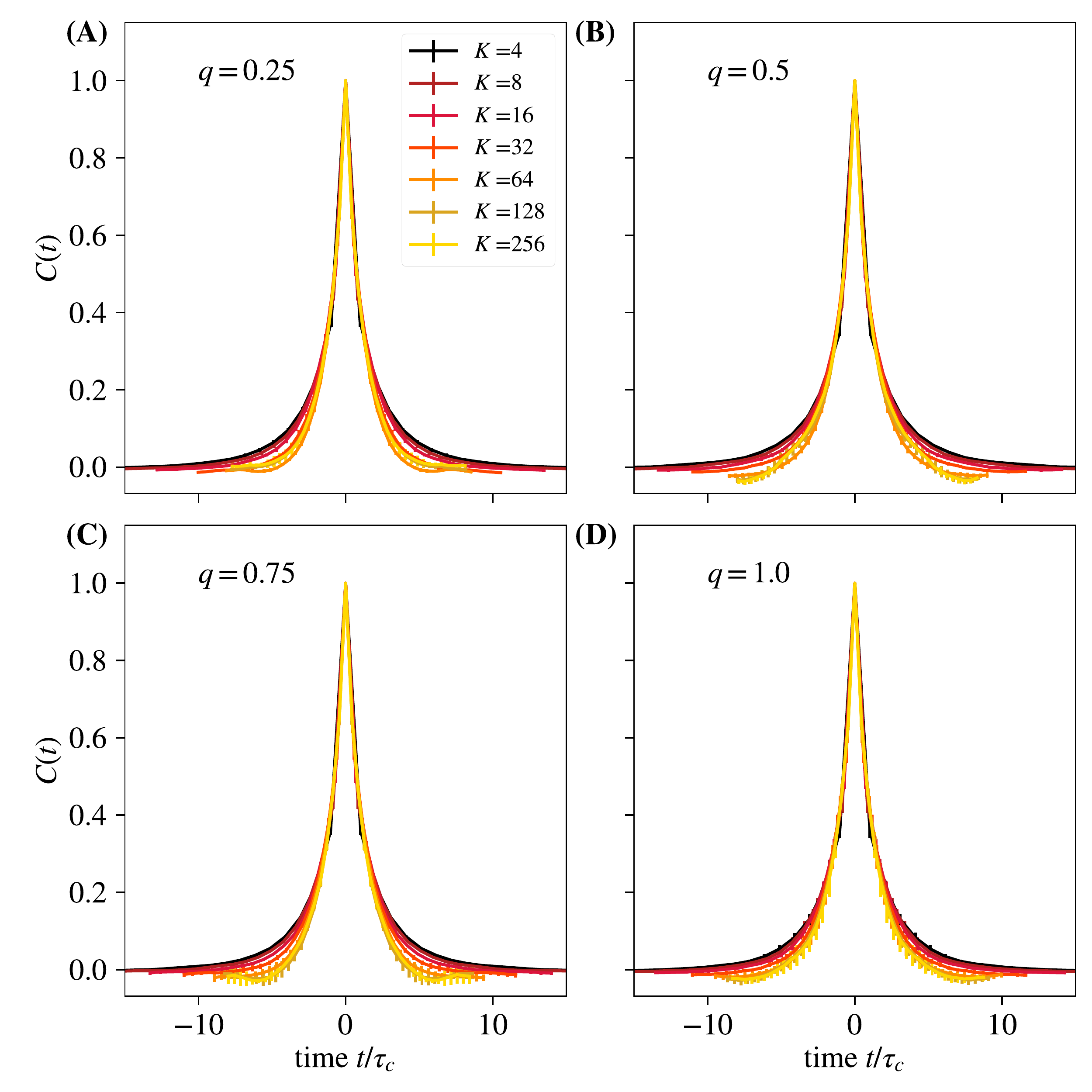}
\caption{\label{fig:fig_27} Average autocorrelation function for cluster sizes $K=2, 4, ..., 256$ where time is rescaled by the appropriate $\tau_c$ for that coarse-graining iteration, for different $q$. Error bars are standard deviations over randomly selected contiguous quarters of the simulation.}
\end{figure}

\begin{figure}[H]
\includegraphics[width=0.47\textwidth]{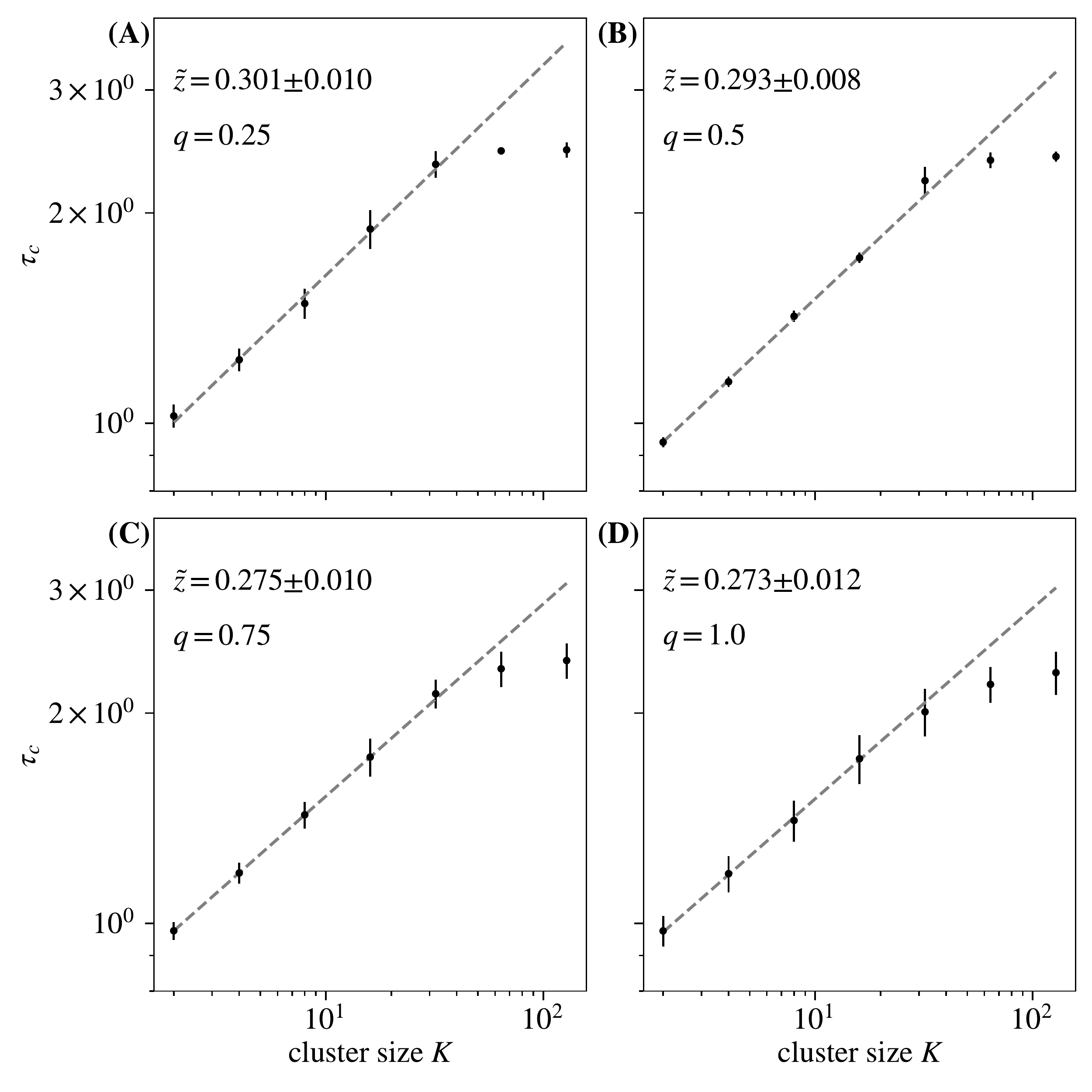}
\caption{\label{fig:fig_28} Time constants $\tau_c$ extracted from each curve in in Fig.~\ref{fig:fig_26}, and observe behavior obeying $\tau_c \propto K^{\tilde{z}}$ for roughly 1 decade. Varying $q$ yields no significant difference in quality of temporal scaling. Error bars are standard deviations over randomly selected contiguous quarters of the simulation.}
\end{figure}

\begin{figure}[H]
\includegraphics[width=0.47\textwidth]{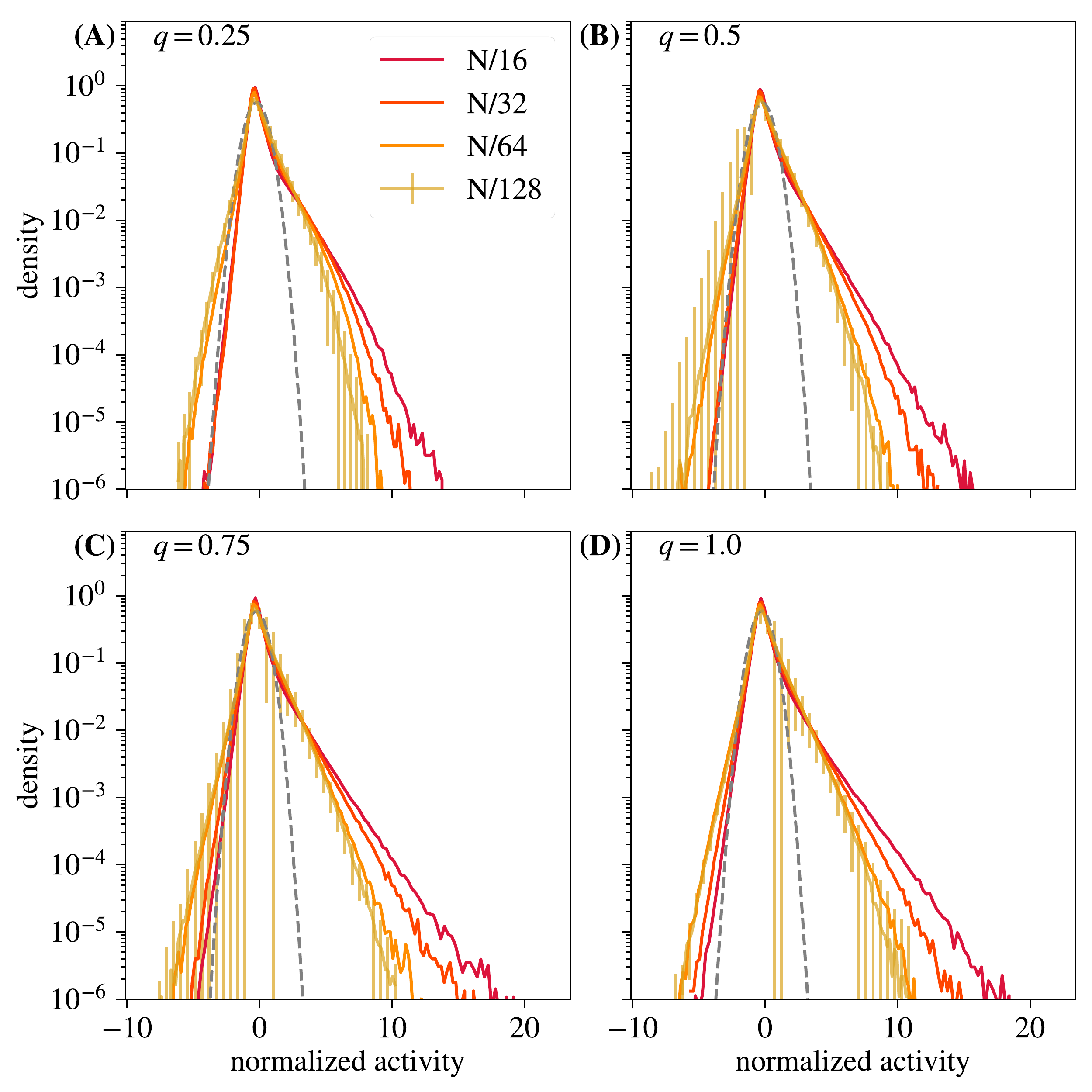}
\caption{\label{fig:fig_29} Distribution of coarse-grained variables for $k= N/16, N/32, N/64, N/128$ modes retained for different $q$. Convergence to a non-Gaussian fixed point is unaffected by varying $q$. Error bars are standard deviations over randomly selected contiguous quarters of the simulation.}
\end{figure}

We find that varying $q$ causes slight deviations in variance scaling for $q < 0.5$ (Fig.~\ref{fig:fig_24}). Eigenvalue scaling (Fig.~\ref{fig:fig_23}), free energy scaling (Fig.~\ref{fig:fig_25}), approach to a non-Gaussian fixed point (Fig.~\ref{fig:fig_29}), and correlation time scaling (Fig.~\ref{fig:fig_26}-\ref{fig:fig_28}) are not significantly affected by variation in $q$ from $0.5$ to $1.0$. We conclude that varying the probability of coupling to a latent field does not have a significant impact on scaling for $q \geq 0.5$, but is deleterious to scaling for $q<0.5$. 
\subsubsection{Varying the penalty term $\epsilon$}
\begin{figure}[H]
\includegraphics[width=0.47\textwidth]{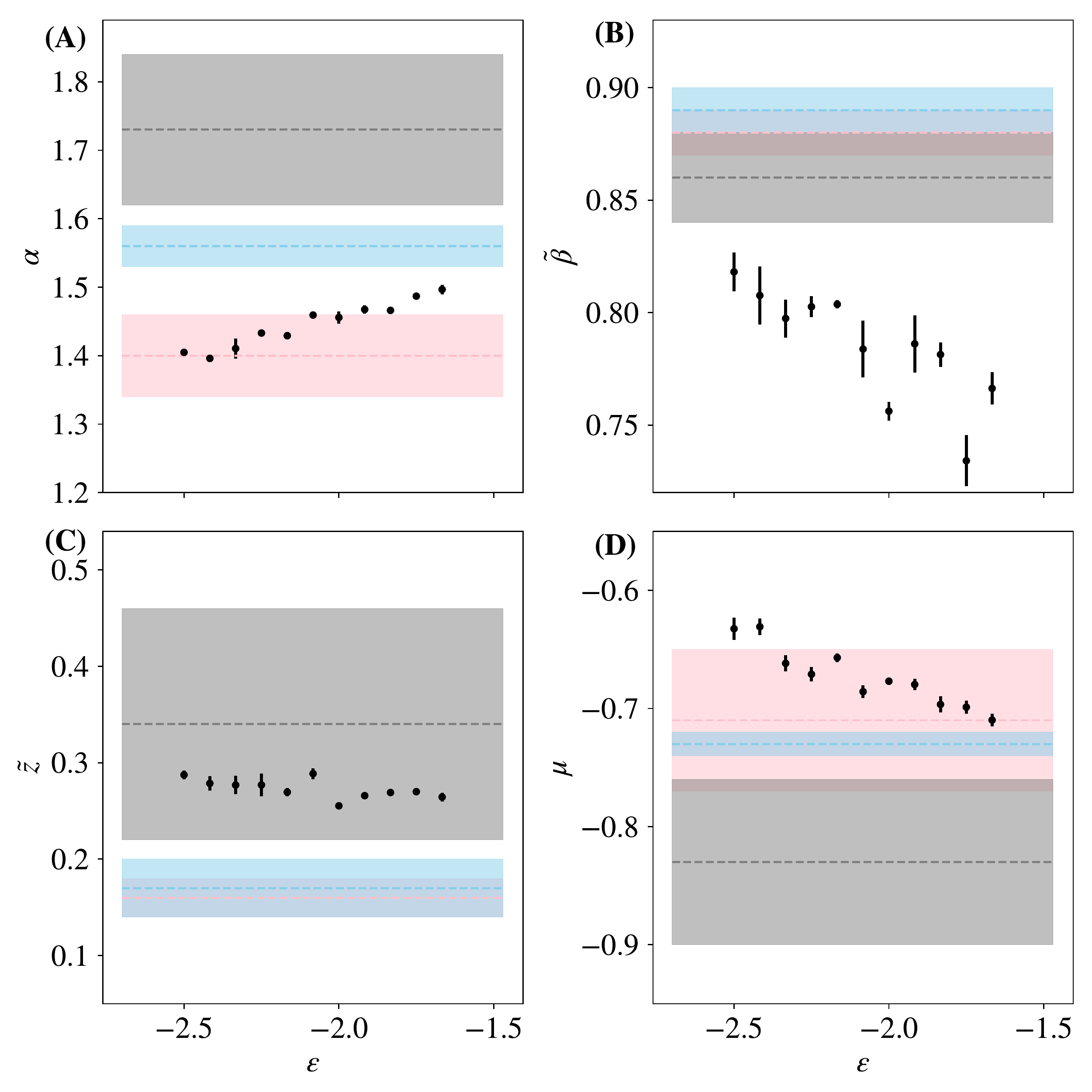}
\caption{\label{fig:fig_30} Each critical exponent, $\alpha, \tilde{\beta}, \tilde{z}, \mu$ vs penalty term $
\epsilon$. Results from \cite{Meshulam2018} marked and shaded in gray, pink, and blue. Error bars are standard deviations over randomly selected contiguous quarters of the simulation}
\end{figure}

We perform simulations sweeping over values of the penalty term $\epsilon$, which controls the sparseness of activity. In Fig.~\ref{fig:fig_30}, we note that our simulations include a regime which quantitatively matches experimental results \cite{Meshulam2018}. 

\begin{figure}[H]
\includegraphics[width=0.47\textwidth]{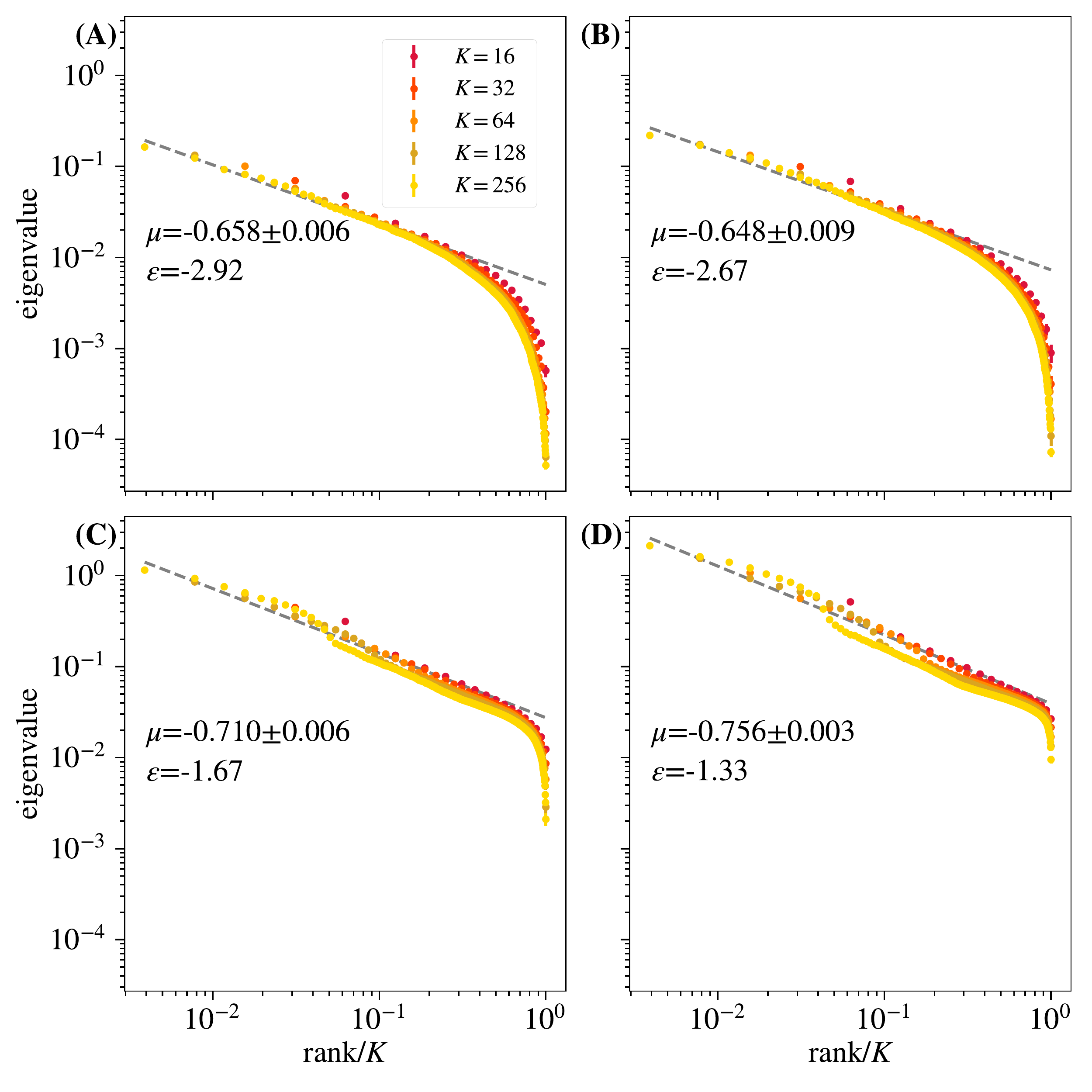}
\caption{\label{fig:fig_31} Average eigenvalue spectrum of cluster covariance for cluster sizes $K=32,64,128$ for different $\epsilon$. For $\epsilon = -1.33$ and $\epsilon = -1.67$, the scaling relationship degrades. Error bars are standard deviations over randomly selected contiguous quarters of the simulation.}
\end{figure}

\begin{figure}[H]
\includegraphics[width=0.47\textwidth]{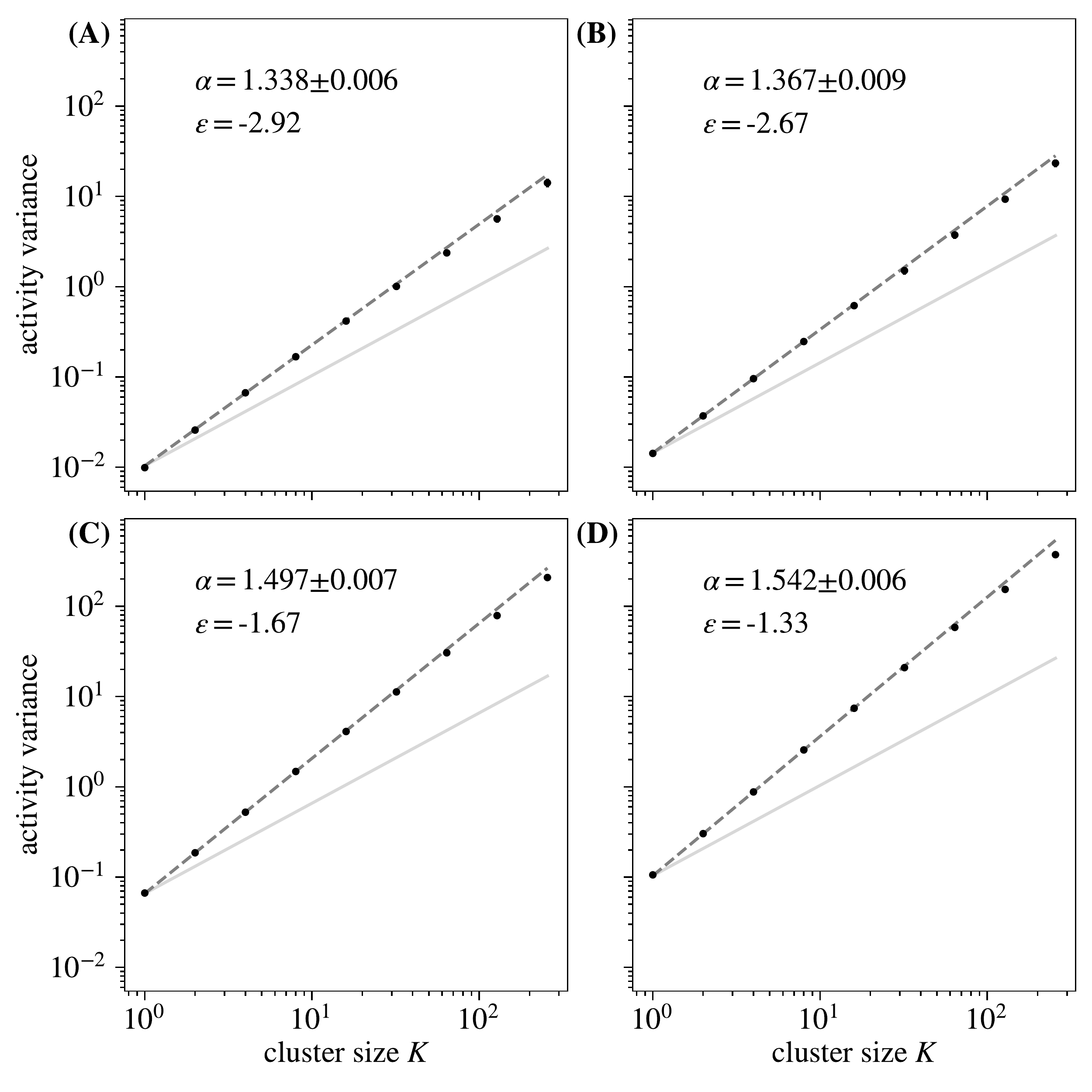}
\caption{\label{fig:fig_32}Activity variance over coarse-grained variables at each coarse-graining iteration for different $\epsilon$. Varying $\epsilon$ yields no significant difference in scaling quality. Error bars are standard deviations over randomly selected contiguous quarters of the simulation.}
\end{figure}

\begin{figure}[H]
\includegraphics[width=0.47\textwidth]{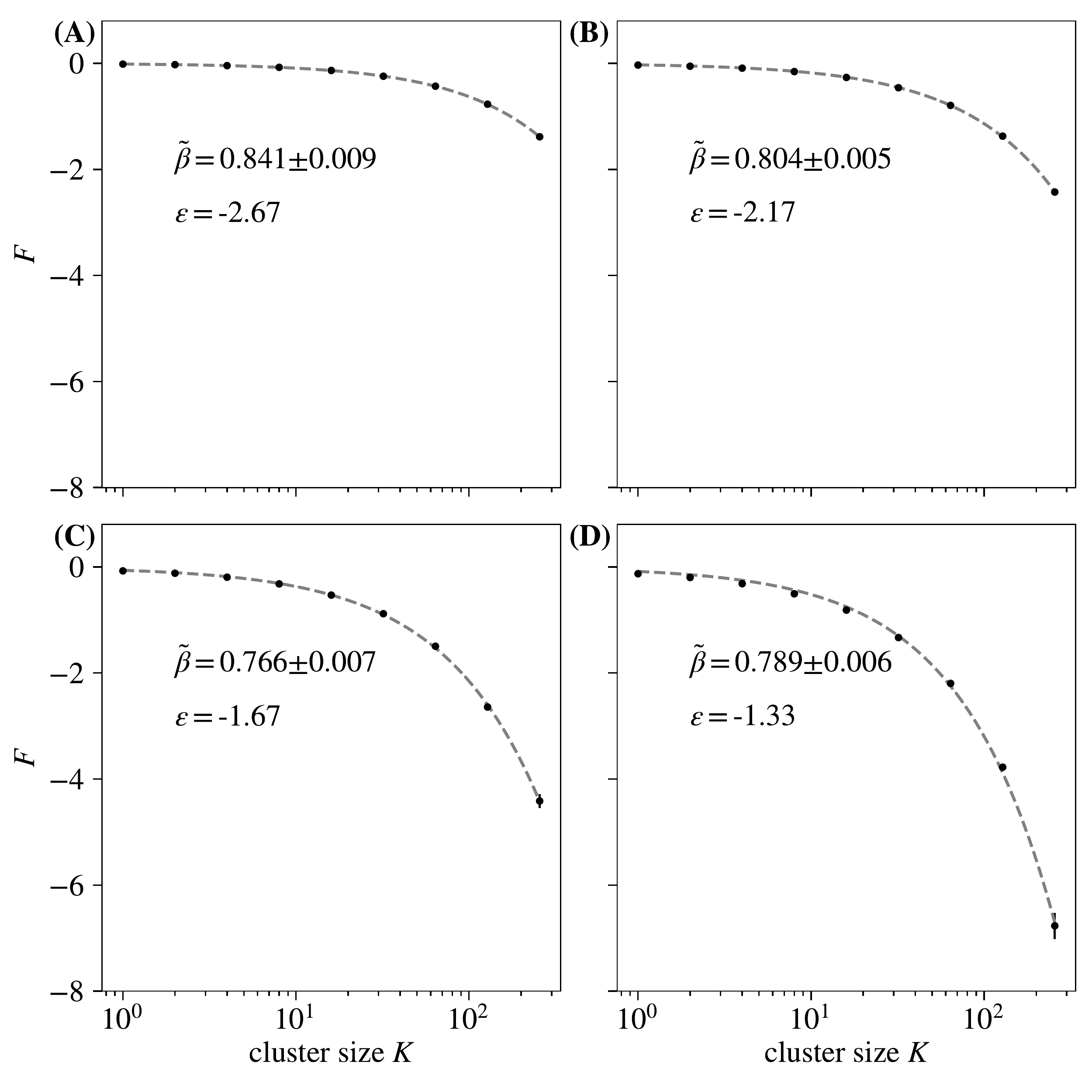}
\caption{\label{fig:fig_33} Average free energy at each coarse-graining iteration for different $\epsilon$. Varying $\epsilon$ yields no significant change in quality of free energy scaling. Error bars are standard deviations over randomly selected contiguous quarters of the simulation.}
\end{figure}

\begin{figure}[H]
\includegraphics[width=0.47\textwidth]{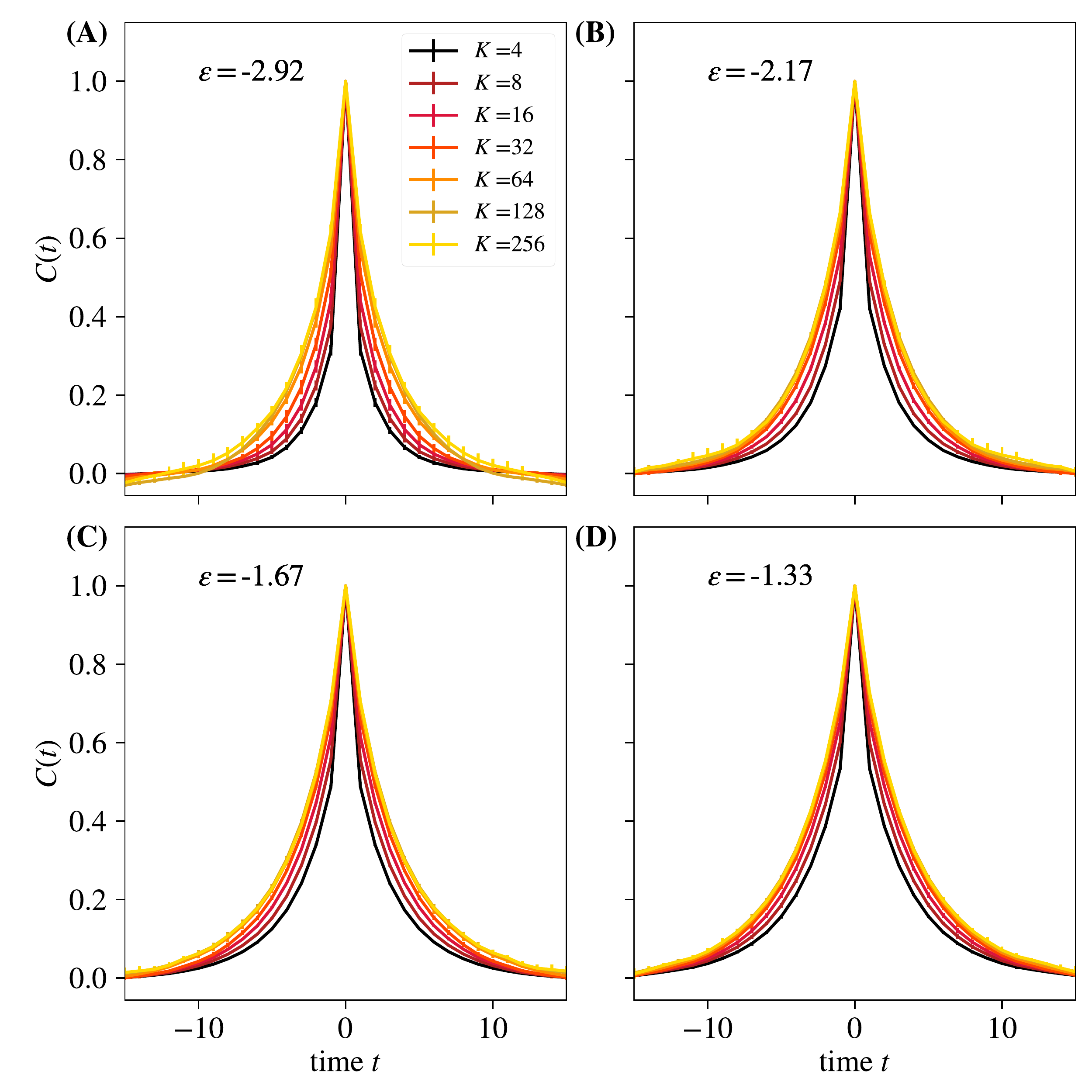}
\caption{\label{fig:fig_34} Average autocorrelation function for cluster sizes $K=2, 4, ..., 256$ for different $\epsilon$. Error bars are standard deviations over randomly selected contiguous quarters of the simulation.}
\end{figure}

\begin{figure}[H]
\includegraphics[width=0.47\textwidth]{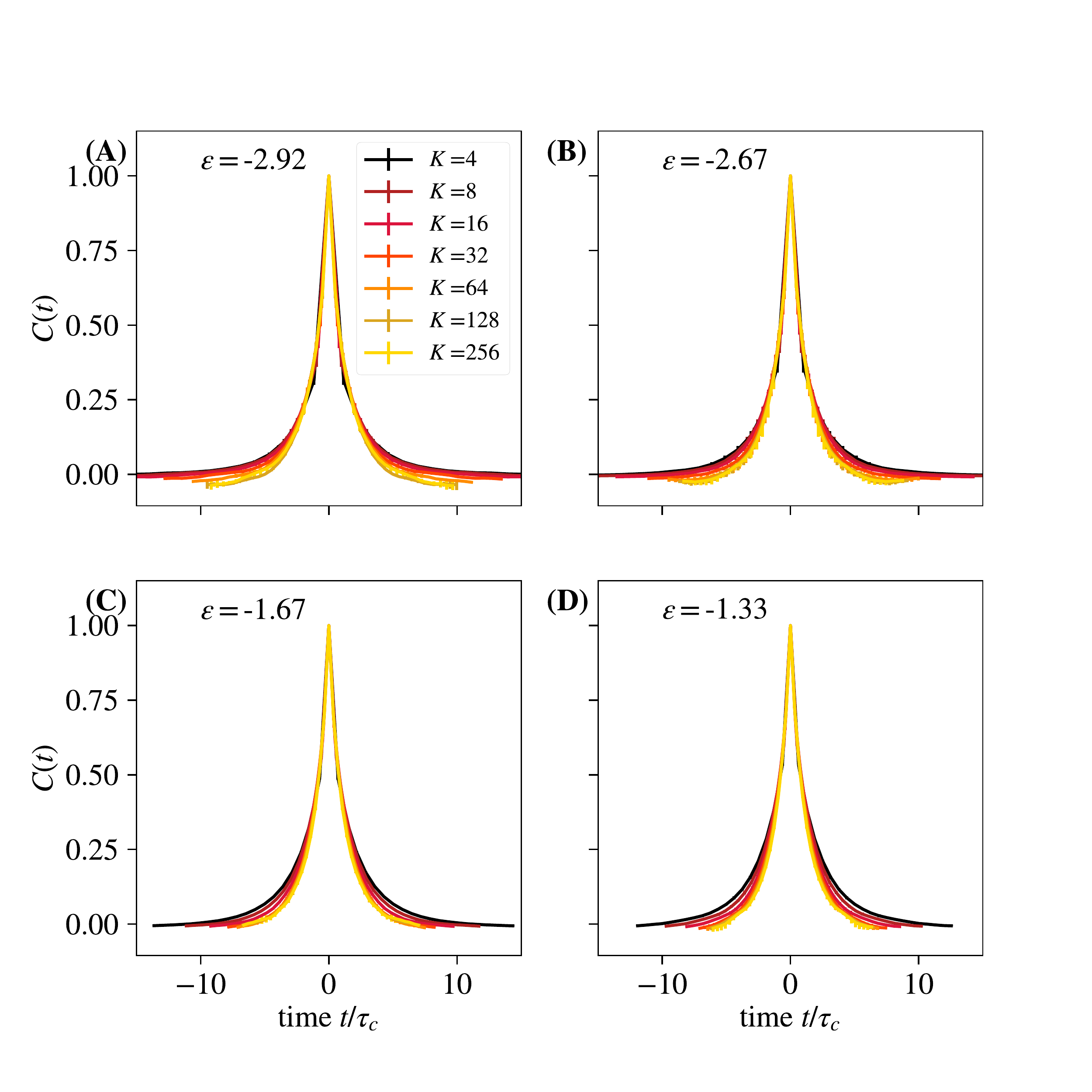}
\caption{\label{fig:fig_35} Average autocorrelation function for cluster sizes $K=2, 4, ..., 256$ where time is rescaled by the appropriate $\tau_c$ for that coarse-graining iteration, for different $\epsilon$. Error bars are standard deviations over randomly selected contiguous quarters of the simulation.}
\end{figure}

\begin{figure}[H]
\includegraphics[width=0.47\textwidth]{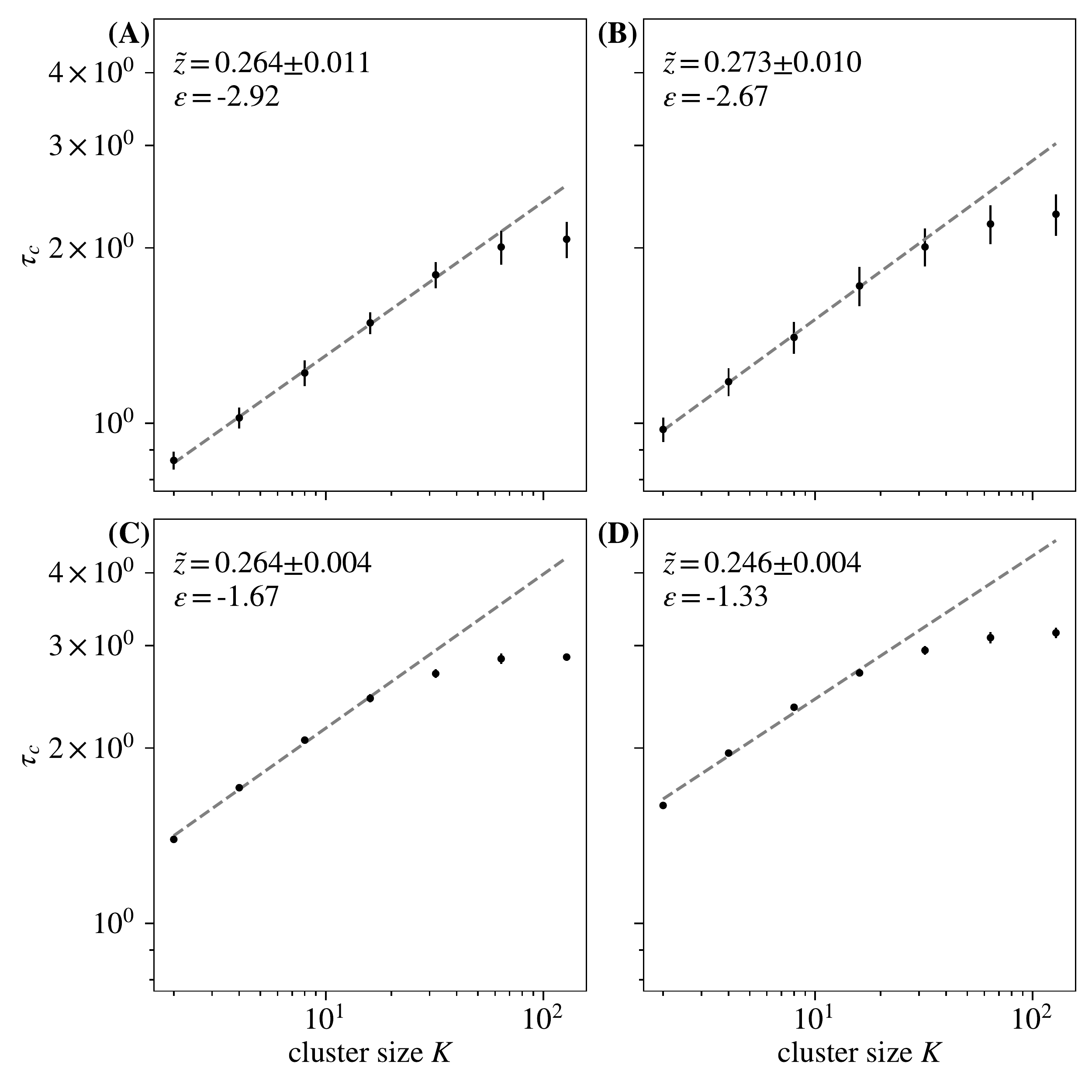}
\caption{\label{fig:fig_36} Time constants $\tau_c$ extracted from each curve in Fig.~\ref{fig:fig_34}. Observe behavior obeying $\tau_c \propto K^{\tilde{z}}$ for roughly 1 decade. Varying $\epsilon$ results in no significant changes in temporal scaling. Error bars are standard deviations over randomly selected contiguous quarters of the simulation.}
\end{figure}

\begin{figure}[H]
\includegraphics[width=0.47\textwidth]{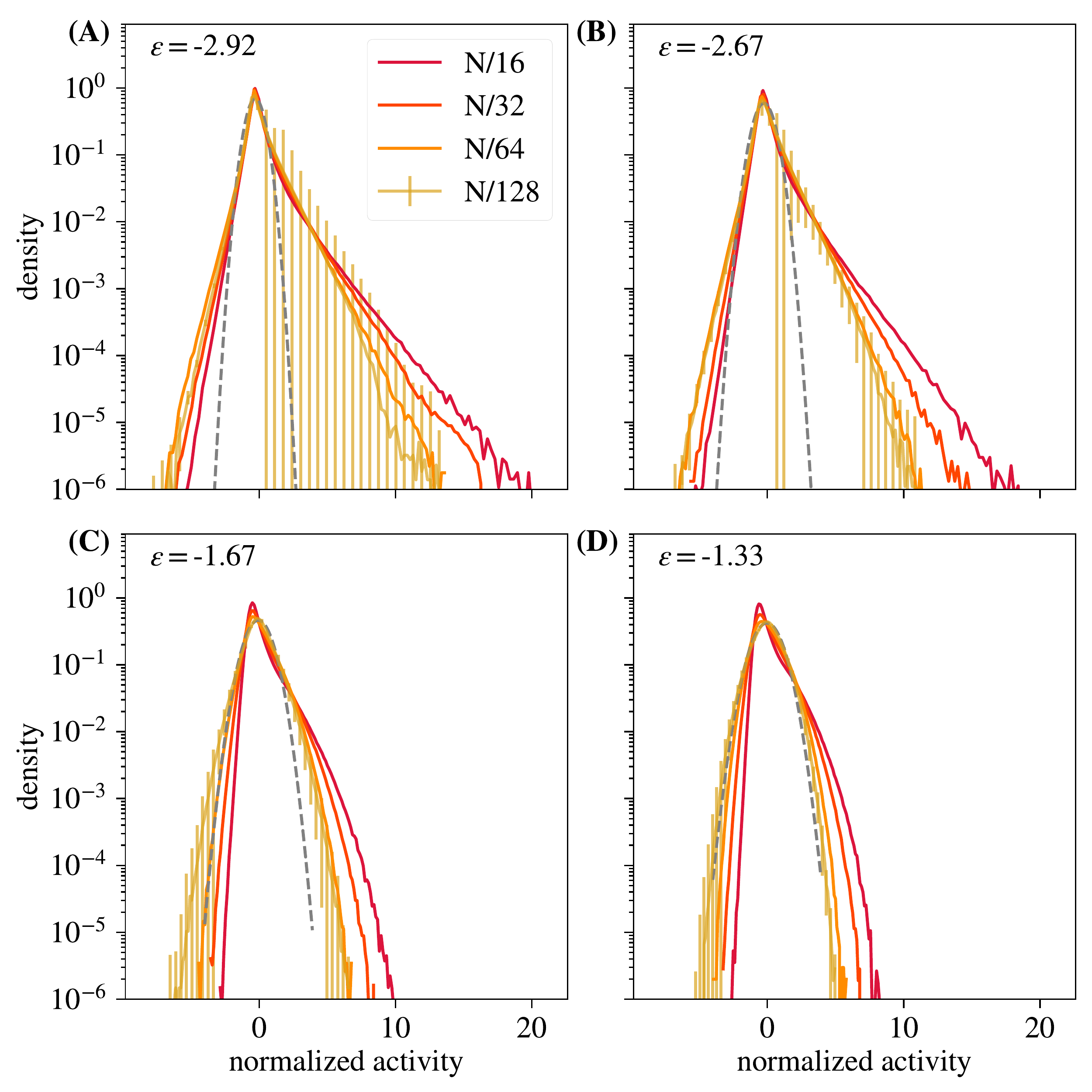}
\caption{\label{fig:fig_37} Distribution of coarse-grained variables for $k= N/16, N/32, N/64, N/128$ modes retained for different $\epsilon$. Coarse-grained distributions of activity approach a non-Gaussian fixed point for $\epsilon > -1.5$. Error bars are standard deviations over randomly selected contiguous quarters of the simulation.}
\end{figure}

Several scaling results are sensitive to $\epsilon$, with damaged scaling for large $\epsilon$, which corresponds to higher overall levels of activity. We find that varying $\epsilon$ significantly damages eigenvalue scaling for $\epsilon > -1.92$ (Fig.~\ref{fig:fig_31}). We observe that coarse-grained distributions of activity from simulations with $\epsilon > -1.5$ approach but to do not reach a Gaussian fixed point (Fig.~\ref{fig:fig_37}). However, free energy scaling (Fig.~\ref{fig:fig_33}), variance scaling (Fig.~\ref{fig:fig_32}), and correlation time scaling (Fig.~\ref{fig:fig_34}-\ref{fig:fig_36}) are not significantly affected by variation in $\epsilon$. We conclude that highly active simulations do not display a clear eigenvalue spectra collapse or approach a non-Gaussian fixed point upon coarse-graining.
\subsubsection{Varying the latent field time constant $\tau$}

\begin{figure}[H]

\includegraphics[width=0.47\textwidth]{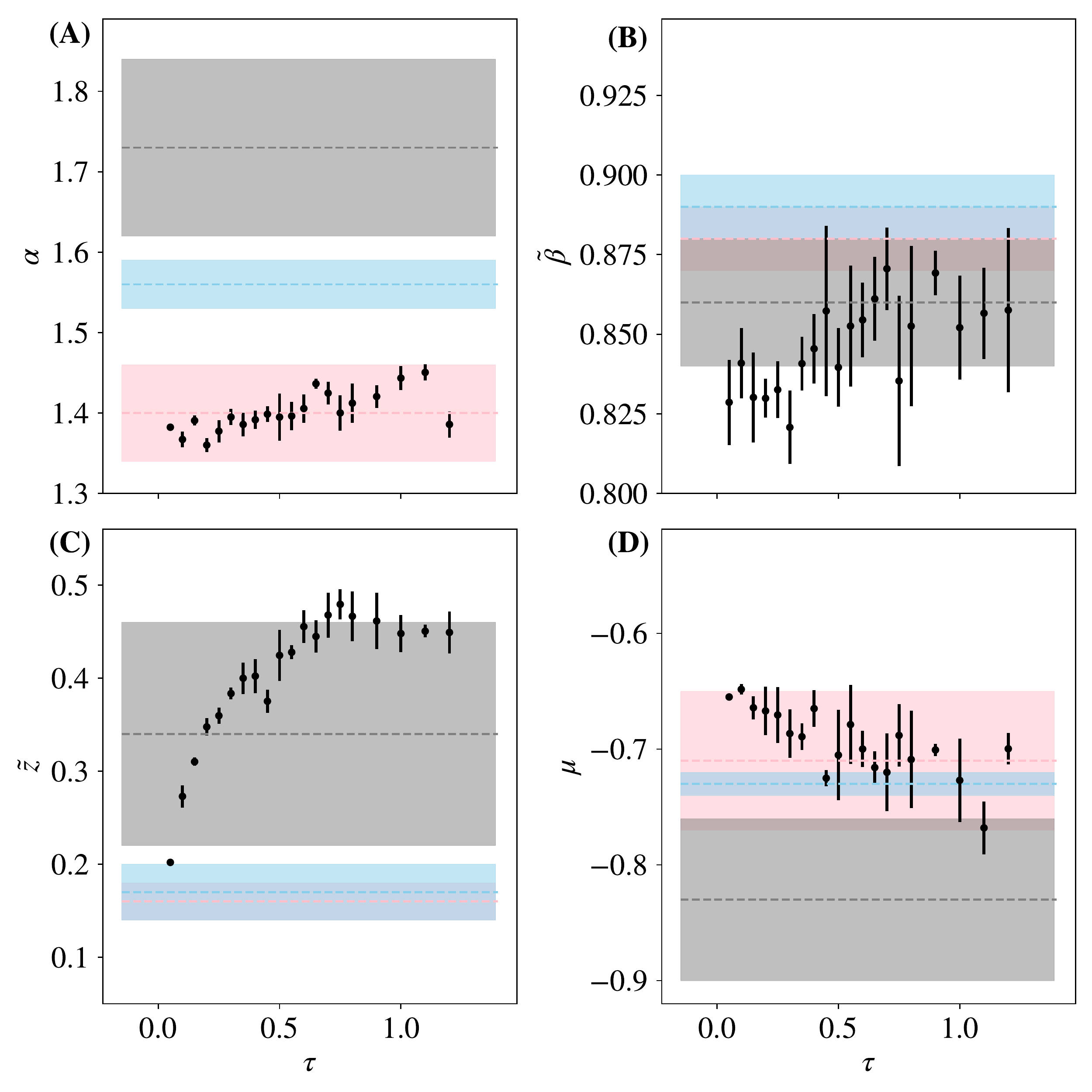}
\caption{\label{fig:fig_38} Each critical exponent, $\alpha, \tilde{\beta}, \tilde{z}, \mu$ vs latent field time constant $\tau$. Results from \cite{Meshulam2018} marked and shaded in gray, pink, and blue. Error bars are standard deviations over randomly selected contiguous quarters of the simulation.}
\end{figure}

We will now consider the effects of varying $\tau$ in our simulation while fixing the other parameters to default values. As in the Main Text, all latent fields have the same value of $\tau$. We vary $\tau$ from $0.05$ to $1.2$, where the time for one track length to be run in simulations is $1$. In Fig.~\ref{fig:fig_38}, we note that our simulations include a regime which quantitatively matches experimental results \cite{Meshulam2018}. 

\begin{figure}[H]
\includegraphics[width=0.47\textwidth]{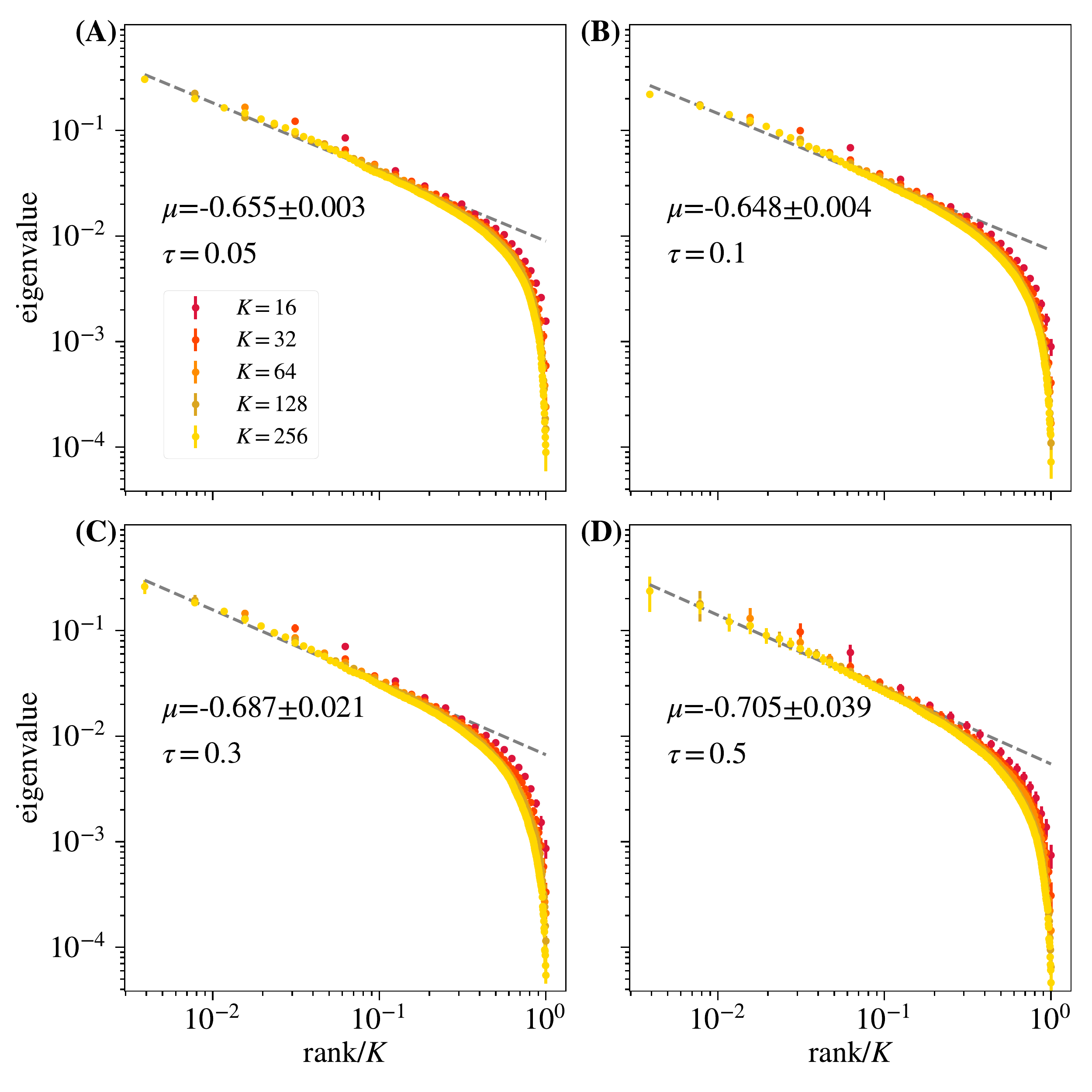}
\caption{\label{fig:fig_39} Average eigenvalue spectrum of cluster covariance for cluster sizes $K=32,64,128$ for different values of $\tau$. Varying $\tau$ does not have a significant effect on eigenvalue scaling. Error bars are standard deviations over randomly selected contiguous quarters of the simulation.}
\end{figure}

\begin{figure}[H]
\includegraphics[width=0.47\textwidth]{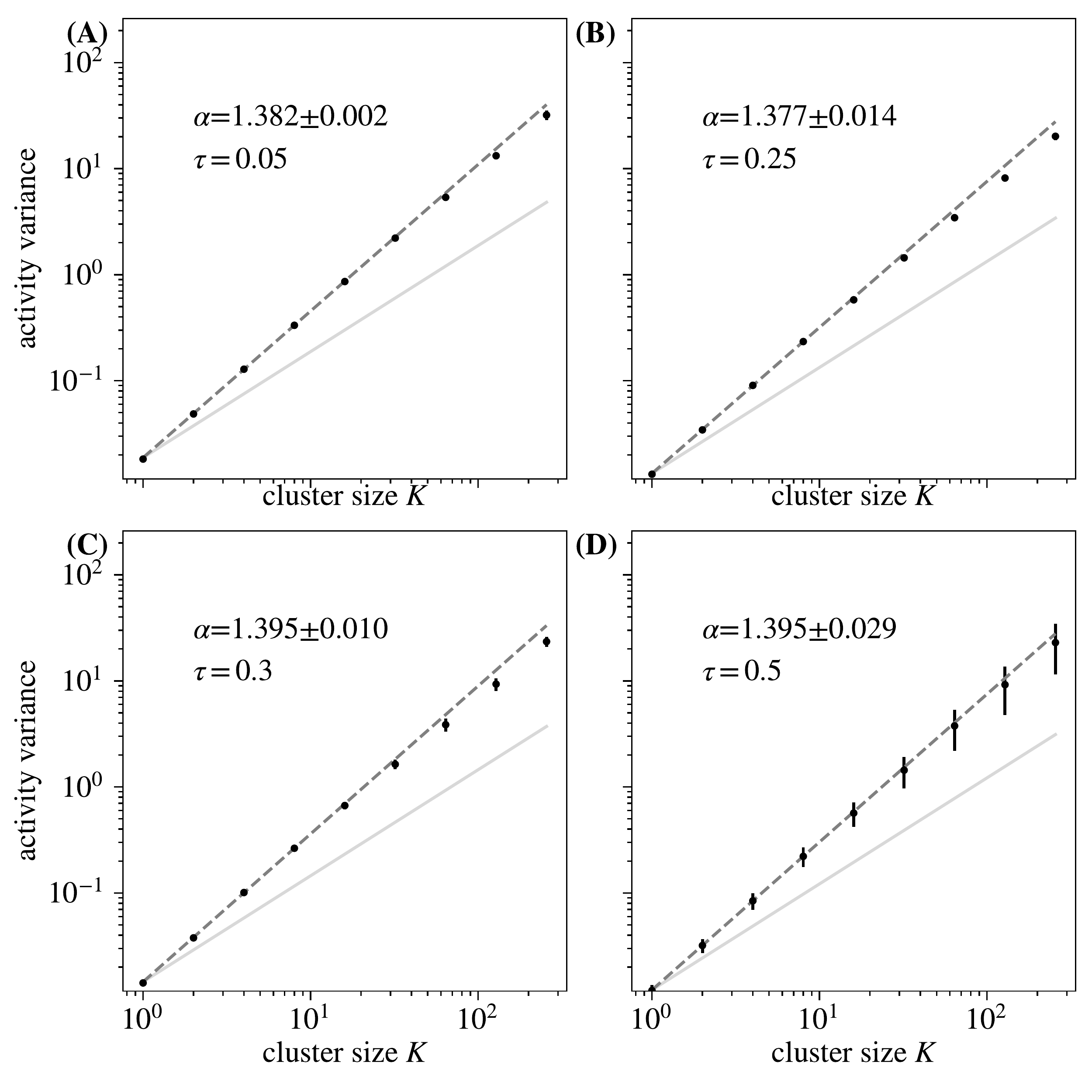}
\caption{\label{fig:fig_40}Activity variance over coarse-grained variables at each coarse-graining iteration for different values of $\tau$. Varying $\tau$ does not have a significant effect on activity variance scaling. Error bars are standard deviations over randomly selected contiguous quarters of the simulation.}
\end{figure}

\begin{figure}[H]
\includegraphics[width=0.47\textwidth]{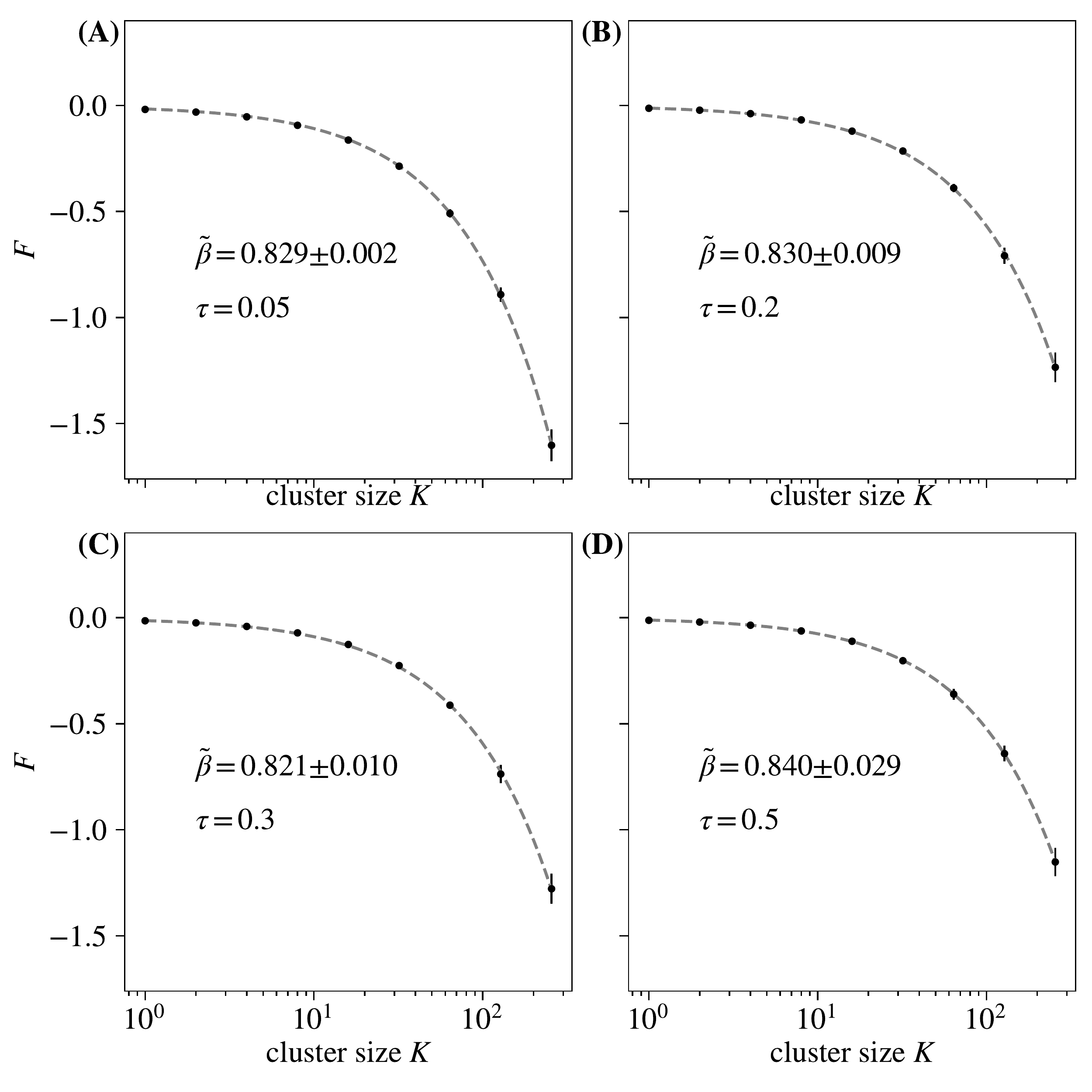}
\caption{\label{fig:fig_41} Average free energy at each coarse-graining iteration for different $\tau$. Varying $\tau$ does not appear to affect free energy scaling. Error bars are standard deviations over randomly selected contiguous quarters of the simulation.}
\end{figure}

\begin{figure}[H]
\includegraphics[width=0.47\textwidth]{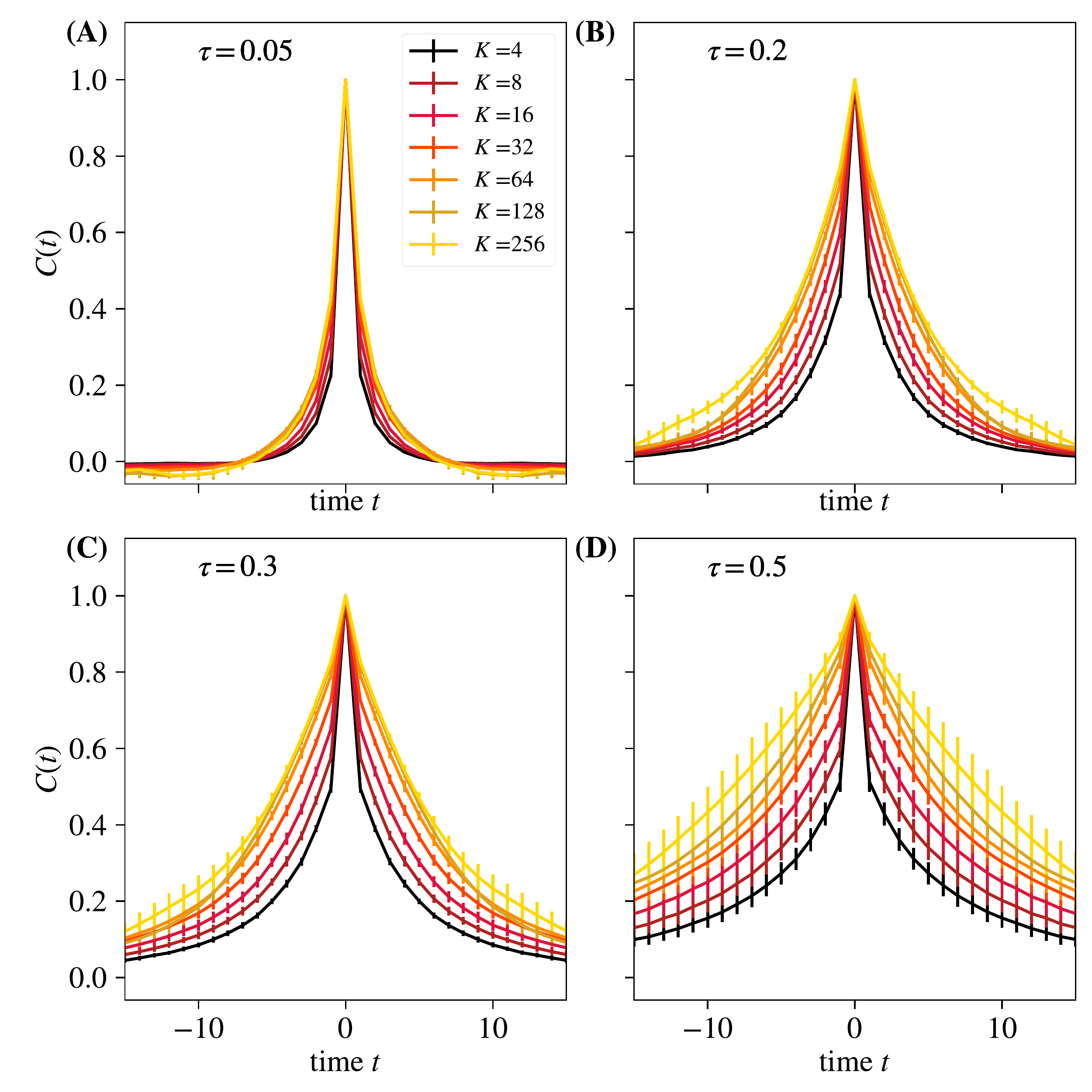}
\caption{\label{fig:fig_42} Average autocorrelation function for cluster sizes $K=2, 4, ..., 256$ as a function of time, for different $\tau$. Error bars are standard deviations over randomly selected contiguous quarters of the simulation.}
\end{figure}

\begin{figure}[H]
\includegraphics[width=0.47\textwidth]{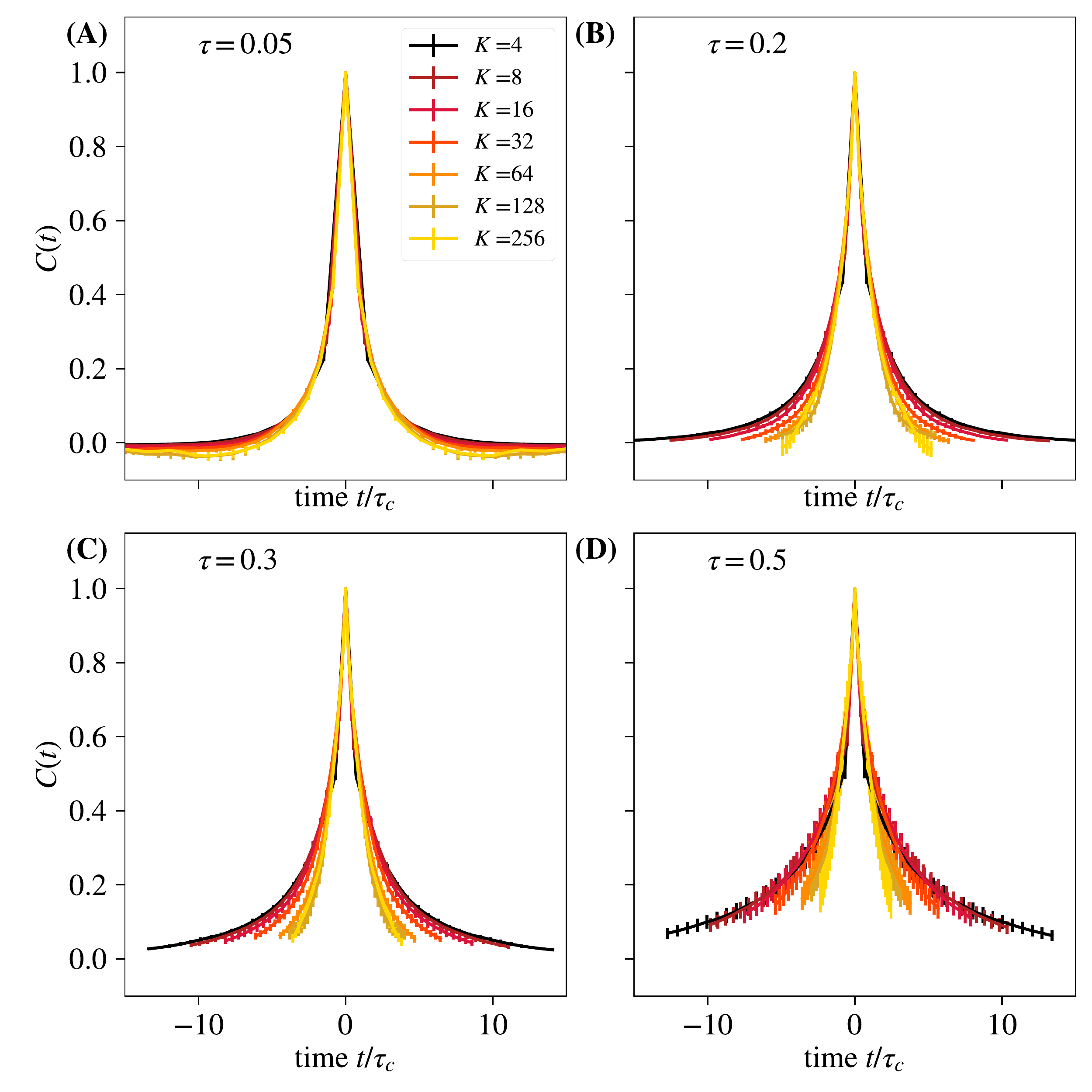}
\caption{\label{fig:fig_43} Average autocorrelation function for cluster sizes $K=2, 4, ..., 256$, where time is rescaled by the appropriate $\tau_c$ for that coarse-graining iteration, for different $\tau$. Error bars are standard deviations over randomly selected contiguous quarters of the simulation.}
\end{figure}

\begin{figure}[H]
\includegraphics[width=0.47\textwidth]{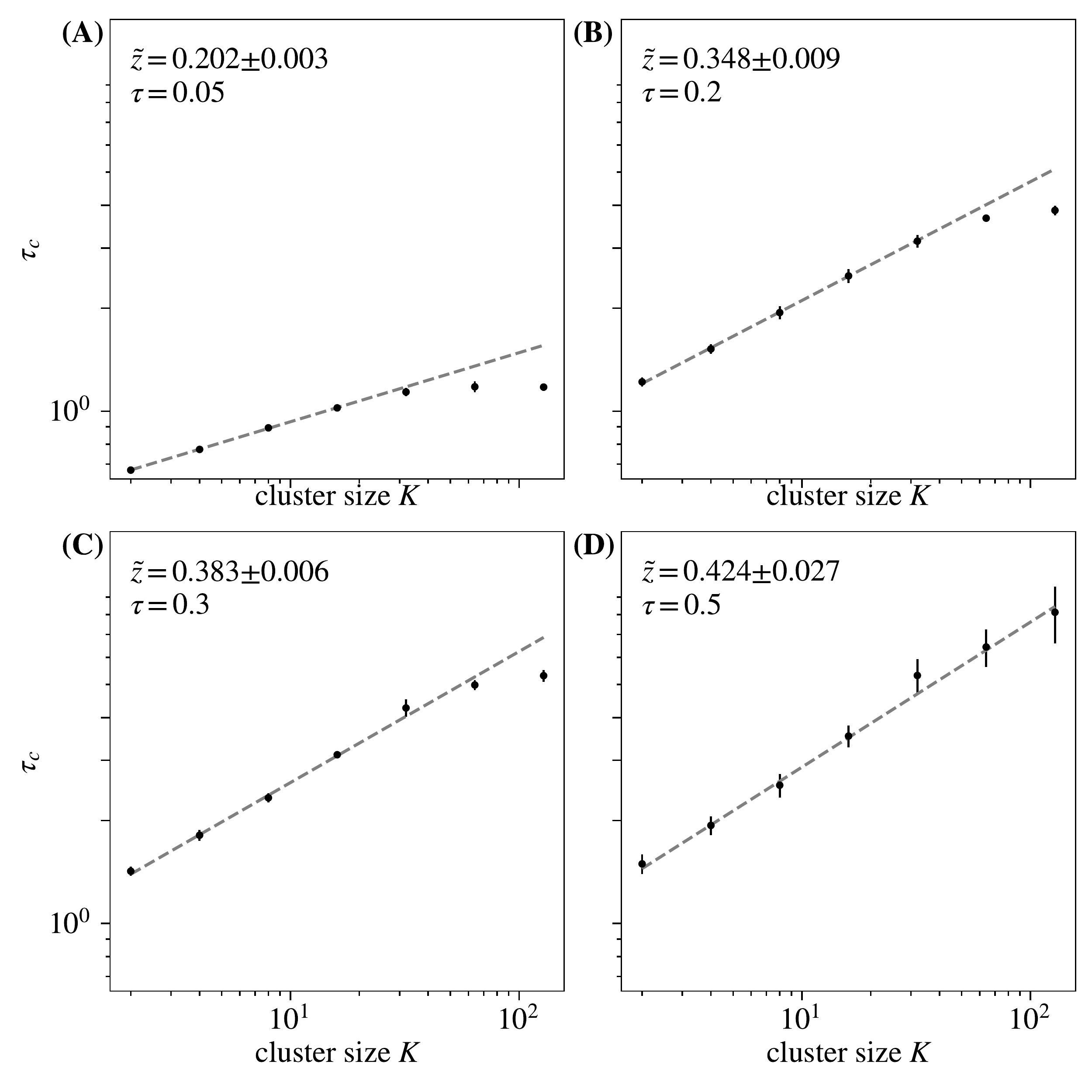}
\caption{\label{fig:fig_44} Time constants $\tau_c$ extracted from each curve in in Fig.~\ref{fig:fig_42}, and observe behavior obeying $\tau_c \propto K^{\tilde{z}}$ for roughly 1 decade. Varying $\tau$ does not have a significant effect on scaling quality, but results in different scaling exponents $\tilde{z}$, with larger $\tau$ resulting in larger $\tilde{z}$. Error bars are standard deviations over randomly selected contiguous quarters of the simulation.}
\end{figure}

\begin{figure}[H]
\includegraphics[width=0.47\textwidth]{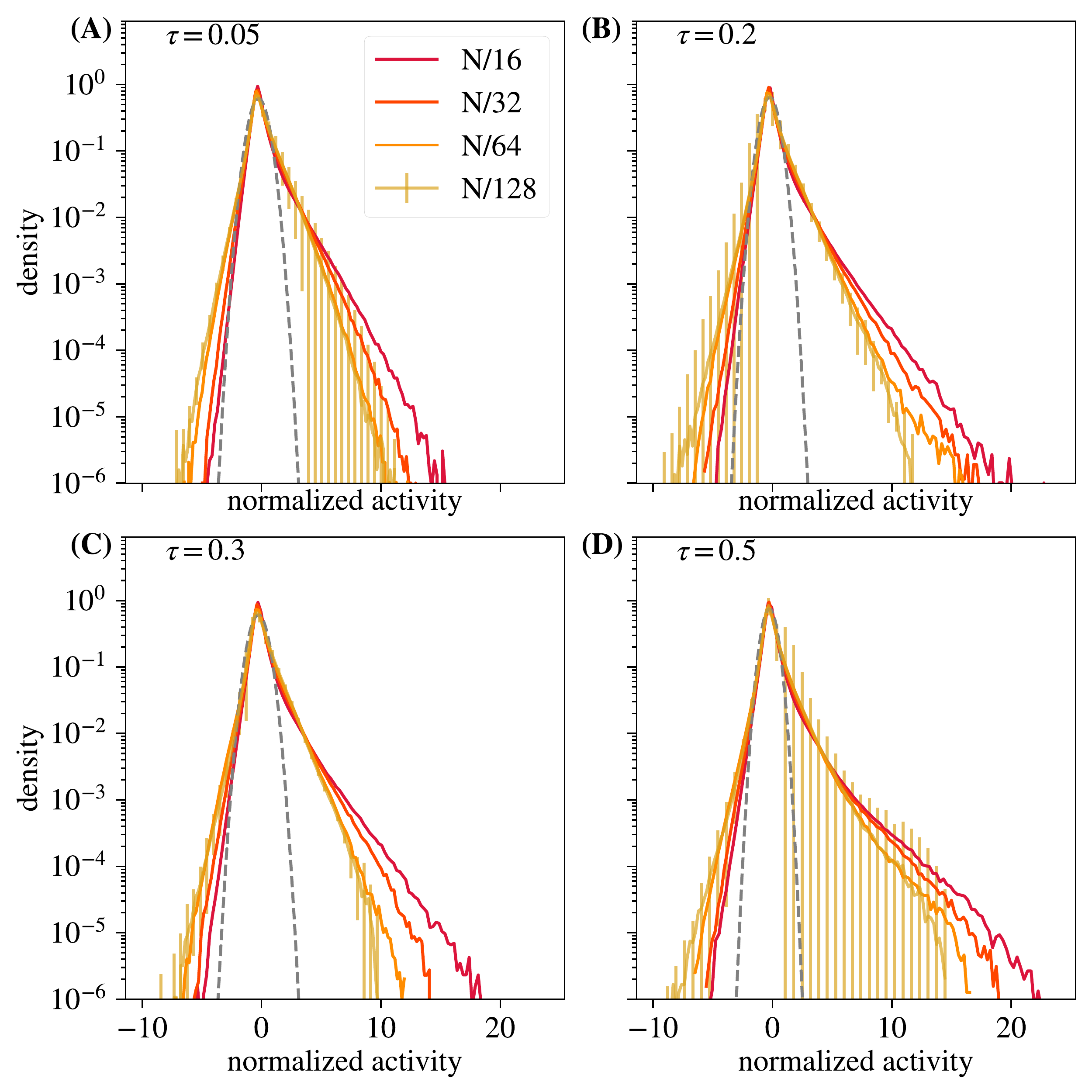}
\caption{\label{fig:fig_45} Distribution of coarse-grained variables for $k= N/16, N/32, N/64, N/128$ modes retained for different $\tau$. Varying $\tau$ has no significant impact on approach to a non-Gaussian fixed point. Error bars are standard deviations over randomly selected contiguous quarters of the simulation.}

\end{figure}

We find that varying $\tau$ changes the exponent $\tilde{z}$, with larger $\tau$ corresponding to larger $\tilde{z}$ and smaller $\tau$ corresponding to small $\tilde{z}$ (Fig.~\ref{fig:fig_38}). However, free energy scaling (Fig.~\ref{fig:fig_41}), variance scaling (Fig.~\ref{fig:fig_40}), eigenvalue scaling (Fig.~\ref{fig:fig_39}, and approach to a non-Gaussian fixed point (Fig.~\ref{fig:fig_45}) are not significantly affected by variation in $\tau_f$. Figure~\ref{fig:fig_42} through Fig.~\ref{fig:fig_45} suggest that dynamic scaling is robust to an increase in $\tau$, but that better quantitative agreement with the experimental $\tilde{z}$ is achieved with smaller $\tau$. 
\subsubsection{Varying the multiplier $\eta$}

\begin{figure}[H]
\includegraphics[width=0.47\textwidth]{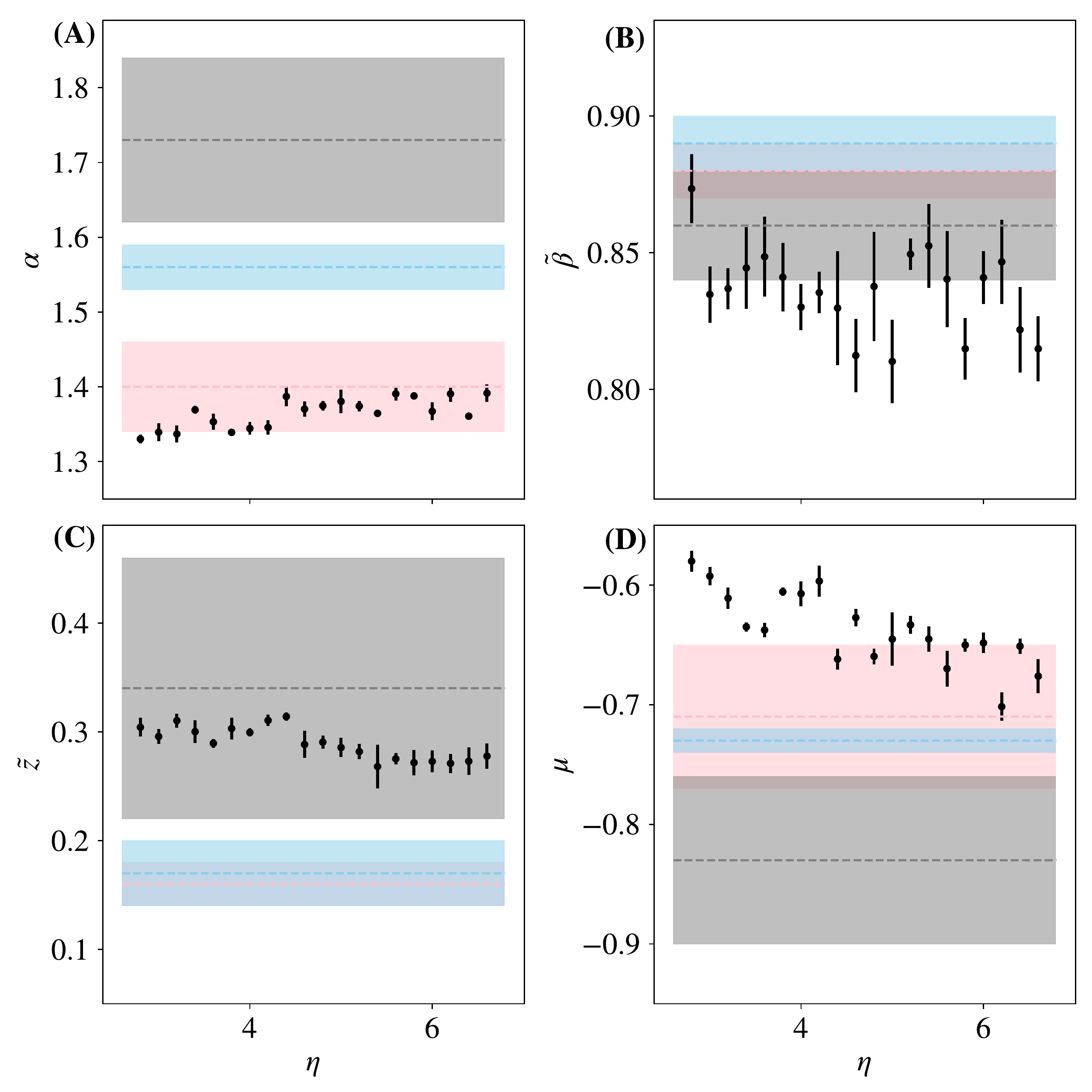}
\caption{\label{fig:fig_46} Each critical exponent, $\alpha, \tilde{\beta}, \tilde{z}, \mu$ vs multiplier $\eta$. Results from \cite{Meshulam2018} marked and shaded in gray, pink, and blue. Error bars are standard deviations over randomly selected contiguous quarters of the simulation.}
\end{figure}

We perform simulations with the default parameters sweeping over values of $\eta$, which is an overall multiplier for the ``energy'' (Eq.~\ref{eq:energy}). In Fig.~\ref{fig:fig_46}, we note that our simulations include a regime which quantitatively matches experimental results \cite{Meshulam2018}. 

\begin{figure}[H]
\includegraphics[width=0.47\textwidth]{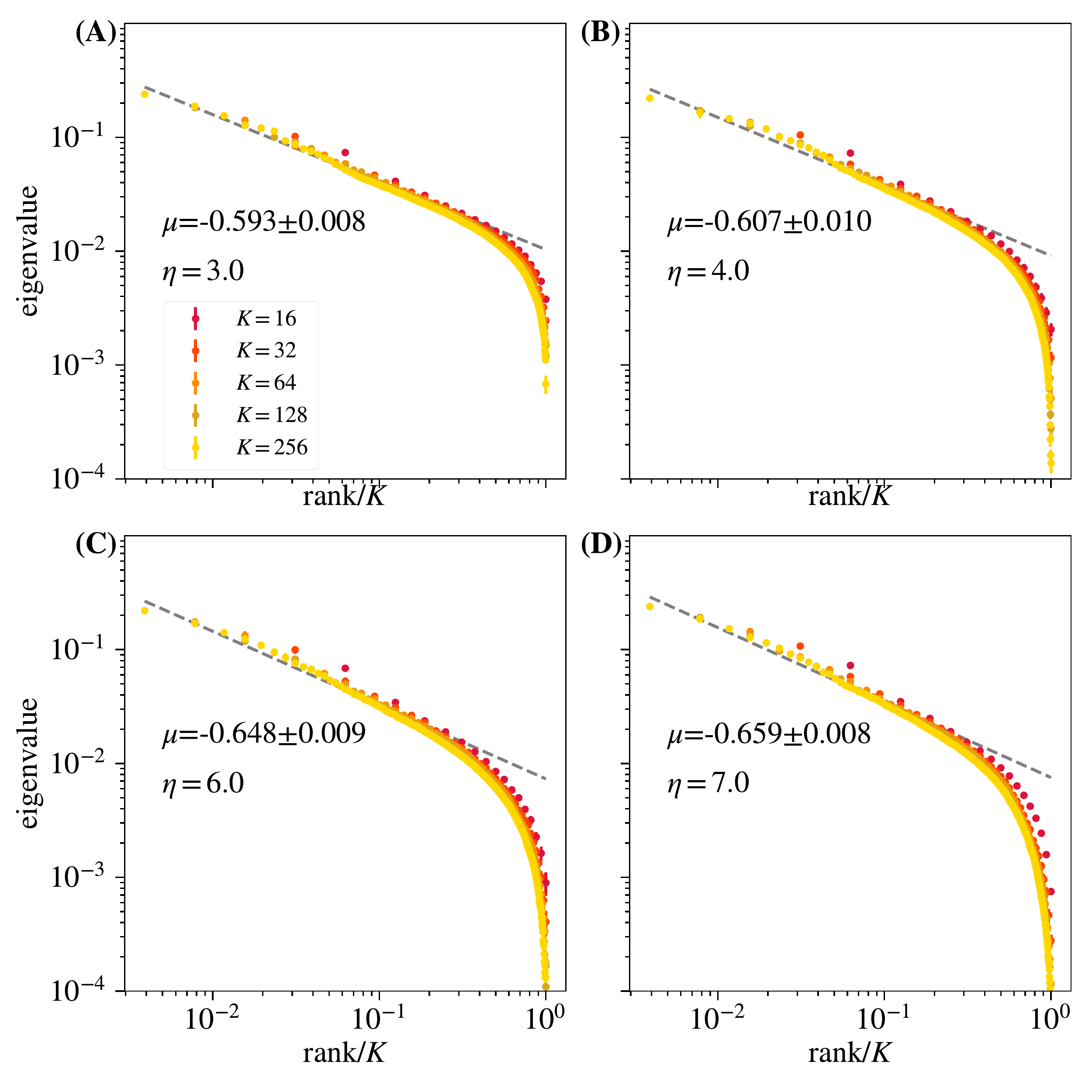}
\caption{\label{fig:fig_47} Average eigenvalue spectrum of cluster covariance for cluster sizes $K=32,64,128$ for different $\eta$. Varying $\eta$ does not appear to impact quality of eigenvalue scaling. Error bars are standard deviations over randomly selected contiguous quarters of the simulation.}
\end{figure}

\begin{figure}[H]
\includegraphics[width=0.47\textwidth]{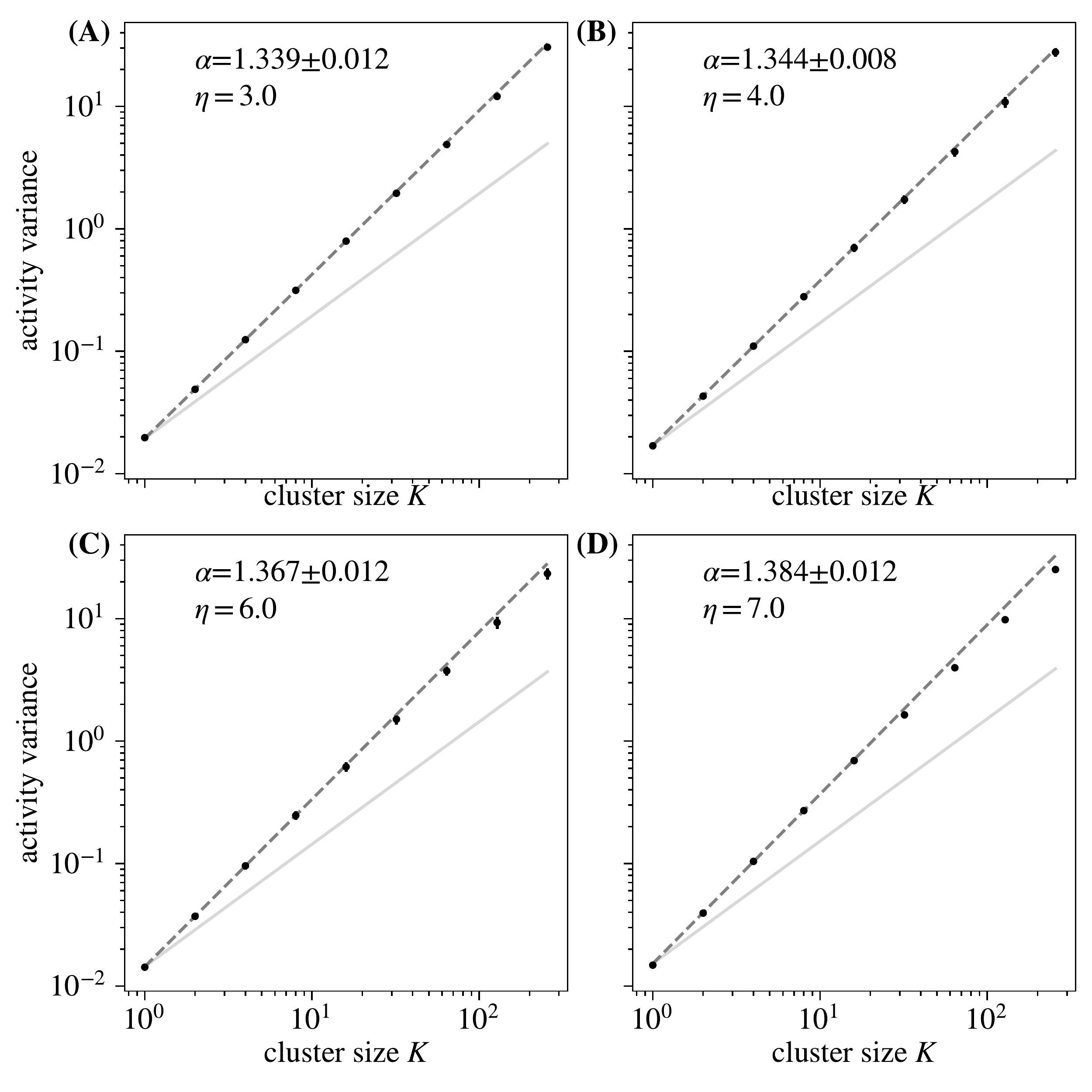}
\caption{\label{fig:fig_48}Activity variance over coarse-grained variables at each coarse-graining iteration for different $\eta$. Varying $\eta$ does not impact activity variance scaling. Error bars are standard deviations over randomly selected contiguous quarters of the simulation.}
\end{figure}

\begin{figure}[H]
\includegraphics[width=0.47\textwidth]{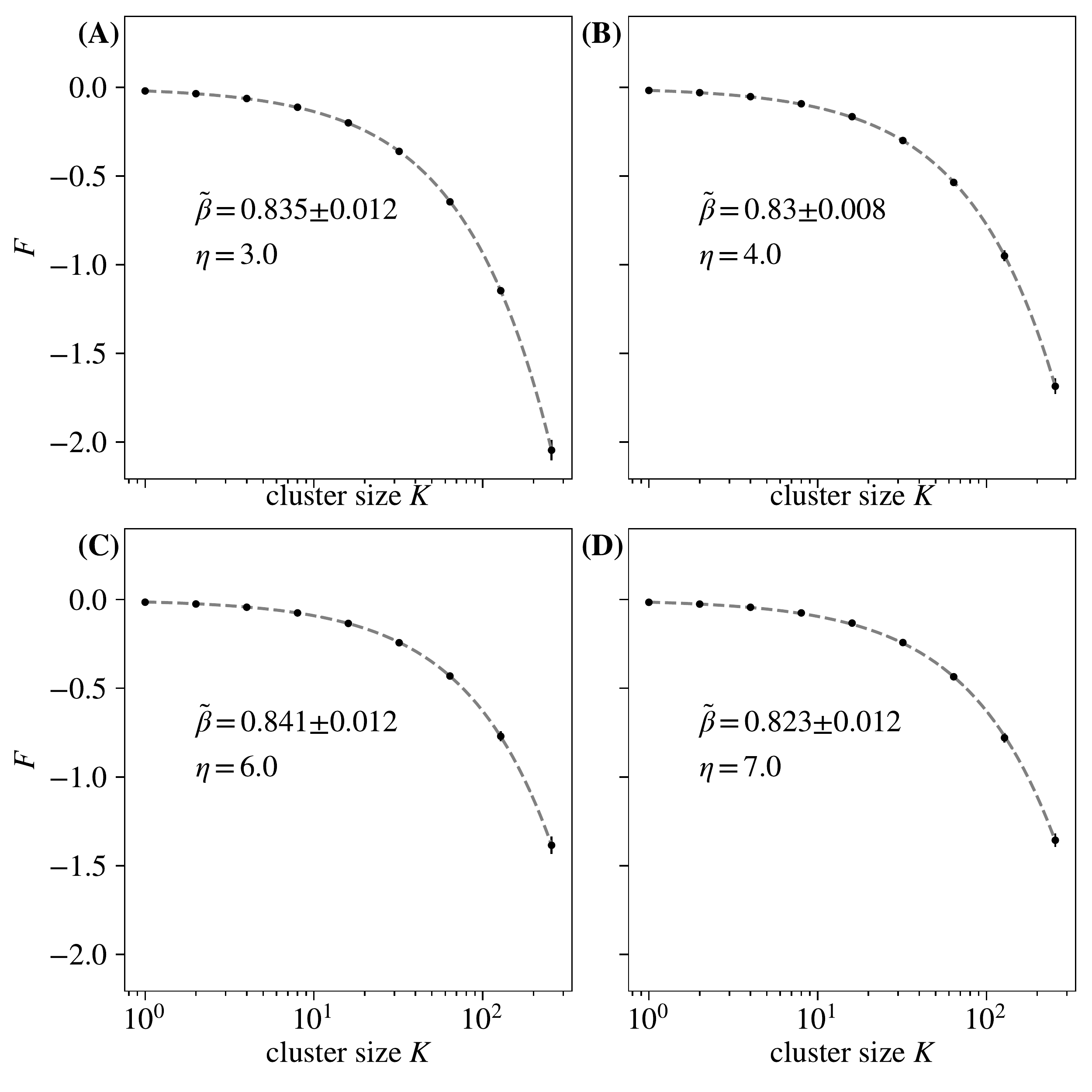}
\caption{\label{fig:fig_49} Average free energy at each coarse-graining iteration for different $\eta$. Varying $\eta$ does not affect quality of free energy scaling. Error bars are standard deviations over randomly selected contiguous quarters of the simulation.}
\end{figure}

\begin{figure}[H]
\includegraphics[width=0.47\textwidth]{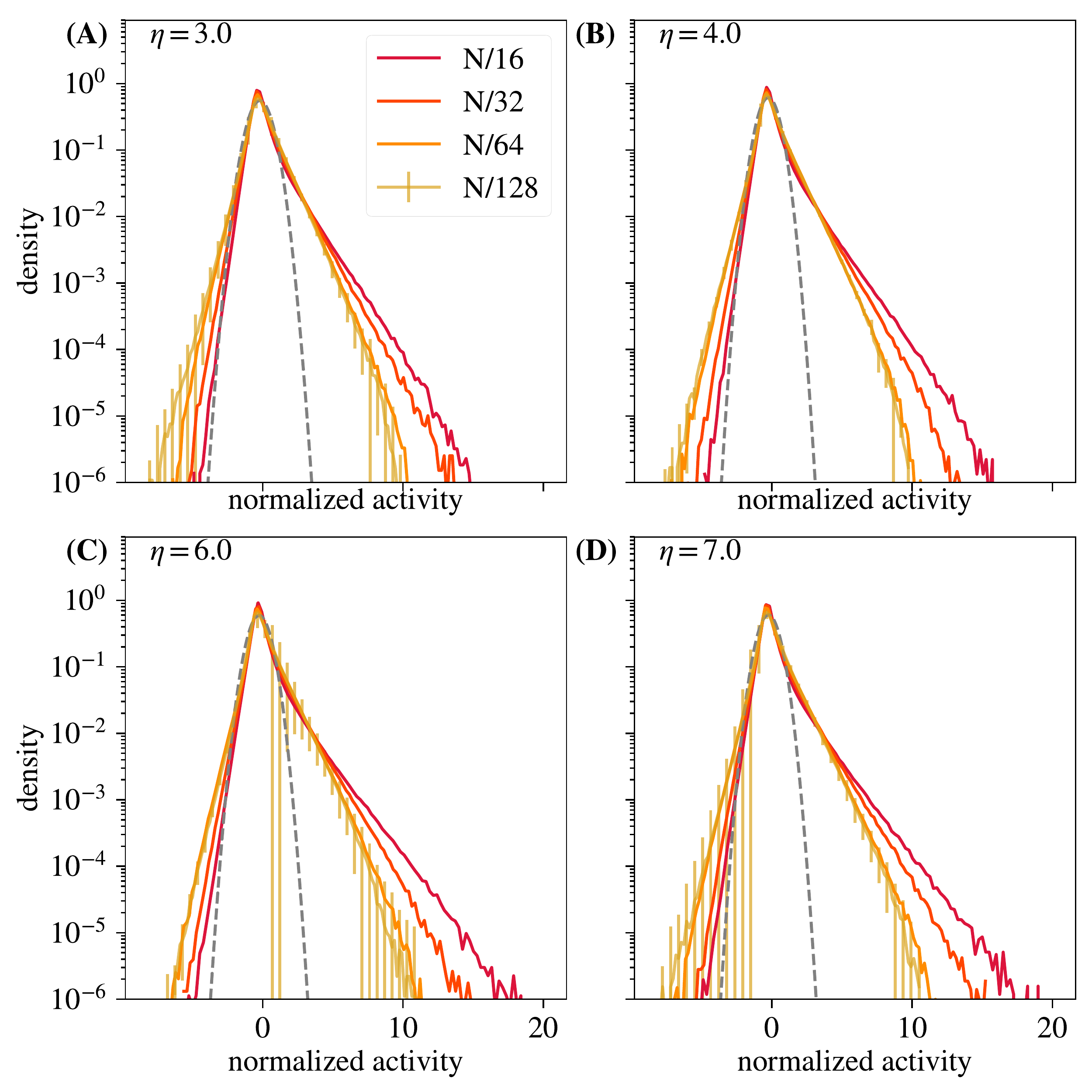}
\caption{\label{fig:fig_53} Distribution of coarse-grained variables for $k= N/16, N/32, N/64, N/128$ retained for different $\eta$. Approach to a non-Gaussian fixed point is not affected by varying $\eta$. Error bars are standard deviations over randomly selected contiguous quarters of the simulation.}
\end{figure}

Free energy scaling (Fig.~\ref{fig:fig_49}), variance scaling (Fig.~\ref{fig:fig_48}), approach to a non-Gaussian fixed point (Fig.~\ref{fig:fig_53}), and dynamic scaling are not significantly affected by variation in $\eta$. The quality of scaling of eigenvalues (Fig.~\ref{fig:fig_47}) is high across all values of $\eta$, although the scaling exponent $\mu$ decreases with $\eta$. Thus, adjusting $\eta$ has little effect on the quality of scaling. 

\subsubsection{Varying the latent fields multiplier $\phi$}

\begin{figure}[H]
\includegraphics[width=0.47\textwidth]{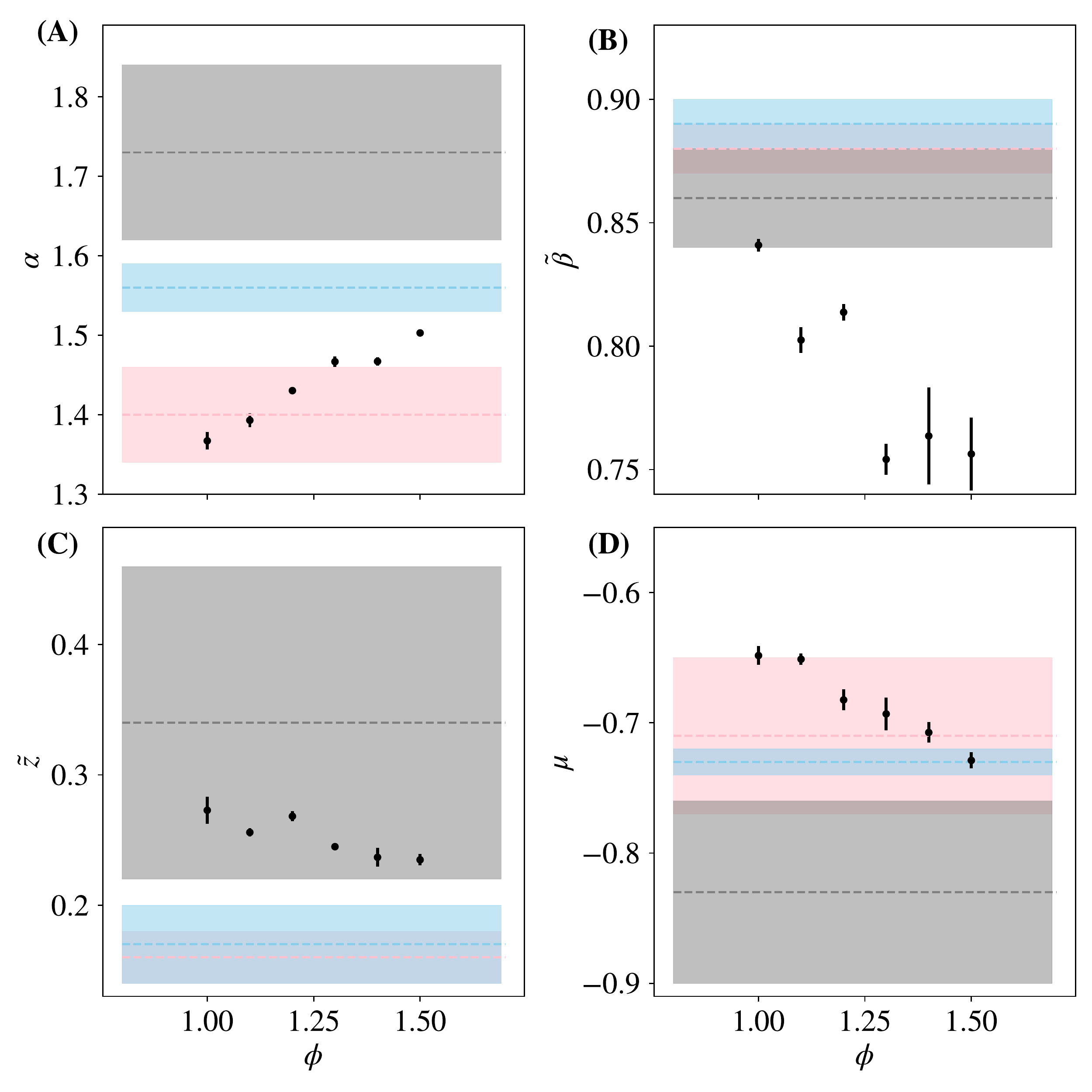}
\caption{\label{fig:fig_55} Each critical exponent, $\alpha, \tilde{\beta}, \tilde{z}, \mu$ vs latent field multiplier $\phi$. Results from \cite{Meshulam2018} marked and shaded in gray, pink, and blue. Error bars are standard deviations over randomly selected contiguous quarters of the simulation.}
\end{figure}

Finally, we perform simulations sweeping over values of $\phi$, which multiplies the latent field term in the energy (Eq.~\ref{eq:energy}). We vary $\phi$ from $0.8$ to $1.5$, with all other parameters fixed to default values. In Fig.~\ref{fig:fig_55}, we note that our simulations include a regime which quantitatively matches experimental results \cite{Meshulam2018}. 

\begin{figure*}
\includegraphics[width=1.0\textwidth]{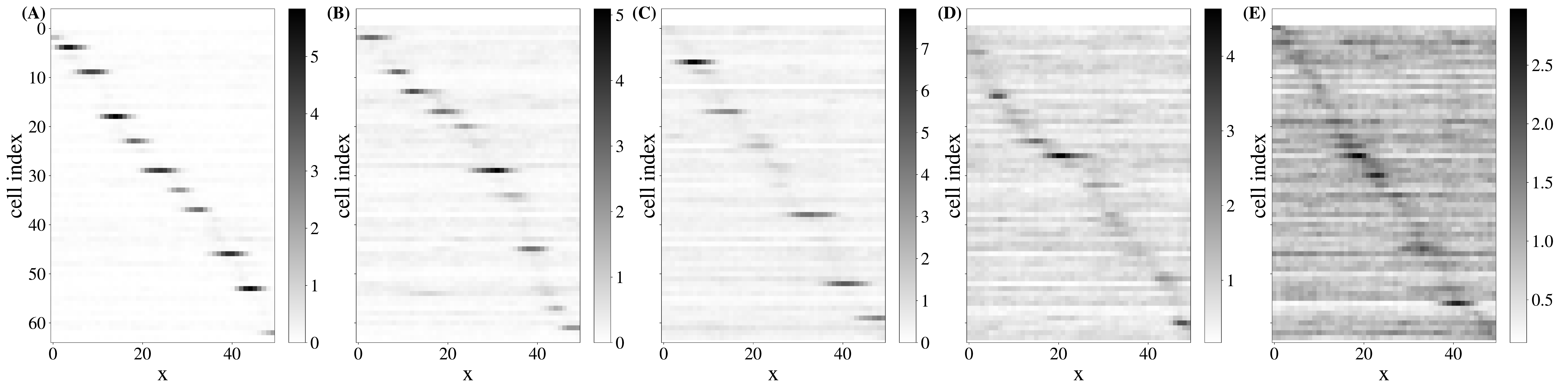}
\caption{\label{fig:fig_54} Average activity at spatial location $x$ for each neuron at coarse-graining step 4. Simulation parameters are those tabulated in Tbl.~\ref{tab:tab_1} with $\phi=0.8$ \textbf{(A)}, $\phi=1.0$ \textbf{(B)}, $\phi=1.2$ \textbf{(C)}, $\phi=1.4$ \textbf{(D)}, and $\phi=1.5$ \textbf{(E)}. Increasing the latent field multiplier $\phi$ decreases the relative strength of place cells compared to cells only coupled to latent fields. }
\end{figure*}

In Fig.~\ref{fig:fig_54} we show that the presence of place cells remains stable over coarse-graining over the full range $\phi \in [0.8,1.5]$, but as $\phi$ increases, the relative strength of place cells compared to background activity is decreased. 

\begin{figure}[H]
\includegraphics[width=0.47\textwidth]{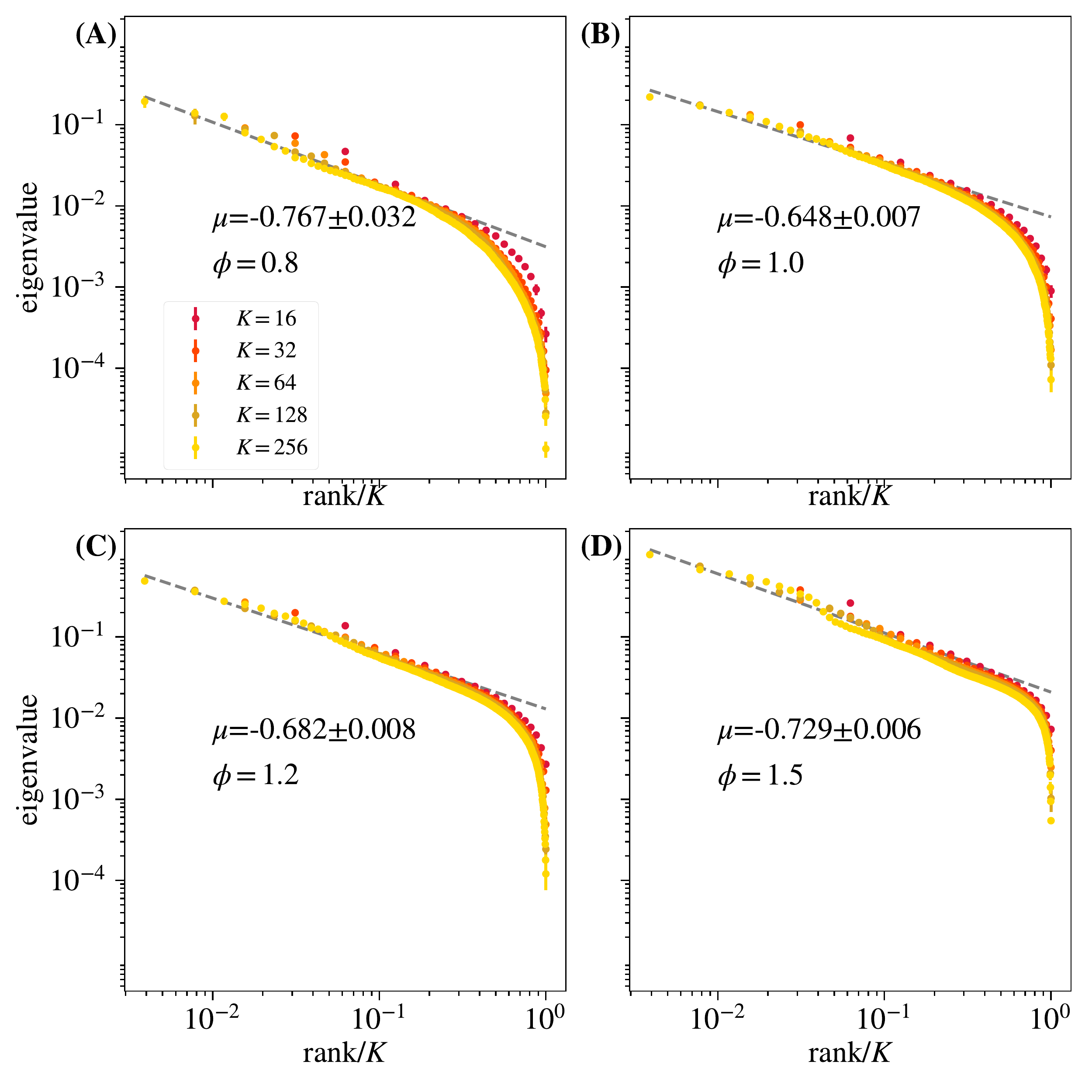}
\caption{\label{fig:fig_56} Average eigenvalue spectrum of cluster covariance for cluster sizes $K=32,64,128$ for different $\phi$. Note that quality of eigenvalue collapse and scaling and value of the exponent $\mu$ is unaffected by varying $\phi$. Error bars are standard deviations over randomly selected contiguous quarters of the simulation.}
\end{figure}

\begin{figure}[H]
\includegraphics[width=0.47\textwidth]{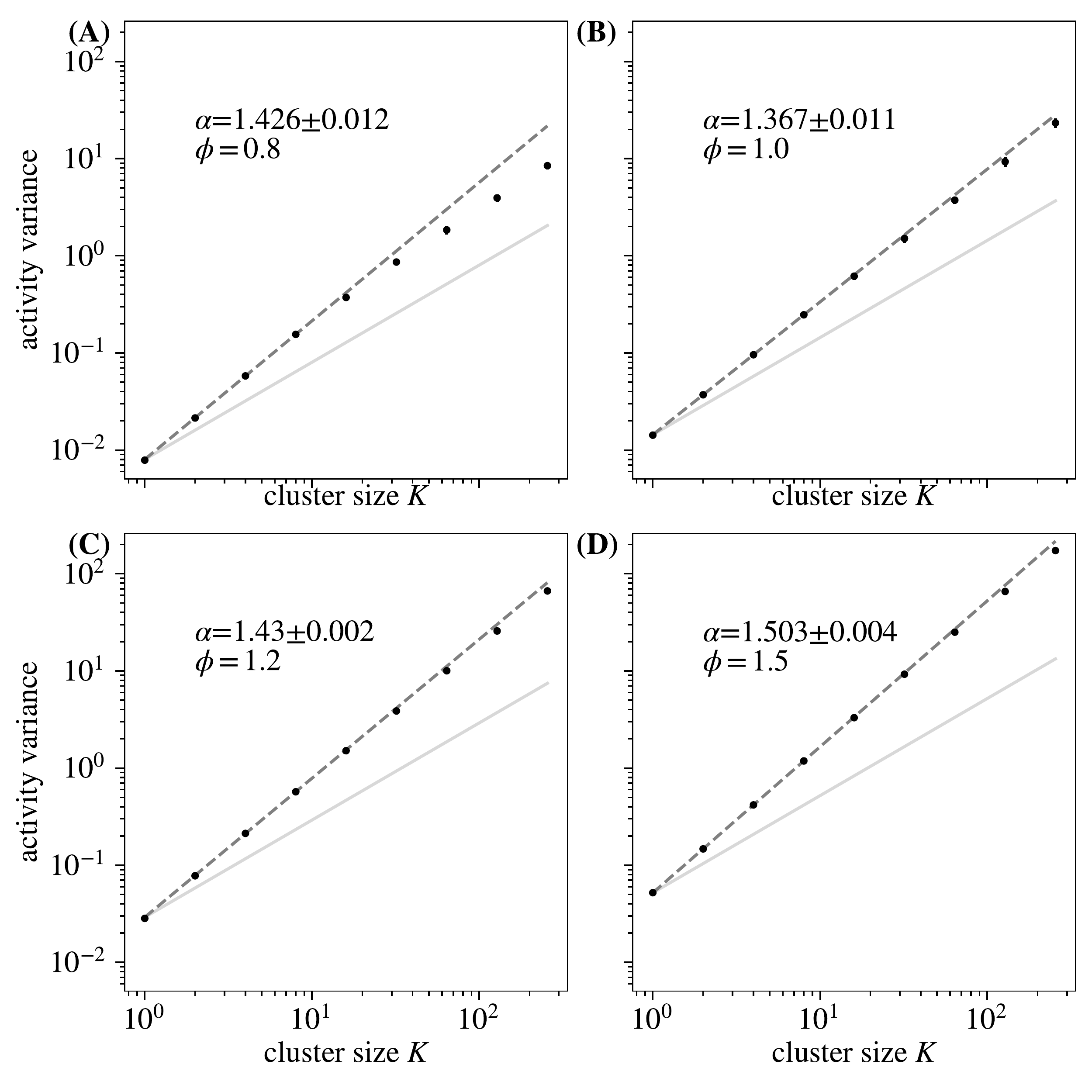}
\caption{\label{fig:fig_57} Activity variance over coarse-grained variables at each coarse-graining iteration for different $\phi$. Note that for $\phi<1.0$, variance scaling is damaged. Error bars are standard deviations over randomly selected contiguous quarters of the simulation.}
\end{figure}

\begin{figure}[H]
\includegraphics[width=0.47\textwidth]{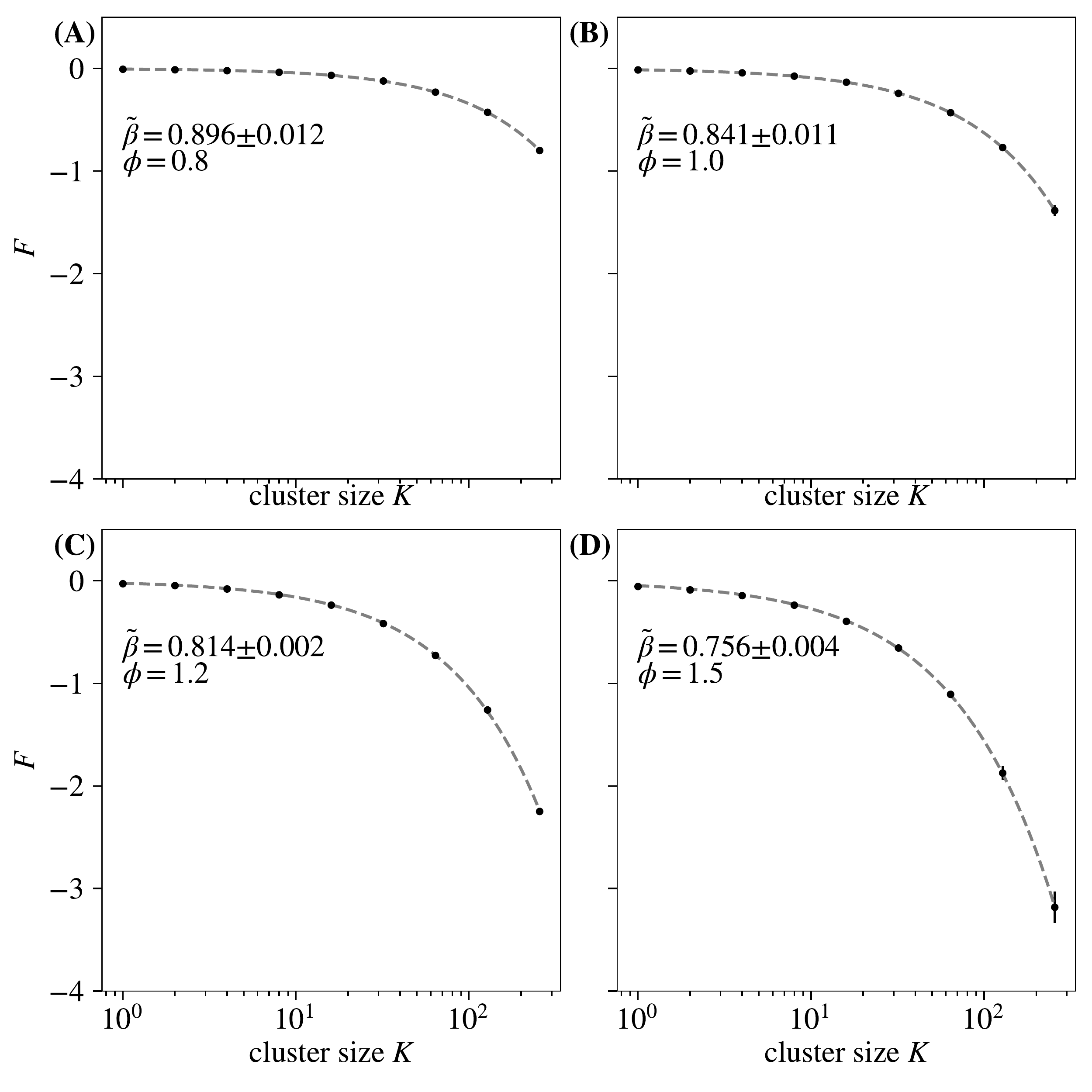}
\caption{\label{fig:fig_58} Average free energy at each coarse-graining iteration for different $\phi$. Quality of free energy scaling is unaffected by varying $\phi$. Error bars are standard deviations over randomly selected contiguous quarters of the simulation.}
\end{figure}

\begin{figure}[H]
\includegraphics[width=0.47\textwidth]{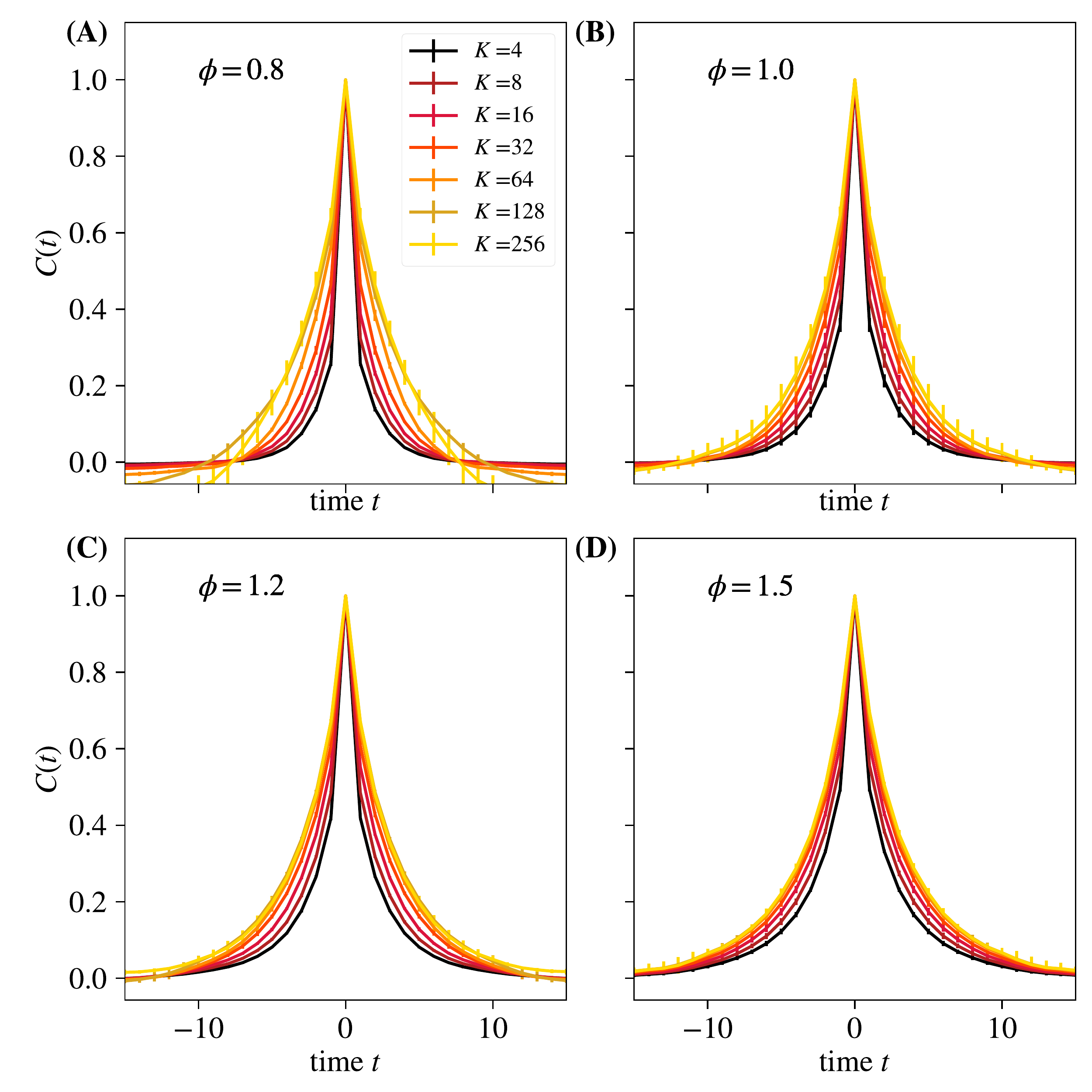}
\caption{\label{fig:fig_59} Average autocorrelation function for cluster sizes $K=2, 4, ..., 256$ as a function of time, for different values of $\phi$. Error bars are standard deviations over randomly selected contiguous quarters of the simulation.}
\end{figure}

\begin{figure}[H]
\includegraphics[width=0.47\textwidth]{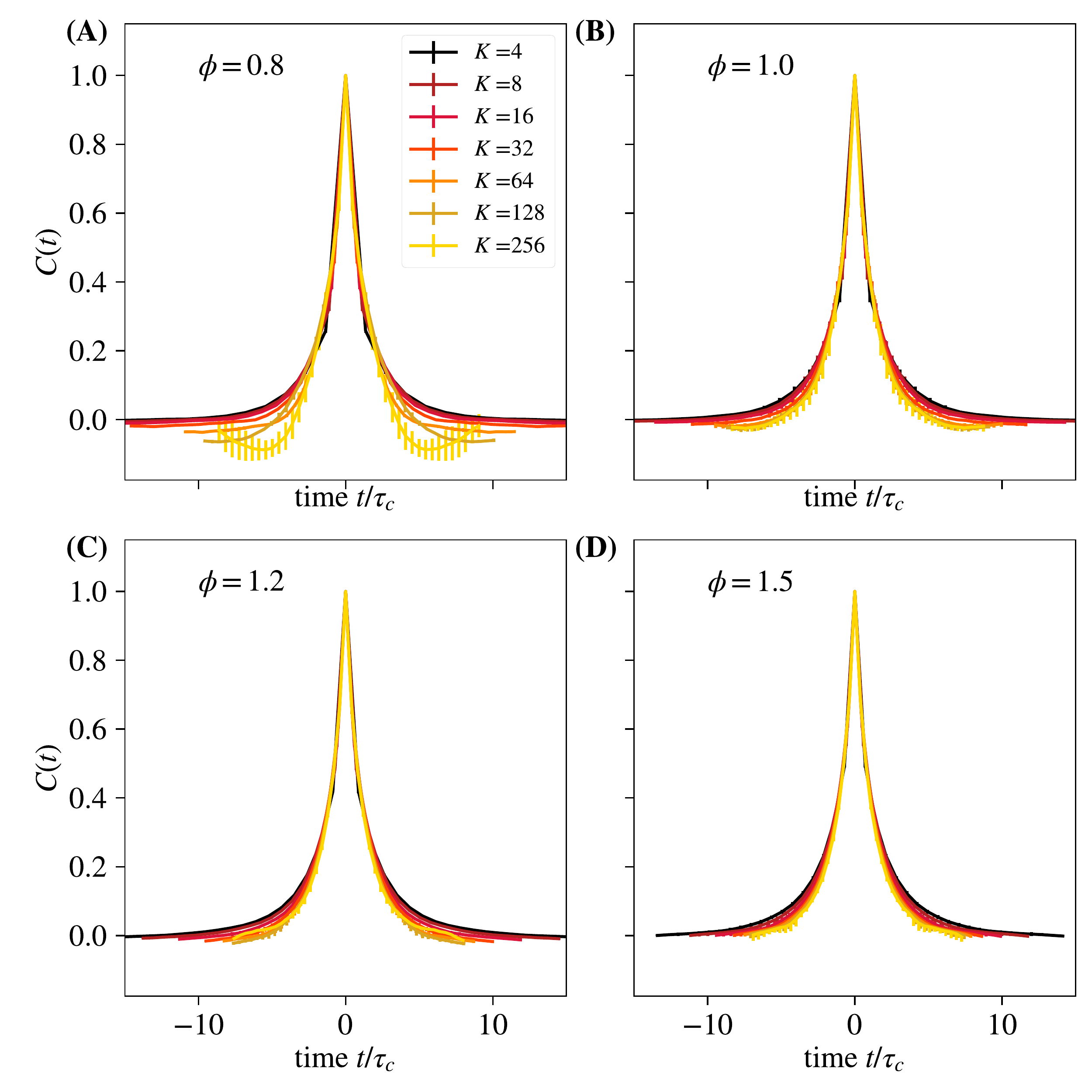}
\caption{\label{fig:fig_60} Average autocorrelation function for cluster sizes $K=2, 4, ..., 256$ where time is rescaled by the appropriate $\tau_c$ for that coarse-graining iteration, for different $\phi$. Error bars are standard deviations over randomly selected contiguous quarters of the simulation.}
\end{figure}

\begin{figure}
\includegraphics[width=0.47\textwidth]{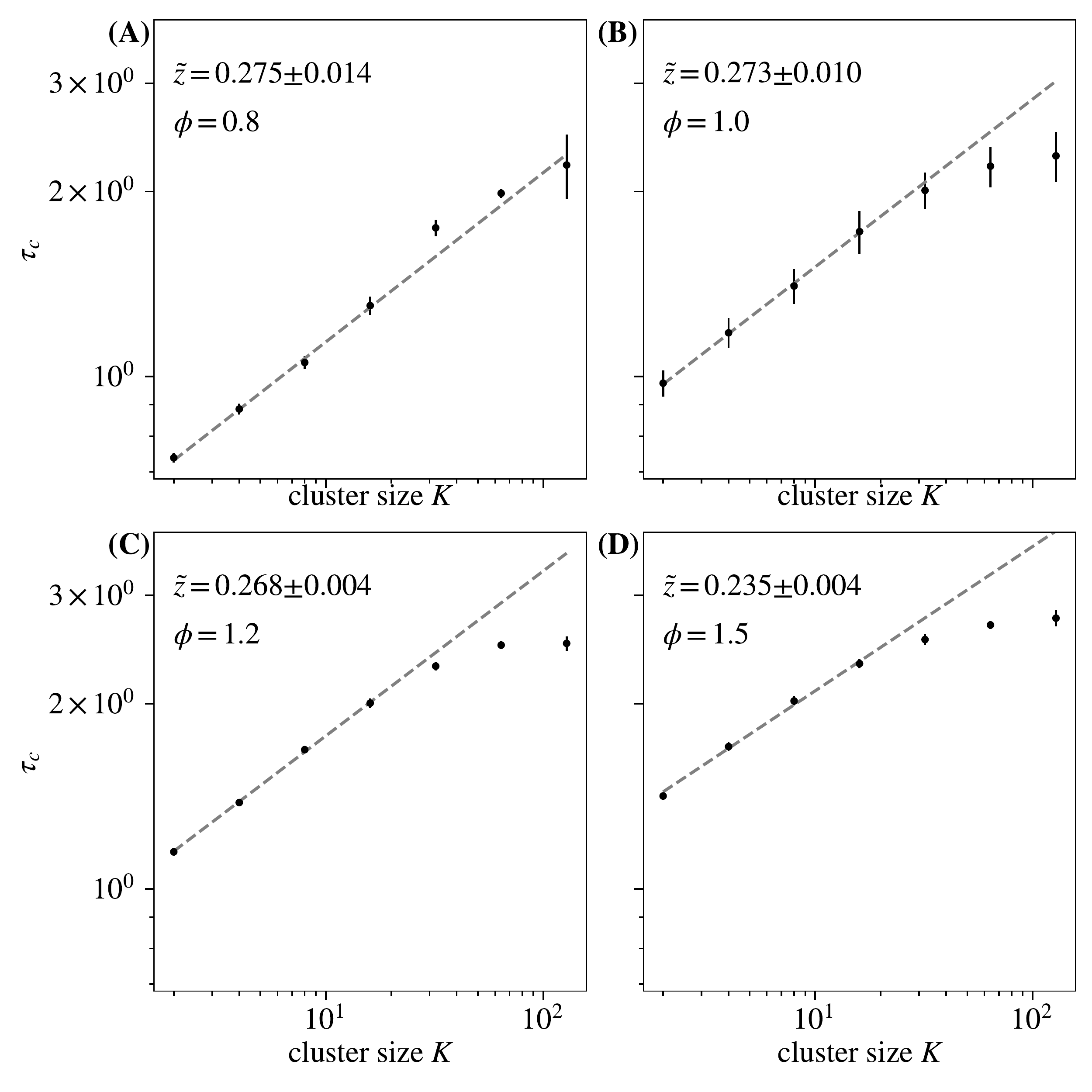}
\caption{\label{fig:fig_61} Time constants $\tau_c$ extracted from each curve in in FIG~\ref{fig:fig_59}, and observe behavior obeying $\tau_c \propto K^{\tilde{z}}$ for roughly 1 decade. Error bars are standard deviations over randomly selected contiguous quarters of the simulation.}

\end{figure}

\begin{figure}[H]
\includegraphics[width=0.47\textwidth]{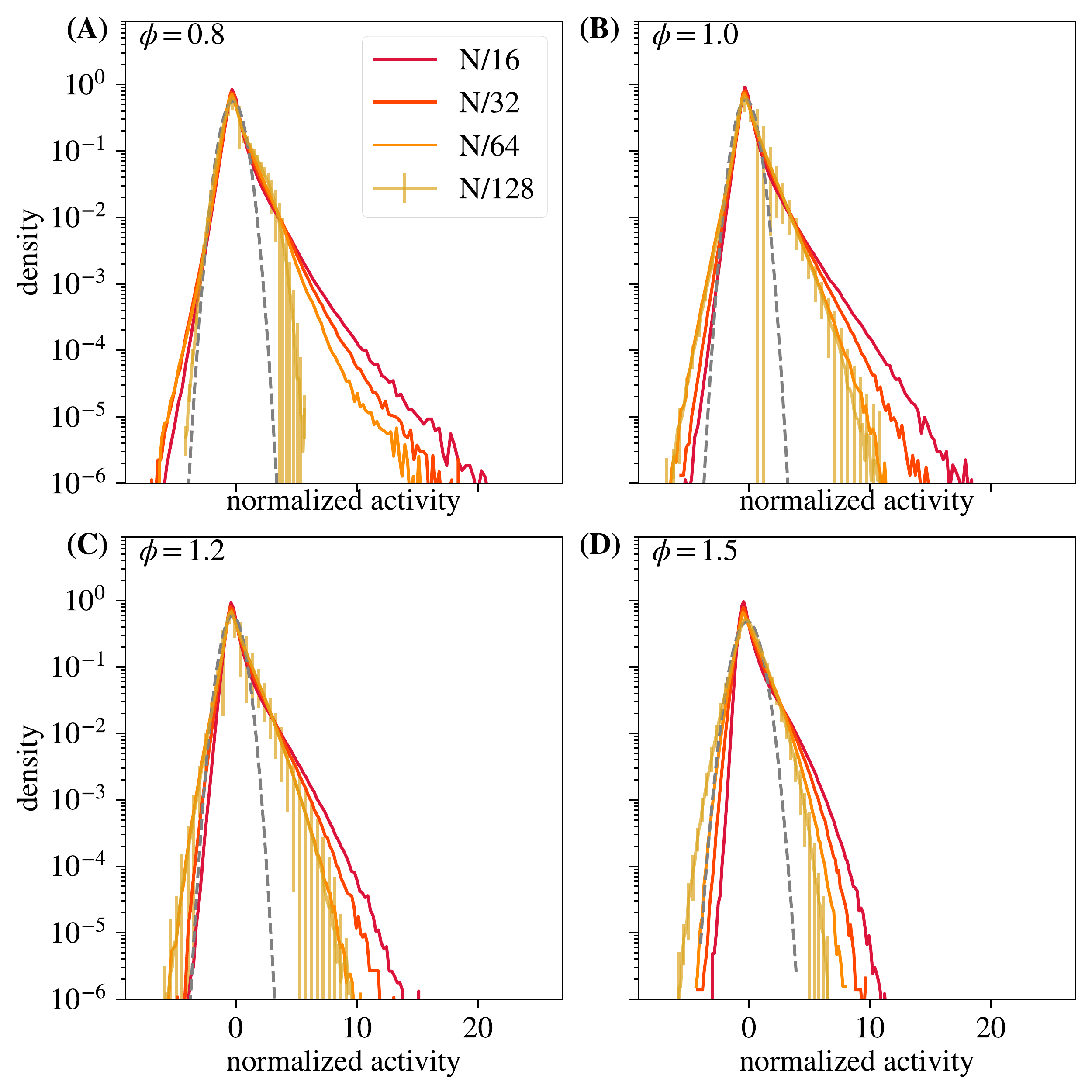}
\caption{\label{fig:phimom} Distribution of coarse-grained variables for $k= N/16, N/32, N/64, N/128$ modes retained for different $\phi$. Note convergence to non-Gaussian fixed point regardless of value of $\phi$. Error bars are standard deviations over randomly selected contiguous quarters of the simulation.}
\end{figure}

We find that varying $\phi$ does not significantly affect the quality of scaling for eigenvalues (Fig.~\ref{fig:fig_56}, free energy (Fig.~\ref{fig:fig_58}), or correlation time (Fig.~\ref{fig:fig_61}), and it does not affect the approach to a non-Gaussian fixed point (Fig.~\ref{fig:phimom}). However, setting $\phi<1.0$ creates slight deviation from power law scaling of activity variance at large cluster size (Fig.~\ref{fig:fig_57}). In summary, Fig.~\ref{fig:fig_54} through Fig.~\ref{fig:phimom} show that weak latent fields are deleterious to scaling behavior.


\clearpage

\clearpage

\clearpage


\clearpage


\clearpage

\begin{figure}[H]
\includegraphics[width=0.47\textwidth]{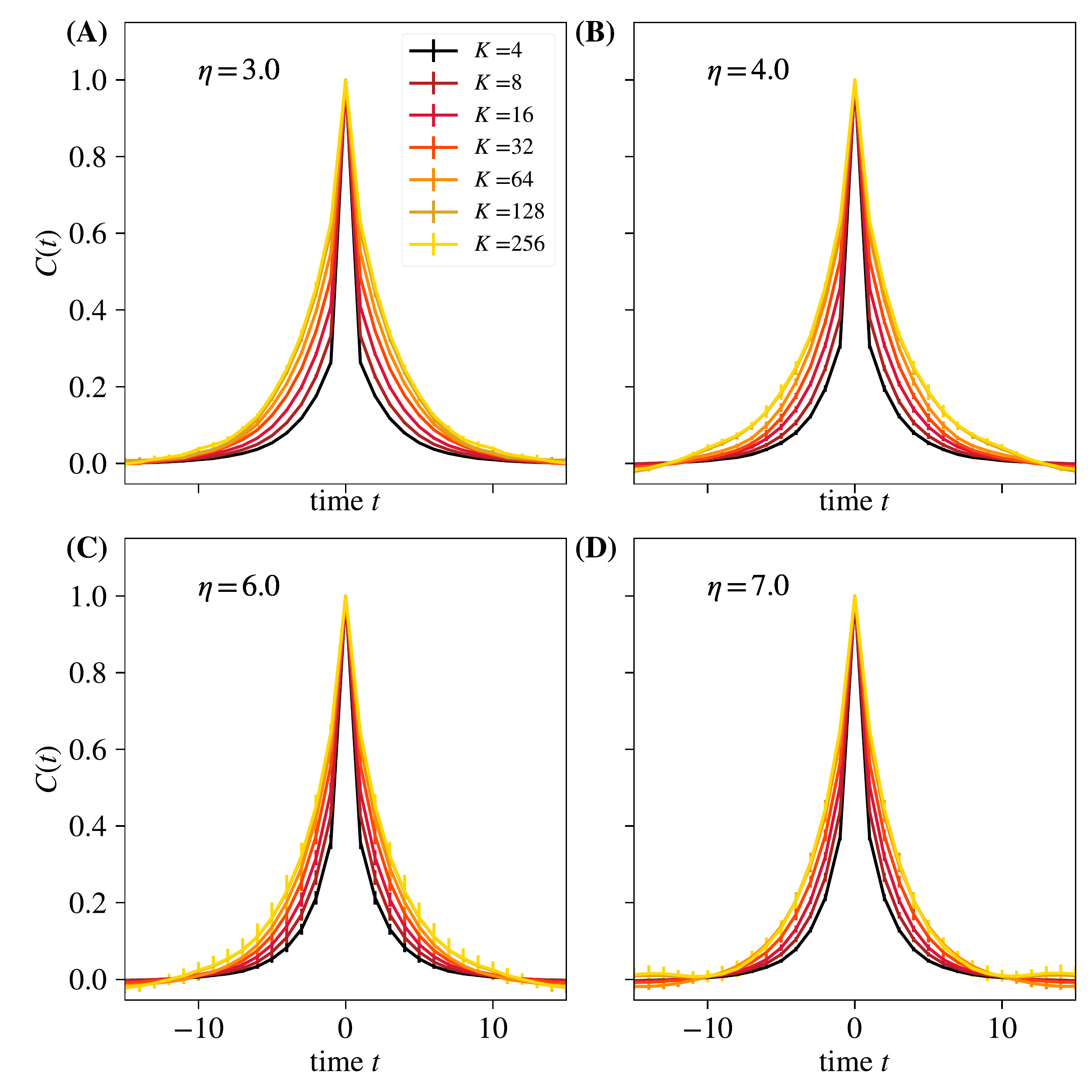}
\caption{\label{fig:fig_50} Average autocorrelation function for cluster sizes $K=2, 4, ..., 256$ as a function of time, for different $\eta$. Error bars are standard deviations over randomly selected contiguous quarters of the simulation.}
\end{figure}
\begin{figure}[H]
\includegraphics[width=0.47\textwidth]{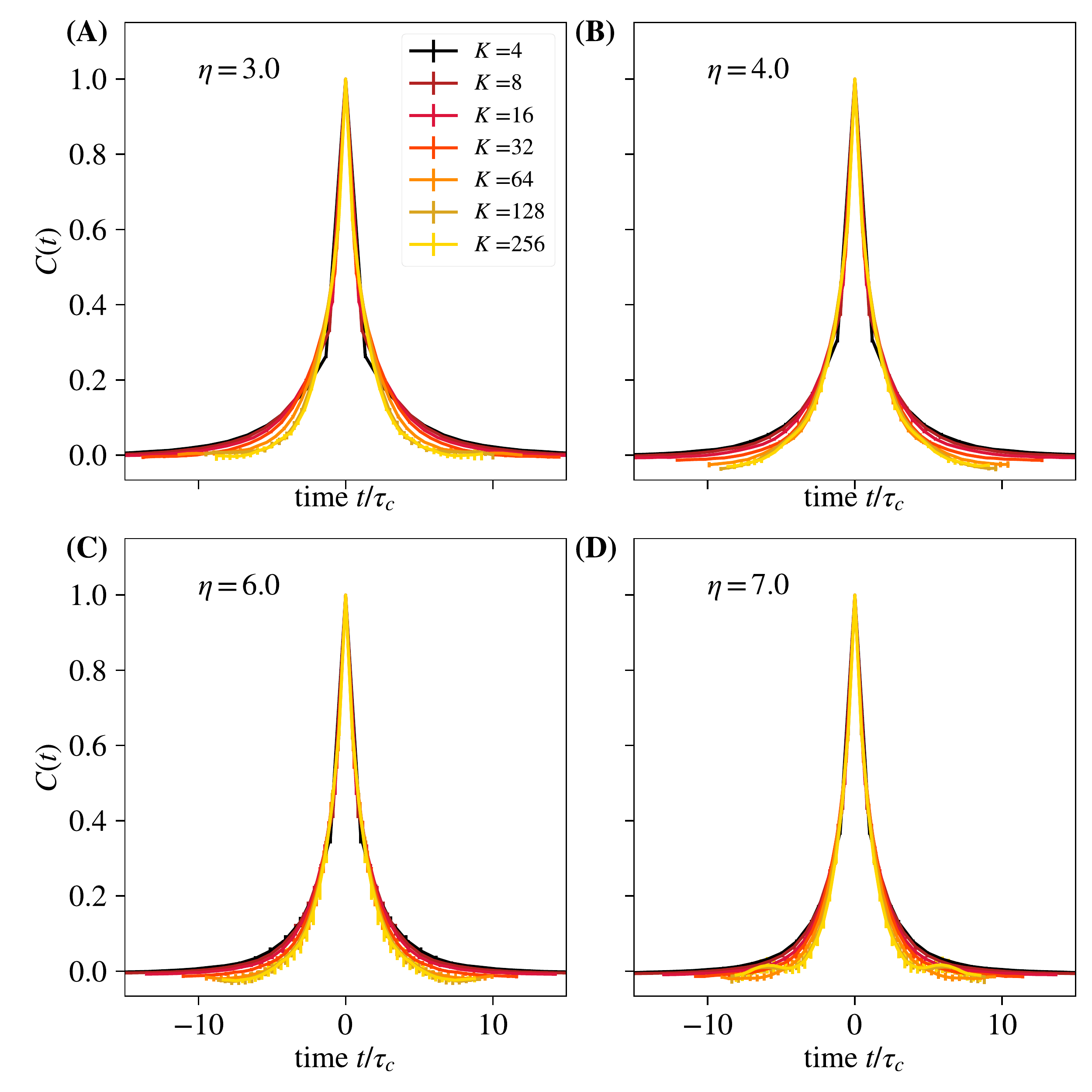}
\caption{\label{fig:fig_51} Average autocorrelation function for cluster sizes $K=2, 4, ..., 256$ where time is rescaled by the appropriate $\tau_c$ for that coarse-graining iteration, for different $\eta$. Error bars are standard deviations over randomly selected contiguous quarters of the simulation.}
\end{figure}
\begin{figure}[H]
\includegraphics[width=0.47\textwidth]{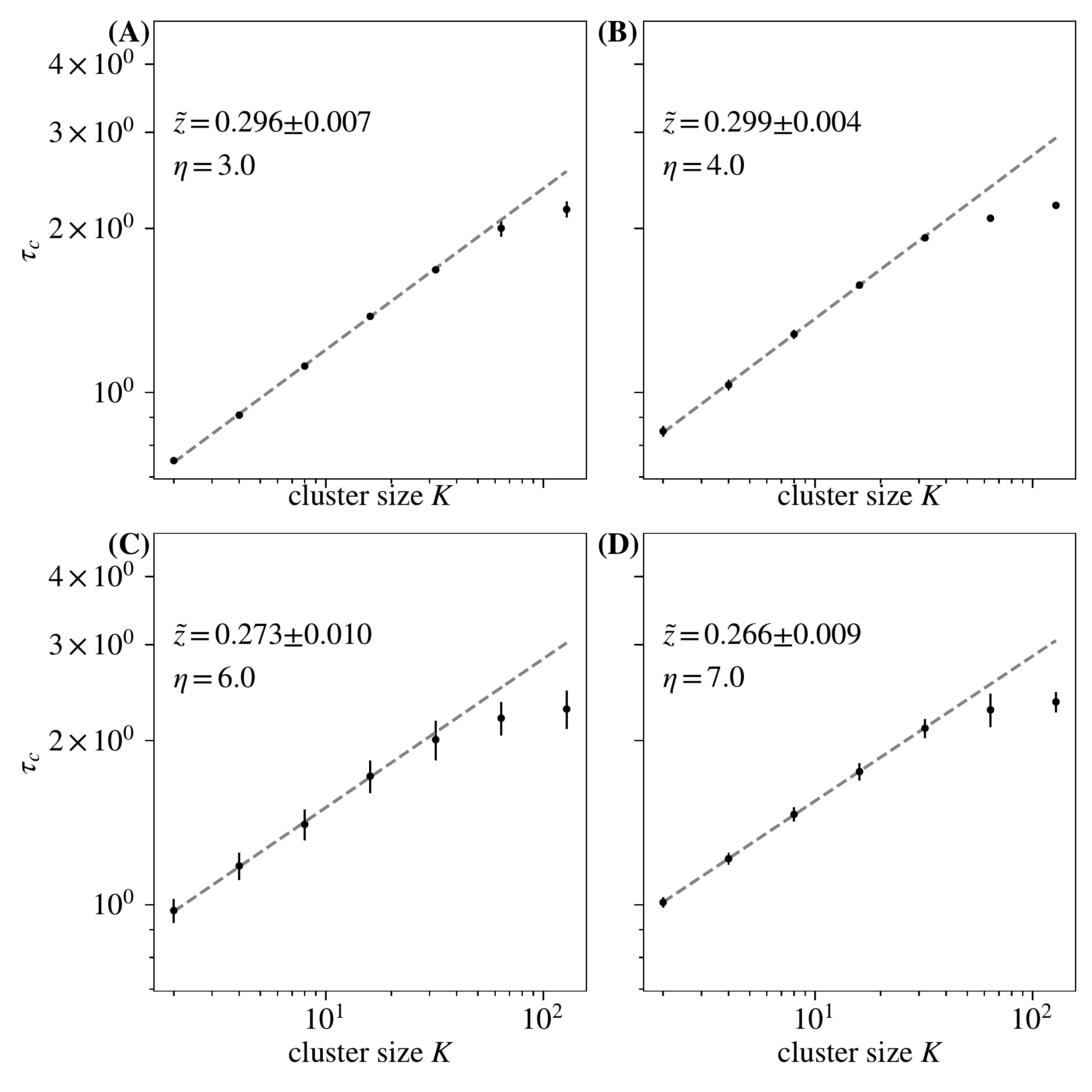}
\caption{\label{fig:fig_52} Time constants $\tau_c$ extracted from each curve in in Fig.~\ref{fig:fig_50}, and observe behavior obeying $\tau_c \propto K^{\tilde{z}}$ for roughly 1 decade. Varying $\eta$ does not significantly affect the quality of temporal scaling or the value of exponent $\tilde{z}$. Error bars are standard deviations over randomly selected contiguous quarters of the simulation.}
\end{figure}



\end{document}